\pdfoutput=1
\newcommand*{\ATLASLATEXPATH}{latex/}

\documentclass[cernpreprint, texlive=2016, UKenglish]{\ATLASLATEXPATH atlasdoc}

\usepackage{\ATLASLATEXPATH atlaspackage}
\usepackage{\ATLASLATEXPATH atlasbiblatex}
\usepackage{\ATLASLATEXPATH atlasphysics}

\graphicspath{{logos/}{figures/}}

\usepackage{note-defs}

\AtlasTitle{Electron and photon energy calibration with the ATLAS detector using 2015--2016 LHC proton--proton collision data}

\author{The ATLAS Collaboration}

\AtlasDOI{10.1088/1748-0221/14/03/P03017}
\AtlasJournalRef{JINST 14 (2019) P03017}

\AtlasRefCode{PERF-2017-03}

\PreprintIdNumber{CERN-EP-2018-296}

\AtlasAbstract{%
This paper presents the electron and photon energy calibration obtained with the ATLAS detector
using about 36 fb$^{-1}$ of LHC proton--proton collision data recorded at $\sqrt{s}=13$~\TeV\ in 2015 and 2016.
The different calibration steps applied to the data and the optimization of the reconstruction of
electron and photon energies are discussed. The absolute energy scale is set using a large sample
of $Z$ boson decays into electron--positron pairs. The systematic uncertainty in the energy scale
calibration varies between 0.03\% to 0.2\% in most of the detector acceptance for electrons with transverse momentum
close to 45~\GeV. For electrons with transverse momentum of 10~\GeV\ the typical uncertainty is
0.3\% to 0.8\% and it varies between 0.25\% and 1\% for photons with transverse momentum around 60~\GeV.
Validations of the energy calibration with $J/\psi \rightarrow e^+e^-$ decays and radiative
$Z$ boson decays are also presented.
}

\hypersetup{pdftitle={ATLAS document},pdfauthor={The ATLAS Collaboration}}

\usepackage{adjustbox}

\begin{document}

\maketitle

\tableofcontents

\section{Introduction}
\label{sec:intro}
A precise calibration of the energy measurement of electrons and photons is required for many analyses
performed at the CERN Large Hadron Collider (LHC), among which the studies of the Higgs
boson in the two-photon and four-lepton decay channels and precise studies of $W$ and $Z$ boson production
and properties.
This paper presents the calibration of the energy measurement of electrons and photons achieved with the
ATLAS detector using 36  fb$^{-1}$ of LHC proton$-$proton collision data collected in 2015 and 2016 at $\sqrt{s}=13$~\TeV. 

The calibration scheme comprises a simulation-based optimization of the energy resolution for electrons and photons, 
corrections accounting for differences between data and simulation,
the adjustment of the absolute energy scale using $Z$ boson decays into $e^+e^-$ pairs,
and the validation of the energy scale universality using $J/\psi$ decays  decays into $e^+e^-$ pairs 
and radiative $Z$ boson decays.
This strategy closely follows the procedure used for the final energy calibration applied to the data collected in 
2011 and 2012 (Run~1)~\cite{PERF-2013-05}, with updates to reflect the changes in data-taking and detector conditions.

This paper is organized as follows. Section~\ref{sec:detector} briefly describes the ATLAS detector and
the reconstruction of electron and photon candidates.
Section~\ref{sec:method} introduces the calibration procedure and the changes relative to the Run~1 calibration.
Section~\ref{sec:samples} gives a list
of the different data and simulated event samples used in these studies.
Section~\ref{sec:mva} explains how the simulated event samples are used to optimize the
estimate of electron and photon energies, as well as the expected resolutions of the energy measurements.
Section~\ref{sec:data_correction} describes the different corrections applied to the data.
Section~\ref{sec:alpha} discusses the extraction 
of the overall energy scale and resolution corrections between data and simulation
from $Z  \rightarrow ee$ decays. Section~\ref{sec:systematics}
describes the different systematic uncertainties affecting the energy scale and resolution. Finally
Section~\ref{sec:checks} presents the cross-checks performed using independent data samples.

\section{ATLAS detector, electron and photon reconstruction}
\label{sec:detector}
\subsection{The ATLAS detector} 
The ATLAS experiment~\cite{PERF-2007-01} at the LHC is a multipurpose particle detector
with a forward$-$backward symmetric cylindrical geometry and a near $4\pi$ coverage in 
solid angle.\footnote{
ATLAS uses a right-handed coordinate system with its origin at the nominal interaction point (IP) in the centre of the detector and the $z$-axis along the beam pipe. The $x$-axis points from the IP to the centre of the LHC ring, and the $y$-axis points upward. Cylindrical coordinates $(r,\phi)$ are used in the transverse plane, $\phi$ being the azimuthal angle around the $z$-axis. The pseudorapidity is defined in terms of the polar angle $\theta$ as $\eta=-\ln\tan(\theta/2)$. The transverse energy is defined as $\ET = E \sin \theta$.}

It consists of an inner tracking detector surrounded by a thin superconducting solenoid, electromagnetic and hadronic calorimeters,
and a muon spectrometer incorporating three large superconducting toroidal magnets with eight coils each.
The inner-detector system (ID) is immersed in a \SI{2}{\tesla} axial magnetic field 
and provides charged-particle tracking in the range $|\eta| < 2.5$.

The high-granularity silicon pixel detector covers the vertex region and typically provides four measurements per track.
It is followed by the silicon microstrip tracker, which usually provides four two-dimensional measurement points per track.
These silicon detectors are complemented by the transition radiation tracker,
which enables radially extended track reconstruction up to $|\eta| = 2.0$. 
The transition radiation tracker also provides electron identification information 
based on the fraction of hits (typically 30 in total) above a higher energy-deposit threshold corresponding to transition radiation.

The electromagnetic (EM) calorimeter is a lead/liquid-argon (LAr) sampling calorimeter with an accordion
geometry. It is divided into a barrel section (EMB), covering the pseudorapidity region $|\eta|<1.475$
and two endcap sections (EMEC), covering $1.375 <  |\eta|< 3.2$.
For  $|\eta|<2.5$, the EM calorimeter is divided into three layers in depth. Each layer is segmented in 
$\eta$--$\phi$ projective readout cells. The first layer is finely segmented in the $\eta$ direction for the regions
$0 <  |\eta|< 1.4$ and $1.5 < |\eta|< 2.4$ with a cell size in $\Delta\eta \times \Delta\phi$ varying from $0.003 \times 0.1$
in the barrel region to 
$0.006 \times 0.1$ in the region $|\eta| > 2.0$.
The fine segmentation in the $\eta$ direction provides event-by-event discrimination between single-photon or single-electron
showers and overlapping showers produced in the decays of neutral hadrons. The first layer's thickness varies
between 3 and 5~radiation lengths, depending on $\eta$.
The second layer collects most of the energy deposited in the calorimeter by electron
and photon showers. Its thickness is between 17 and 20~radiation lengths and the cell size is $0.025 \times 0.025$
in  $\Delta\eta \times \Delta\phi$. A third layer with cell size of $0.050 \times 0.025$ and thickness of 2 to 10~~radiation lengths is used to correct
for the leakage beyond the EM calorimeter. 
A high-voltage system generates an electric field of about 1~kV/mm between the lead
absorbers and copper electrodes located at the middle of the liquid-argon gaps. It induces ionization
electrons to drift in the gap.
In the region $|\eta|<1.8$,  a thin presampler layer, located in front
of the accordion calorimeter, is used to correct for energy loss upstream of the calorimeter. It consists of
an active liquid-argon layer with a thickness of 1.1~cm (0.5~cm) in the barrel (endcap) with a cell size of
$0.025 \times 0.1$ in  $\Delta\eta \times \Delta\phi$.

In the transition region between the EMB and the EMEC, 
$1.37 < |\eta| < 1.52$, a large amount of material is located in front of the first calorimeter 
layer, ranging from 5 to almost 10~radiation lengths. This section is instrumented with scintillators
located between the barrel and endcap cryostats, and extending up to $|\eta|=1.6$. 

Hadronic calorimetry is provided by the steel/scintillator-tile calorimeter, divided into three barrel
structures within $|\eta|<1.7$ and two copper/LAr hadronic endcap calorimeters. The solid angle coverage
is completed in the region
$3.2 < |\eta| <4.9$
with forward copper/LAr and tungsten/LAr calorimeter modules optimized for electromagnetic
and hadronic measurements respectively.

The muon spectrometer comprises separate trigger and
high-precision tracking chambers measuring the deflection of muons in a magnetic field generated by superconducting air-core toroid magnets.
The field integral ranges between \num{2.0} and \SI{6.0}{\tesla\metre}
across most of the detector. 
A set of precision chambers covers the region $|\eta| < 2.7$ with three layers of monitored drift tubes,
complemented by cathode-strip chambers in the forward region, where the background is highest.
The muon trigger system covers the range $|\eta| < 2.4$ with resistive-plate chambers in the barrel and thin-gap chambers in the endcap regions.

A two-level trigger system is used to select interesting events~\cite{TRIG-2016-01}.
The first-level trigger is implemented in hardware and uses a subset of detector information to reduce the event rate to a design value of at
most \SI{100}{\kilo\hertz}. This is followed by a software-based high-level trigger 
which reduces the event rate to about \SI{1000}{\hertz}.

\subsection{EM calorimeter cell energy estimate}

The deposit of energy in the liquid-argon gap induces an electric current proportional to the deposited
energy. For a uniform energy deposit in the gap, the signal has a triangular shape as a function of time with a
length corresponding to the maximum drift time of the ionization electrons, typically 450~ns. This signal is amplified and shaped by
a bipolar filter in the front-end readout boards~\cite{1748-0221-5-09-P09003} to reduce the effect of out-of-time energy deposits
from collisions in the following or previous bunch crossings. To accommodate the required dynamic range, three
different gains (high, medium and low) are used. The shaped and amplified signals are sampled at 40~MHz
and, for each first-level trigger, digitized by a 12-bit analogue-to-digital (ADC) converter. The medium
gain for the sample corresponding to the maximum expected amplitude is digitized first to choose the most
suited gain. Four time samples for the selected gain are then digitized
and sent to the off-detector electronics via optical fibres. The position of the maximum of the signal is in the third sample
for an energy deposit produced in the same bunch crossing as the triggered event.

From the digitized time samples ($s_i$), the total energy deposited in a calorimeter cell can be estimated as
\begin{equation}
E = F_{\mu\mathrm{A} \rightarrow \MeV} \times F_{\mathrm{ADC} \rightarrow \mu\mathrm{A}} \times \Sigma_{i=1}^4 a_i(s_i-p).
\label{eqn:cell_energy}
\end{equation}
\begin{itemize}
\item $p$ is the readout electronics pedestal. It is measured for each gain in dedicated electronics calibration runs~\cite{1748-0221-5-09-P09003}.
\item $a_i$ are optimal filtering coefficients~\cite{Cleland:2002rya}  used
to estimate the amplitude of the pulse. They are derived from the predicted pulse shape and the noise correlation functions between time samples 
so as to minimize the total noise arising from the electronics and the fluctuations of energy deposits from additional
interactions in the same bunch crossing as the triggered event or in neighbouring crossings.
\item $F_{\mathrm{ADC} \rightarrow \mu\mathrm{A}}$ is the conversion factor from ADC counts to input current. It is determined from dedicated
electronics calibration runs and takes into account the difference in the response between the injected
current from the pulser in calibration runs and the ionization current created by energy deposited in the
gap~\cite{AHARROUCHE2007429}.
\item $F_{\mu\mathrm{A} \rightarrow \MeV}$ converts the ionization current to the total deposited energy in one cell. It is determined
from test-beam studies~\cite{AHARROUCHE2007429}.
\end{itemize}


\subsection{Electron and photon reconstruction and identification}

The reconstruction of electrons and photons in the region $|\eta|<2.47$ starts from clusters of energy deposits 
in the EM calorimeter~\cite{ATL-LARG-PUB-2008-002}. Clusters matched to a reconstructed ID track, consistent with originating from
an electron produced in the beam interaction region, are classified as electrons. Clusters without matching
tracks are classified as unconverted photons. 
Converted photon candidates are defined as clusters matched to a track consistent with originating
from a photon conversion in the material of the ID or matched to a two-track vertex consistent with
the photon conversion hypothesis~\cite{ATLAS-PERF-2017-02}.
The definition of converted photon candidate includes 
requirements on the number of hits in the innermost pixel detector layer and on the fraction of 
high-threshold hits in the transition radiation tracker. The energy of the electron or photon is estimated using
an area corresponding to $3 \times 7$ ($5 \times 5$) second-layer cells in the barrel (endcap) region.

Photon identification is based primarily on shower shapes in the calorimeter.
Two levels of selection, Loose and Tight, are defined~\cite{ATLAS-PERF-2017-02}.
The Tight identification efficiency ranges from 50\% to 95\% for photons of \ET\ between 10 and 100 GeV.
To further reduce the background from jets, isolation
selection criteria are used. 
They are based on topological clusters of energy deposits in the calorimeter~\cite{PERF-2014-07} and 
on reconstructed tracks in a direction close to that of the photon candidate, as described in Ref.~\cite{ATLAS-PERF-2017-02}.

Electrons are identified using a likelihood-based method combining information from the
EM calorimeter and the ID. Different identification levels, Loose, Medium and Tight
are defined~\cite{ATLAS-PERF-2017-01},
with typical efficiencies for electrons of \ET\ around 40~GeV of 92\%, 85\% and 75\% respectively.
Electrons are required to be
isolated using both calorimeter-based and track-based isolation variables.
More details are given in Ref.~\cite{ATLAS-PERF-2017-01}.

\section{Overview of the calibration procedure}
\label{sec:method}

The energy calibration discussed in this paper covers the region $|\eta|<2.47$, which corresponds to the
acceptance of the ID and the highly segmented EM calorimeter.

The different steps performed in the procedure to calibrate the energy
response of electrons and photons from the energy of a cluster of cells in the EM
calorimeter are the following:
\begin{itemize}
\item The estimation of the energy of the electron or photon from the energy deposits in the calorimeter:
The properties of the shower development are used to optimize the energy resolution and to minimize  the
impact of material in front of the calorimeter. The multivariate regression algorithm used for this estimation is trained on
samples of simulated events. The same algorithm is applied to data and simulation.
This step relies on an accurate description of the material in front of the calorimeter in the simulation.
\item The adjustment of the relative energy scales of the different layers of the EM calorimeter:
This adjustment is based on studies of muon energy deposits and electron showers. It is applied as a correction
to the data before the estimation of the energy of the electron or photon. This step is required for the
correct extrapolation of the energy calibration to the full energy range of electrons and photons.
\item The correction for residual local non-uniformities in the calorimeter response affecting the data:
This includes geometric effects at the boundaries between calorimeter modules and improvements
of the corrections for non-nominal HV settings in some regions of the calorimeter.
This is studied using the ratio of the measured calorimeter energy to the track momentum for electrons and positrons from $Z$ boson decays.
\item The adjustment of the overall energy scale in the data: This is done using a large sample of $Z$ boson decays
to electron--positron pairs. At the same time, a correction to account for the difference in energy
resolution between data and simulation is derived, and applied to the simulation. These correction factors are assumed
to be universal for electrons and photons. 
\item Checks of the results comparing data and simulation with independent samples:
$J/\psi \rightarrow ee$ decays probe the energy response for low-energy electrons. Radiative $Z$ boson decays are used to check
the energy response for photons.
\end{itemize}

Compared with the Run~1 calibration~\cite{PERF-2013-05}, the main differences are:
\begin{itemize}
\item The data were collected with 25~ns spacing between the proton bunches instead of 50~ns. 
In addition the number of readout samples was reduced from five to four. This reduction was
required in order to increase the maximum first-level trigger rate.
The optimal filtering coefficients for the cell energy estimate (see Section 2.2) were derived to minimize 
the total noise for a pile-up of 25~interactions per bunch crossing with 25~ns spacing between bunches, using four readout samples.
For the Run~1 dataset, the noise minimization was performed for 20~interactions per bunch crossing with 50~ns spacing, using five readout samples.
These changes can affect the energy scale of the different layers of the calorimeter.
\item The data were collected with a higher number of pile-up interactions. This significantly impacts 
the measurements of muon-induced energy deposits in the calorimeter.
\item The material in front of the calorimeter is mostly the same, with the exception of the addition of a new
innermost pixel detector layer together with a thinner beam pipe and changes in the layout of the services of the pixel detector.
\item In the data reconstruction, the calorimeter area used to collect the energy of unconverted photons was changed in order to be
same size as for electrons and converted photons. This simplifies the estimate of the impact on the energy calibration
of uncertainties in the conversion
reconstruction efficiency.
The corresponding increase of the cluster size for unconverted photons implies an increase in the noise which 
has a limited impact on the energy resolution: for \ET~$>$20~(50)~GeV, the energy resolution for unconverted photons is degraded by less than 10~(5)\%.
\end{itemize}

\section{Data and simulation samples}
\label{sec:samples}
\subsection{Data samples}

The results presented in this article are based on proton$-$proton collision data at $\sqrt{s}~=~13$~\TeV,
recorded in 2015 and 2016 with the ATLAS detector.
During the period relevant to this paper, the LHC circulated 6.5~\TeV\ proton beams with a
25~ns bunch spacing.
The peak instantaneous luminosity was $1.37 \cdot 10^{34}$~cm$^{-2}$s$^{-1}$.
Only data collected while all the detector components were
operational are used.
The integrated luminosity of this dataset is 36.1~fb$^{-1}$. 
The mean number of  proton$-$proton interactions per bunch crossing 
is 23.5.

To select $Z\rightarrow ee$ events, a trigger requiring two electrons is used.
For the 2015 (2016) dataset, the transverse energy (\ET) threshold applied at the first-level
trigger is 10 (15)~\GeV. It is 12 (17)~\GeV\ at the high-level trigger, which uses an energy calibration
scheme close to the one applied in the offline reconstruction. At the high-level trigger,
the electrons are required to fulfil the Loose (Very Loose) likelihood-based identification criteria
for 2015 (2016) data.

To select $J/\psi\rightarrow ee$ events, three dielectron triggers with different thresholds are used.
At the first-level trigger, \ET thresholds of either 3, 7 or 12~\GeV\ were applied for the candidate with highest
\ET, and a 3~\GeV\ threshold was applied on the second candidate.
At least one electron was required to fulfil the Tight identification criteria at the high-level trigger with
\ET larger than 5, 9 and 14~\GeV\ depending on the trigger. The second electron
was only required to have \ET above 4~\GeV. The integrated luminosity collected
with these prescaled triggers varies from 4~pb$^{-1}$ to 640~pb$^{-1}$ depending on the trigger
threshold used. The total luminosity collected is 710~pb$^{-1}$.

To select $Z\rightarrow \mu\mu$ events, two main triggers are used. The first
one requires two muons with transverse momentum (\pt)
above 14 (10)~\GeV\ at the high-level (first-level) trigger. The second one requires
one muon with \pt above 26 (20)~\GeV\ with isolation criteria applied at the high-level trigger.

For the samples of radiative $Z$ boson decays ($ee \gamma$ and $\mu\mu \gamma$), the same 
triggers as for the  $Z\rightarrow ee$ and $Z\rightarrow \mu\mu$ samples are used.

To select a sample of inclusive photons, a single-photon trigger is
used, with an \ET threshold of 22~\GeV\ at the first-level trigger and the Loose photon identification
criteria with \ET larger than 140~\GeV\ applied at the high-level trigger.

\subsection{Simulation samples}

Monte Carlo (MC) samples of $Z\rightarrow ee$ and $Z\rightarrow \mu\mu$ decays were simulated at next-to-leading order (NLO) in QCD using POWHEG-BOX v2~\cite{Alioli:2008gx} interfaced to the PYTHIA8~\cite{Sjostrand:2007gs} version 8.186 parton shower model.
The CT10~\cite{PhysRevD.82.074024} parton distribution function (PDF)  set was used in the matrix element.
The AZNLO set of tuned parameters~\cite{AZNLO:2014} was used, with PDF set CTEQ6L1~\cite{Pumplin:2002vw},
for the modelling of non-perturbative effects.
The EvtGen 1.2.0 program~\cite{EvtGen} was used to model $b$- and $c$-hadron decays.
Photos++~3.52~\cite{Davidson:2010ew} was used for QED emissions from electroweak vertices and charged leptons.

Samples of  $Z\rightarrow ee\gamma$ and $Z\rightarrow \mu\mu\gamma$ events with transverse momentum of the photon
above 10~\GeV\ were generated with SHERPA version 2.1.1~\cite{Gleisberg:2008ta} using QCD leading-order (LO) matrix elements with up
to three additional partons in the final state.
The CT10 PDF set was used in conjunction with the dedicated parton shower tuning developed by the SHERPA authors.

Both non-prompt (originating from $b$-hadron decays) and prompt (not originating from $b$-hadron decays)
$J/\psi\rightarrow ee$  samples were generated using PYTHIA8.
The A14 set of tuned parameters~\cite{ATL-PHYS-PUB-2014-021} was used together with the CTEQ6L1 PDF set.
The EvtGen program was used to model the $b$- and $c$-hadron decays. Three different samples
were produced with different selections on the transverse momenta of the electrons produced in the $J/\psi$ decay.

Samples of inclusive photon production were generated using PYTHIA8.
The PYTHIA8 simulation of the signal includes LO photon-plus-jet events from the
hard subprocesses $qg \rightarrow  q\gamma$ and $qq \rightarrow g \gamma$ and photon bremsstrahlung
in LO QCD dijet events (called the ``bremsstrahlung component''). The bremsstrahlung component was
 modelled by final-state QED radiation arising from calculations of all 2 $\rightarrow$ 2 QCD processes.
The A14 set of tuned parameters was used together with the NNPDF23LO PDF set~\cite{Ball:2012cx}.

Backgrounds affecting the $Z \rightarrow ee$ sample
were generated with POWHEG-BOX v2 interfaced to PYTHIA8 for the $Z \rightarrow \tau\tau$ process,
with SHERPA version 2.2.1 for the vector-boson pair-production processes and with SHERPA version 2.1.1
for top-quark pair production in the dilepton final state.

For the optimization of the MC-based response calibration, samples of 40 million single electrons
and single photons were simulated. Their transverse momentum distribution covers the range
from 1~\GeV\ to 3~\TeV.

The generated events were processed through the full ATLAS detector simulation~\cite{SOFT-2010-01} based
on GEANT4~\cite{AGOSTINELLI2003250}. The MC events were simulated with additional interactions
in the same or neighbouring bunch crossings to match the pile-up conditions during LHC operations
and were weighted to reproduce the distribution of the average number of interactions per bunch crossing in
data.
The overlaid proton$-$proton collisions were generated with the soft QCD processes of PYTHIA8 version 8.186
using the A2 set of tuned parameters~\cite{ATL-PHYS-PUB-2012-003} and the MSTW2008LO PDF set~\cite{Martin:2009iq}.

The detector description used in the GEANT4 simulation was improved using data collected in Run~1~\cite{PERF-2013-05}.
Compared with this improved description, the changes for the results presented in this paper are: 
the addition of the new innermost pixel layer and the new beam pipe in Run~2~\cite{ATLAS-TDR-19,ibl-2}, the modification of the pixel detector
services at small radius~\cite{PERF-2015-07} and a re-tuning in the simulation of the amount of material in the transition
region between the barrel and endcap calorimeter cryostats to agree better with the measurement performed with Run~1 data.
The amount of material in front of the presampler detector is about 1.8~radiation lengths at small values of $|\eta|$,
reaching $\approx$~4~radiation lengths at the end of the EMB acceptance and up to 6~radiation lengths close
to $|\eta|=1.7$.  The amount of material located between the presampler and the first layer of the calorimeter
is typically 0.5 to 1.5~radiation lengths except in the transition region between the EMB and EMEC, where it is larger.
For $|\eta|>1.8$, the total amount of material in front of the calorimeter is typically 3~radiation lengths.
The simulation models the details of the readout electronics response following the same ingredients
as described in Eq.~(\ref{eqn:cell_energy}).

For studies of systematic uncertainties related to the detector description in the simulation,
samples with additional passive material in front of the EM calorimeter were simulated. The samples
vary by the location of the additional material: in the inner-detector volume, in the first pixel detector layer,
in the services of the
pixel detector at small radius, in the regions close to the calorimeter cryostats, between the
presampler and the electromagnetic calorimeter or in the transition region between the barrel and
endcap calorimeters.

\subsection{Event selection}

Table~\ref{table:samples} lists the kinematic selections applied to the different samples and the
number of events recorded in 2015 and 2016. The average electron transverse energy is around 40--50~\GeV\
in the $Z\rightarrow ee$ sample and 10~\GeV\ in the $J/\psi\rightarrow ee$  sample.
For photons, the average transverse energy is about 25~\GeV\ in the
$Z\rightarrow ee\gamma$ and $Z\rightarrow \mu\mu\gamma$  samples.

To select $Z \rightarrow ee$ candidates, both electrons are required to satisfy the
Medium selection of the likelihood discriminant and to fulfil the Loose isolation criteria,
based on both ID- and calorimeter-related variables~\cite{ATLAS-PERF-2017-01}.
In the inclusive photon selection, the photons are required to fulfil the Tight identification
selection and to be isolated, using the Tight criterion based only on calorimetric variables.
To select muons in the  $Z\rightarrow \mu\mu$ sample, the Medium muon identification working
point~\cite{PERF-2015-10} is used.

To select $J/\psi\rightarrow ee$ candidates, both electrons are required to fulfil the Tight
identification and the Loose isolation criteria.

For the  $Z\rightarrow ee\gamma$ sample ($Z\rightarrow \mu\mu\gamma$), the electrons (muons) are required to satisfy the Loose (Medium) identification level while the photon candidate is required to fulfil the Tight identification and the Loose isolation criteria~\cite{ATLAS-PERF-2017-02}. 
The dilepton invariant mass is restricted to the range 40--80~\GeV~to enhance the sample
in radiative $Z$ decays. The photon candidate is required to be significantly separated from any 
charged-lepton candidate, $\Delta R$($\ell$,$\gamma$) $>$ 0.4, with $\Delta R = \sqrt{(\Delta\phi)^2 + (\Delta\eta)^2}$.

\begin{table}
\begin{center}
\caption{Summary of the kinematic selections applied to the main samples used in the calibration studies
and number of events fulfilling all the requirements described in the text in the 2015--2016 dataset, except for the $Z\rightarrow \ell\ell\gamma$ and inclusive photon
samples, which use only data collected
in 2016. The symbol $\ell$ denotes an electron or a muon.}
\label{table:samples}
\begin{tabular}{lll}
\toprule
Process  &      Selections &   $N$(events)  \\
\midrule
$Z\rightarrow ee$       &    $\ET^e >27$~\GeV, $|\eta^e|<2.47$  & 17.3~M \\
                        &    $m_{\mathrm{ee}} > 50$~GeV &  \\
$Z\rightarrow \mu\mu$   &    $\pt^{\mu} >27$~\GeV, $|\eta^{\mu}|<2.5$  & 29.4~M \\
                        &    $80<m_{\mathrm{\mu\mu}}<105$~GeV & \\
$J/\psi\rightarrow ee$   &    $\ET^e > 5$~\GeV, $|\eta^e|<2.4$, $2.1<m_{ee}<4.1$~\GeV &   60~k \\
$Z\rightarrow \ell\ell\gamma$ &    $\ET^e >18$~\GeV, $|\eta^e|<1.37$ or $1.52<|\eta^e|<2.47$,  &    27~k ($ee\gamma$) \\
                            &    $\pt^{\mu} >15$~\GeV, $|\eta^{\mu}|<2.7$,  &      50~k ($\mu\mu\gamma$)    \\
                            &    $\ET^{\gamma}>15$~\GeV, $|\eta^{\gamma}|<1.37$ or $1.52<|\eta^{\gamma}|<2.37$ &   \\
                            &    $\Delta R$($\ell$,$\gamma$) $>$ 0.4 &   \\
                            &    $40<m_{\ell \ell}<80$~\GeV & \\
Inclusive photons           &  $\ET^{\gamma}>147$~\GeV, $|\eta^{\gamma}|<1.37$ or $1.52<|\eta^{\gamma}|<2.37$ & ~3.6~M \\
\bottomrule
\end{tabular}
\end{center}
\end{table}

\section{Electron and photon energy estimate and expected resolution from the simulation}
\label{sec:mva}
\subsection{Algorithm for estimating the energy of electrons and photons}
\label{sec:mva1}

The energy of electrons and photons is computed from the energy of the reconstructed cluster,
applying a correction for the energy lost in the material upstream of the calorimeter,
for the energy deposited in the cells neighbouring the cluster in $\eta$ and $\phi$,
and the energy lost beyond the LAr calorimeter.
A single correction for all of these effects is computed using multivariate regression algorithms tuned on samples of simulated single particles without pile-up,
separately for electrons, converted photons and unconverted photons.
The training of the algorithm, based on Boosted Decision Trees, is done in intervals of $|\eta|$ and of transverse energy.
An updated version of the method described in Ref.~\cite{PERF-2013-05} is implemented.
The set of input variables is refined and the procedure is extended
to the whole EM calorimeter up to $|\eta|=2.5$, including the transition region between the barrel and the endcap.

The variables considered in the regression algorithm are: the energy deposited in the calorimeter, the energy deposited in the presampler, 
the ratio of the energies deposited in the first and second layers ($E_1/E_2$) of the EM calorimeter,
the $\eta$ impact point of the shower in the calorimeter,
and the distances in $\eta$ and in $\phi$ between the impact point of the shower and the centre of the
closest cell in the second calorimeter layer.
The impact point of the shower is computed from the energy-weighted barycentre of the positions of the cells in the cluster.
For converted photon candidates, the estimated radius of the photon conversion in the transverse plane as well as the properties of the tracks
associated with the conversion are added.
These variables are identical to those used in the Run~1 version 
except that $E_1/E_2$ is used instead of the longitudinal shower depth.
The ratio $E_1/E_2$ is strongly correlated with the longitudinal shower depth but it has been studied in more detail, comparing data with simulations as described in Section~\ref{sec:E1E2}.

\begin{figure}[htb]
 \centering
 \includegraphics[width=0.45\textwidth]{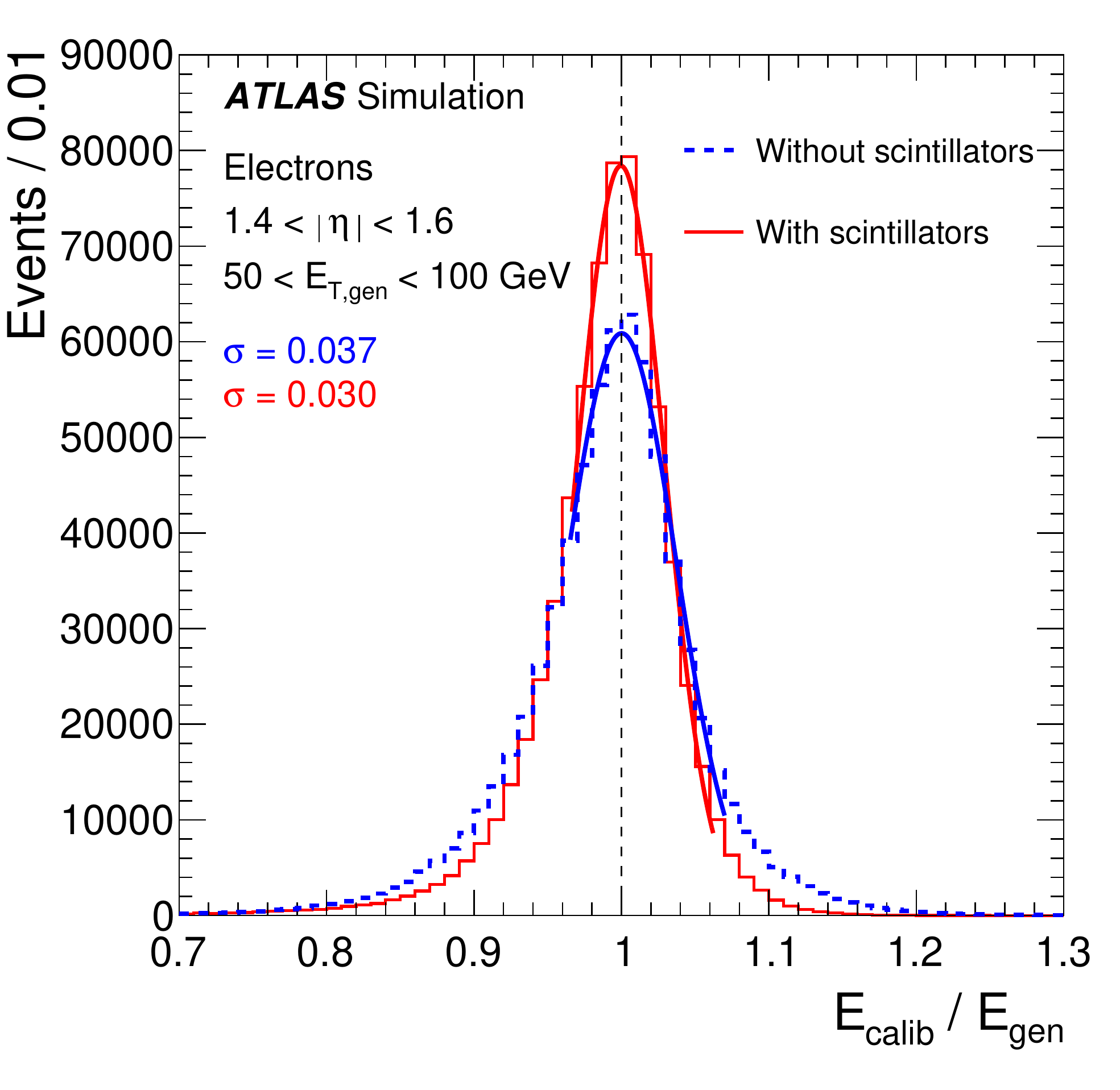}\label{fig:mva_crack_a}
 \caption{Distributions of the calibrated energy, $E_\text{calib}$, divided by the generated energy,
 $E_\text{gen}$, for electrons
 with $1.4 < |\eta| < 1.6$ and $50~< E_{\text{T,gen}} <~100$~\GeV.
 The dashed (solid) histogram shows the results based on the energy calibration without (with)
 the scintillator information.
 The curves represent Gaussian fits to the cores of the distributions.}
 \label{fig:mva_crack}
\end{figure}

In the transition region between the barrel and endcap calorimeters, 
$1.4 < |\eta| < 1.6$, the amount of material traversed by the particles before reaching the first active layer of the calorimeter is large
and the energy resolution is degraded.
To mitigate this effect, information from the $E_4$ scintillators~\cite{PERF-2007-01} installed in the transition region is used.
The $E_4$ scintillators are part of the intermediate tile calorimeter (ITC).
The ITC is located in the gap region, between the long barrel and the extended barrels of the tile calorimeter and it was designed to correct for the energy lost in the passive material that fills the gap region.
Electrons and photons in the gap region deposit energy in the barrel and the endcap of the EM calorimeter, as well as in the $E_4$ scintillators. In this region, the energy deposited in the $E_4$ cells (each of size $\Delta\eta \times \Delta\phi = 0.1 \times 0.1$) and the
difference between the cluster position and the centre of the $E_4$ cell are added to the set of input variables for the regression algorithm.
Due to this additional information the energy resolution is improved as shown in Figure~\ref{fig:mva_crack} for simulated electrons generated with transverse energy between \SI{50}{\GeV} and \SI{100}{\GeV}.
In this range, the improvement is largest for electrons (around 20\%), while it is smaller for unconverted photons (5\%). Such behaviour is expected, as the degradation of the energy resolution due to inactive material in front of the calorimeter is much higher for electrons.

\subsection{Energy resolution in the simulation}

\begin{figure}[htb]
 \centering
          \mbox{
             \begin{tabular}{c}
             \subfloat[\label{fig:mcres_elec}]{\includegraphics[width=0.53\figwidth]{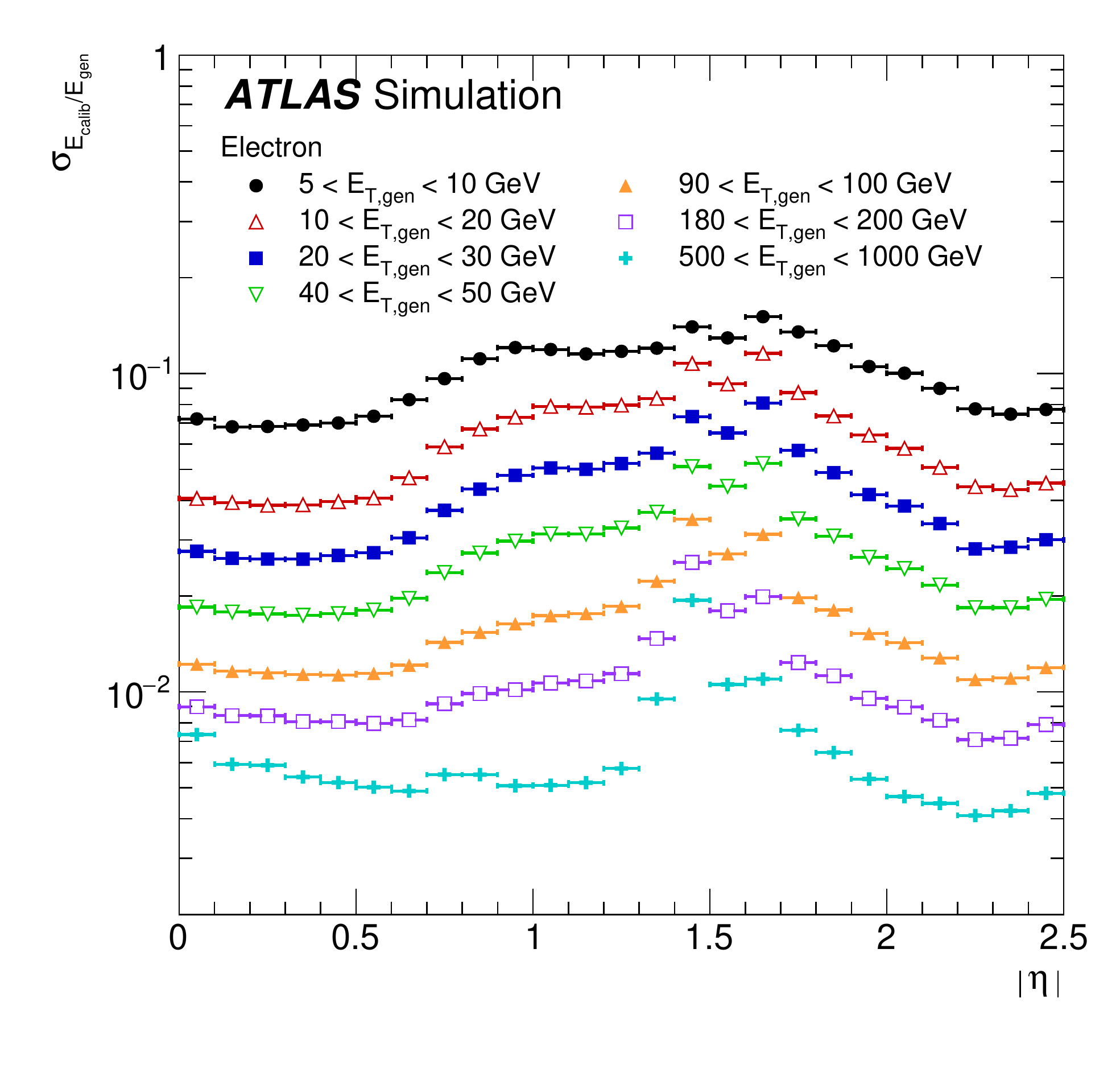}} \\
              \end{tabular}
           }
          \mbox{
             \begin{tabular}{cc}
             \subfloat[\label{fig:mcres_conv}]{\includegraphics[width=0.53\figwidth]{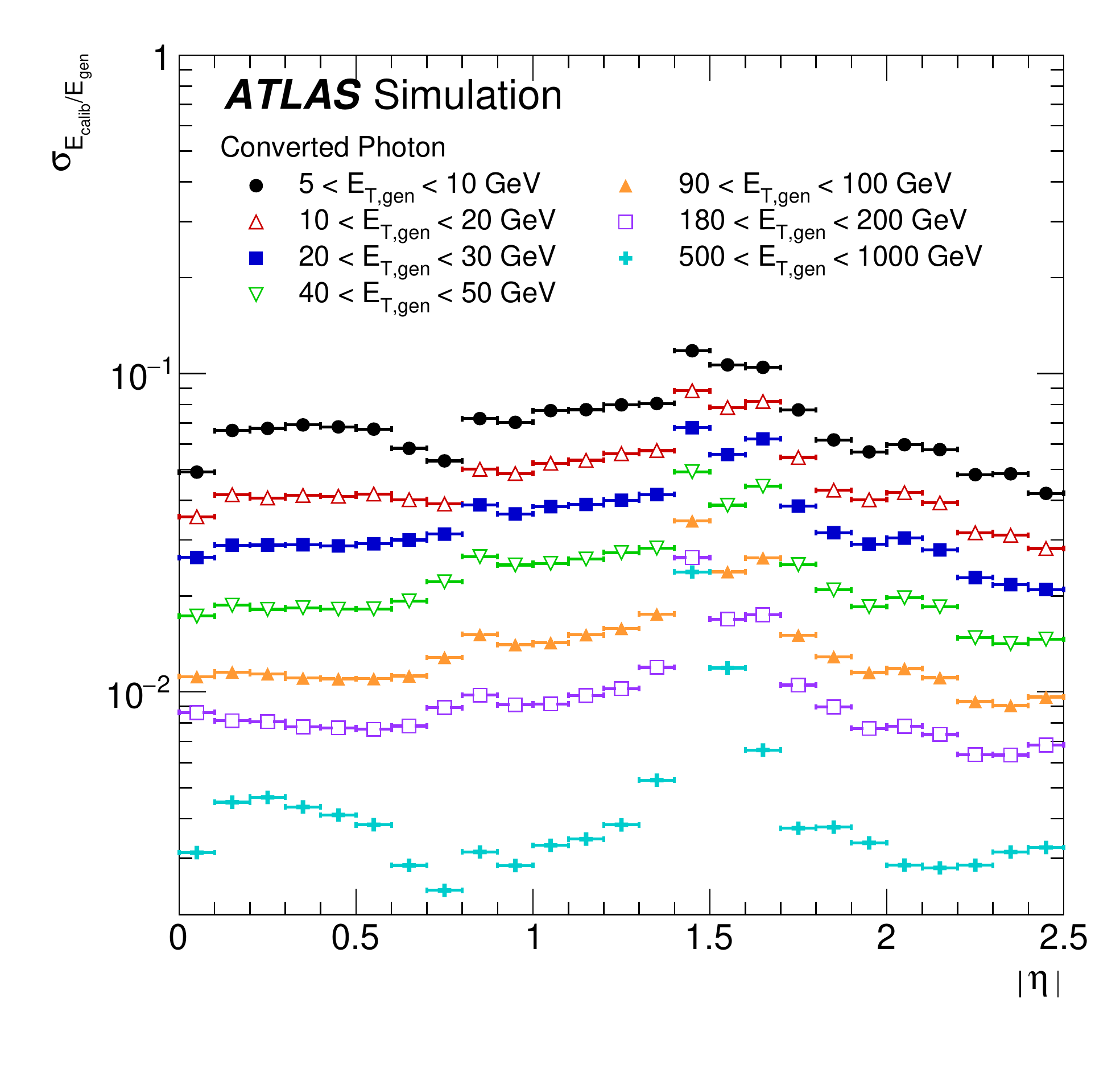}} &
             \subfloat[\label{fig:mcres_unco}]{\includegraphics[width=0.53\figwidth]{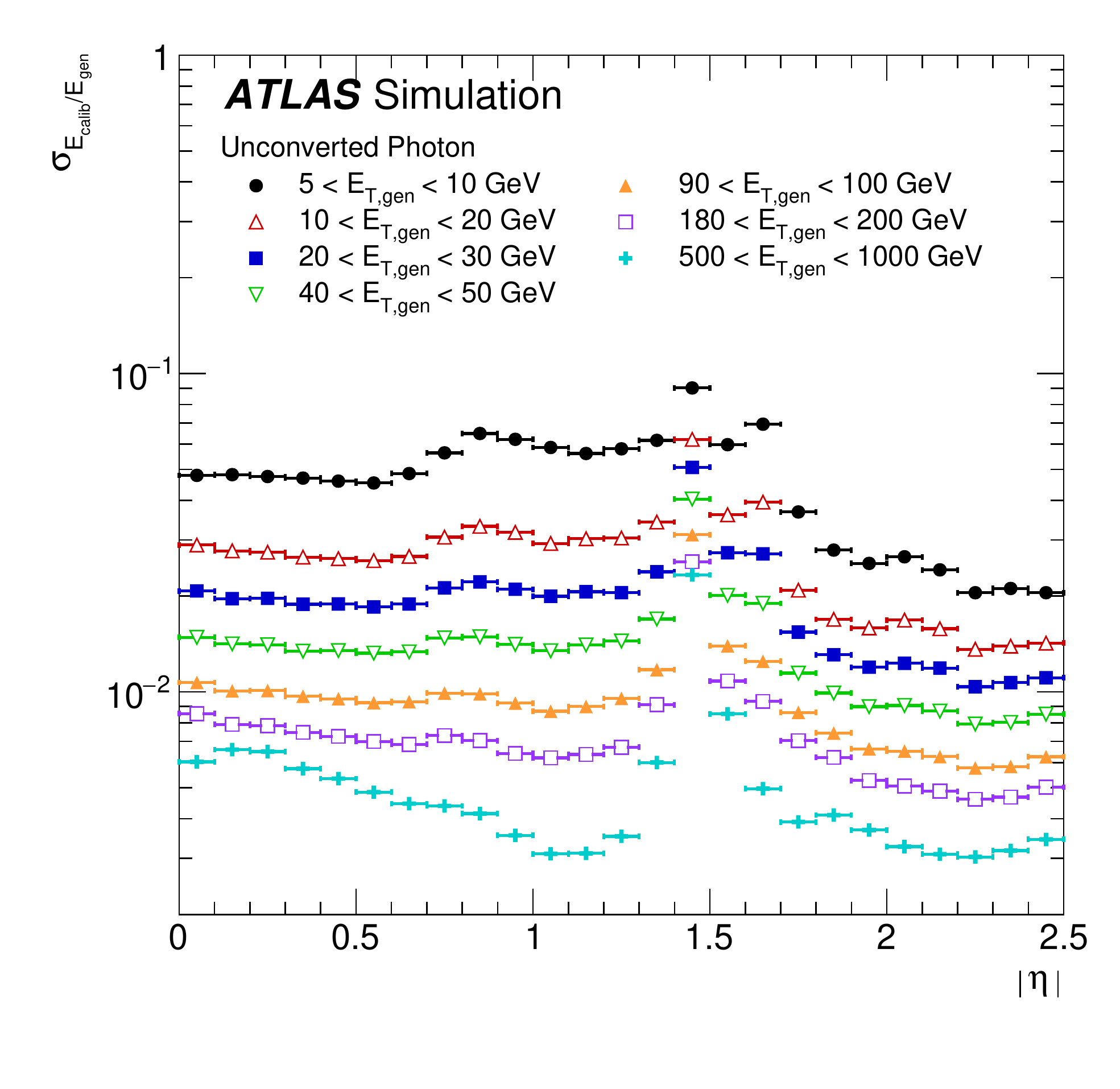}}
             \end{tabular}
           }
  \caption{Energy resolution, $\sigma_{E_{\mathrm{calib}}/E_{\mathrm{gen}}}$, estimated from the interquartile range of $E_{\mathrm{calib}}/E_{\mathrm{gen}}$ as a function
of $|\eta|$ for (a) electrons, (b) converted photons and (c) unconverted photons,
for different \ET\ ranges.}
  \label{fig:resolution_mc}
\end{figure}

The energy resolution after application of the regression algorithm in the MC samples
is illustrated in Figure~\ref{fig:resolution_mc}, using simulated single-particle samples.
The resolution is defined as the interquartile range of $E_{\mathrm{calib}}/E_{\mathrm{gen}}$, i.e. the interval excluding 
the first and last quartiles of the  $E_{\mathrm{calib}}/E_{\mathrm{gen}}$ distribution in each bin, 
divided by 1.35, to convert to the equivalent standard deviation of a Gaussian distribution. The quantity $E_{\mathrm{gen}}$ 
is the true energy of the generated particle and $E_{\mathrm{calib}}$ is the reconstructed energy after applying the 
regression algorithm.

For unconverted photons, the energy resolution in these MC samples, which do not have any simulated pile-up, closely 
follows the expected sampling term of the calorimeter ($\approx$~10\%/$\sqrt{E/\GeV}$ in the barrel and 
$\approx$~15\%/$\sqrt{E/\GeV}$ in the endcap). 
For electrons and converted photons, the degraded energy resolution at low energies reflects
the presence of significant tails induced by interactions with the material upstream of the calorimeter. This degradation
is largest in the regions with the largest amount of material upstream of the calorimeter, i.e.\ for $1.2 < |\eta| < 1.8$.

\section{Corrections applied to data }
\label{sec:data_correction}
In this section, the corrections applied to the data to account for residual differences between data and simulation are discussed.
They include the
intercalibration of the different calorimeter layers, corrections for energy shifts induced by pile-up
and corrections to improve the uniformity of the energy response.
Since the absolute energy scale is set with $Z\rightarrow ee$ decays, only the relative calibration of the energy scales 
of the first two layers and the presampler is needed. Given the small fraction of the energy deposited in the third layer of the calorimeter, no dedicated corrections for its intercalibration are applied. 

\subsection{Intercalibration of the first and second calorimeter layers}
\label{sec:E1E2}

Muon energy deposits, which are insensitive to the amount of passive material in front of the
EM calorimeter, are used to study the relative calibration of the first and second calorimeter layers. This relative
calibration is derived by comparing the energy deposits in data with simulation predictions. The deposited muon energy, expressed on the same cell-level energy scale as described by Eq.~(\ref{eqn:cell_energy}),
is about 30 to 60~\MeV\ depending on $\eta$ in the first layer and 240 to 300~\MeV\ in the second layer.
The signal-to-noise ratio varies from about 2 to 0.5 (4 to 3) as a function of $|\eta|$ for the first (second) layer.
A significant contribution to the noise, especially in the first layer of the endcap calorimeter, is due to
fluctuations in the pile-up energy deposit.

The analysis uses muons from $Z \rightarrow \mu\mu$ decays, requiring $\pt^{\mu} > 27$~\GeV.
The calorimeter cells crossed by the muon tracks are determined by extrapolating the track to each layer
of the calorimeter, taking into account the geometry of the calorimeter, the misalignment between the inner
detector and the calorimeter (up to a few millimetres) and the magnetic field encountered by the muon along its path.

In the first layer, where the cell size in the $\eta$ direction is small,
the muon signal is estimated by summing the energies measured  in three adjacent cells
along $\eta$ centred around the cell crossed by the extrapolated muon trajectory. Using three cells instead
of only one gives a measurement that is less sensitive to the detailed modelling of the cross-talk between
neighbouring cells and to the exact geometry of the calorimeter.
In the second layer, due to the accordion geometry, the energy is most often shared between two adjacent cells
along $\phi$ and the signal is estimated from the sum of the energies in the cell crossed by the extrapolated muon trajectory
and in the neighbouring cell in $\phi$ with higher energy.

The observed muon energy distribution in each layer can be described by the convolution of a Landau distribution, representing the energy deposit, and a noise distribution. The most probable value (MPV) of the deposited muon energy is extracted using a fit of the convolution function to the observed muon energy distribution (``fit method''). Alternatively, the deposited energy can be estimated using a truncated-mean approach, where the mean is computed over a restricted window to minimize the sensitivity to the tails of the distribution (``truncated-mean method'').
The same procedure is applied to data and MC samples and the relative calibration of the two layers
is computed as $\alpha_{1/2} = ( \left<E_1\right>^{\mathrm{data}}/\left<E_1\right>^{\mathrm{MC}} ) / ( \left<E_2\right>^{\mathrm{data}} / \left<E_2\right>^{\mathrm{MC}} )$ with $\left<E_1\right>$ ($\left<E_2\right>$) denoting the MPV in the first (second) layer. The relative calibration of the two layers is computed as a function of $|\eta|$, since within the uncertainties all measured values are consistent between positive and negative $\eta$ values.

\begin{figure}[htb]
          \centering
          \mbox{
            \begin{tabular}{cc}
            \includegraphics[width=0.53\figwidth]{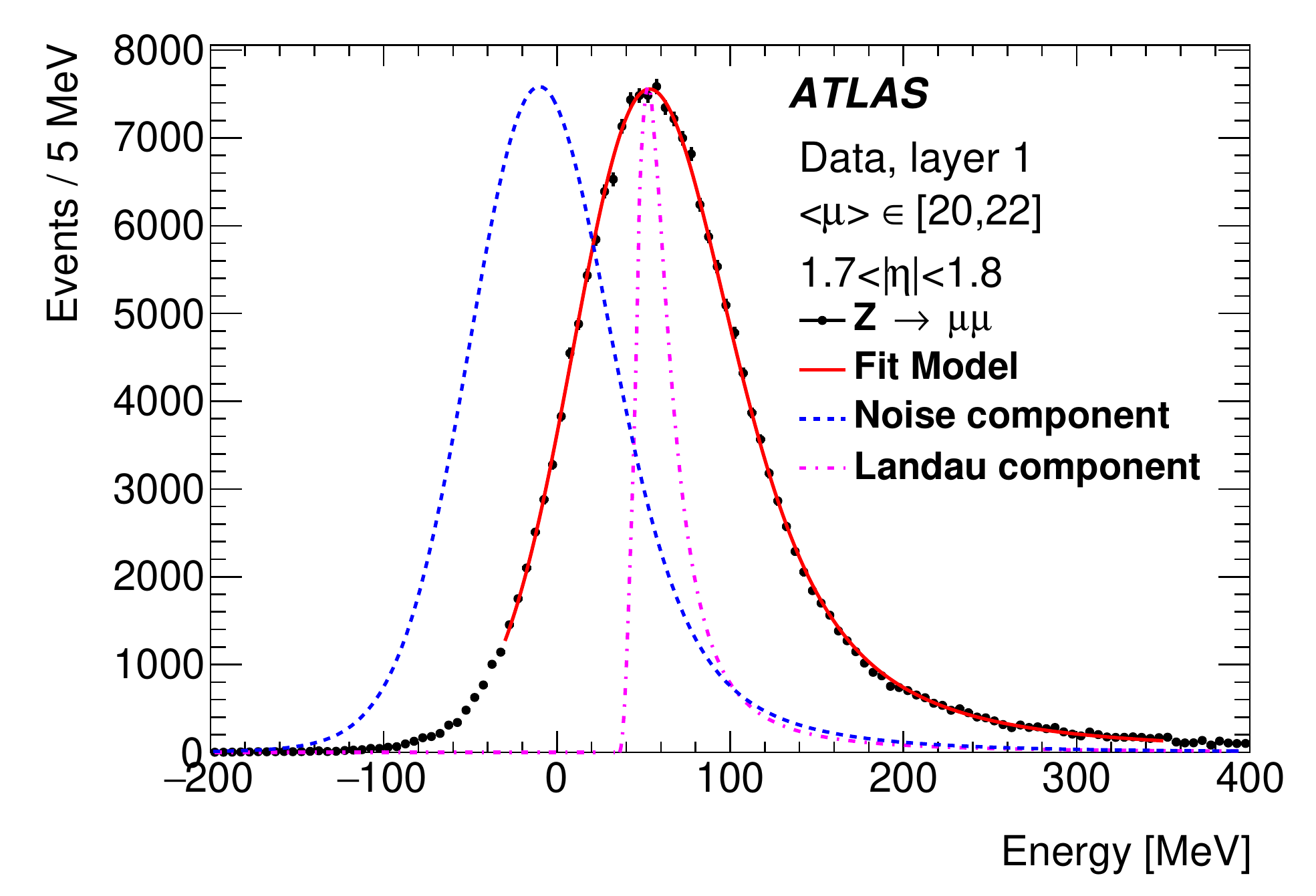}  &
            \includegraphics[width=0.53\figwidth]{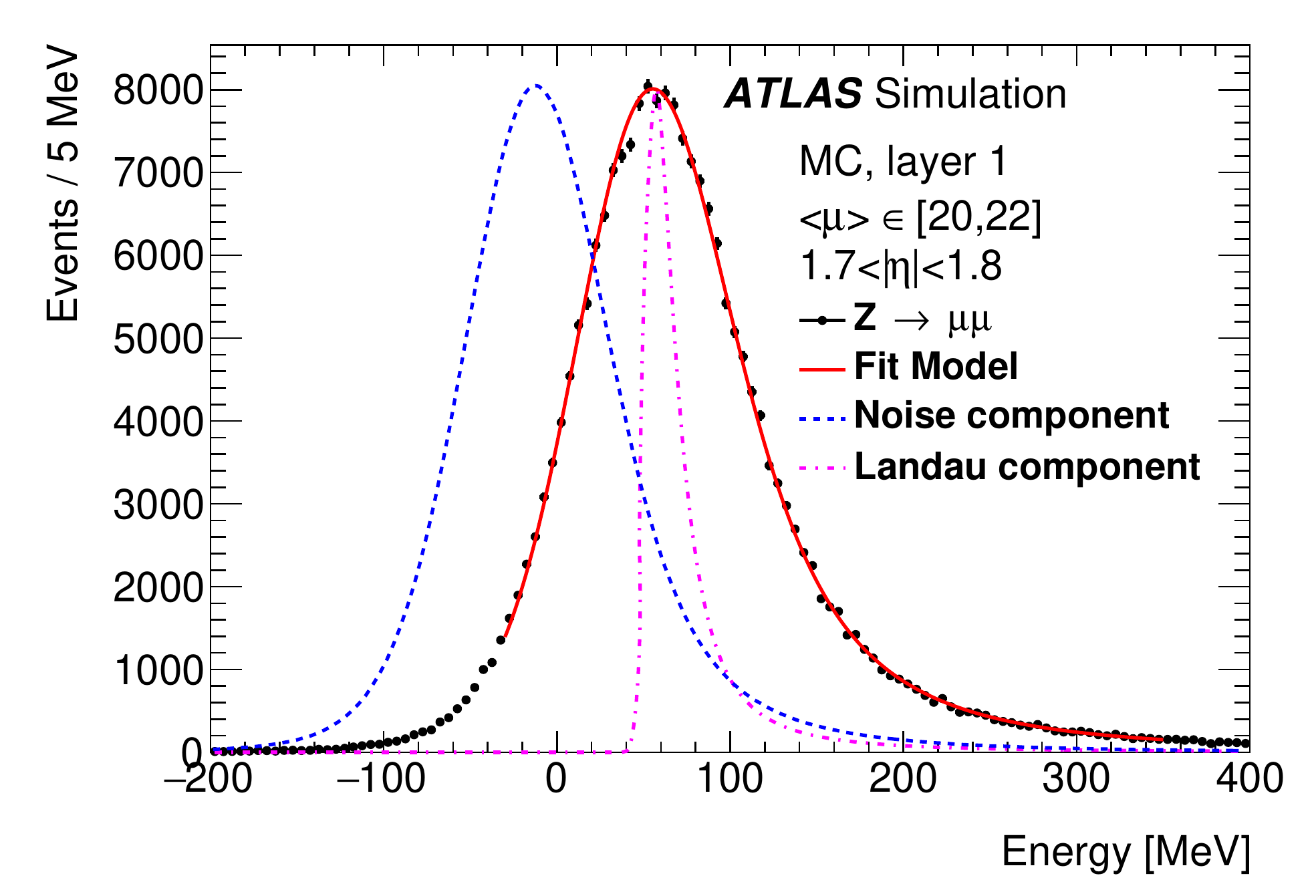} \\
            \includegraphics[width=0.53\figwidth]{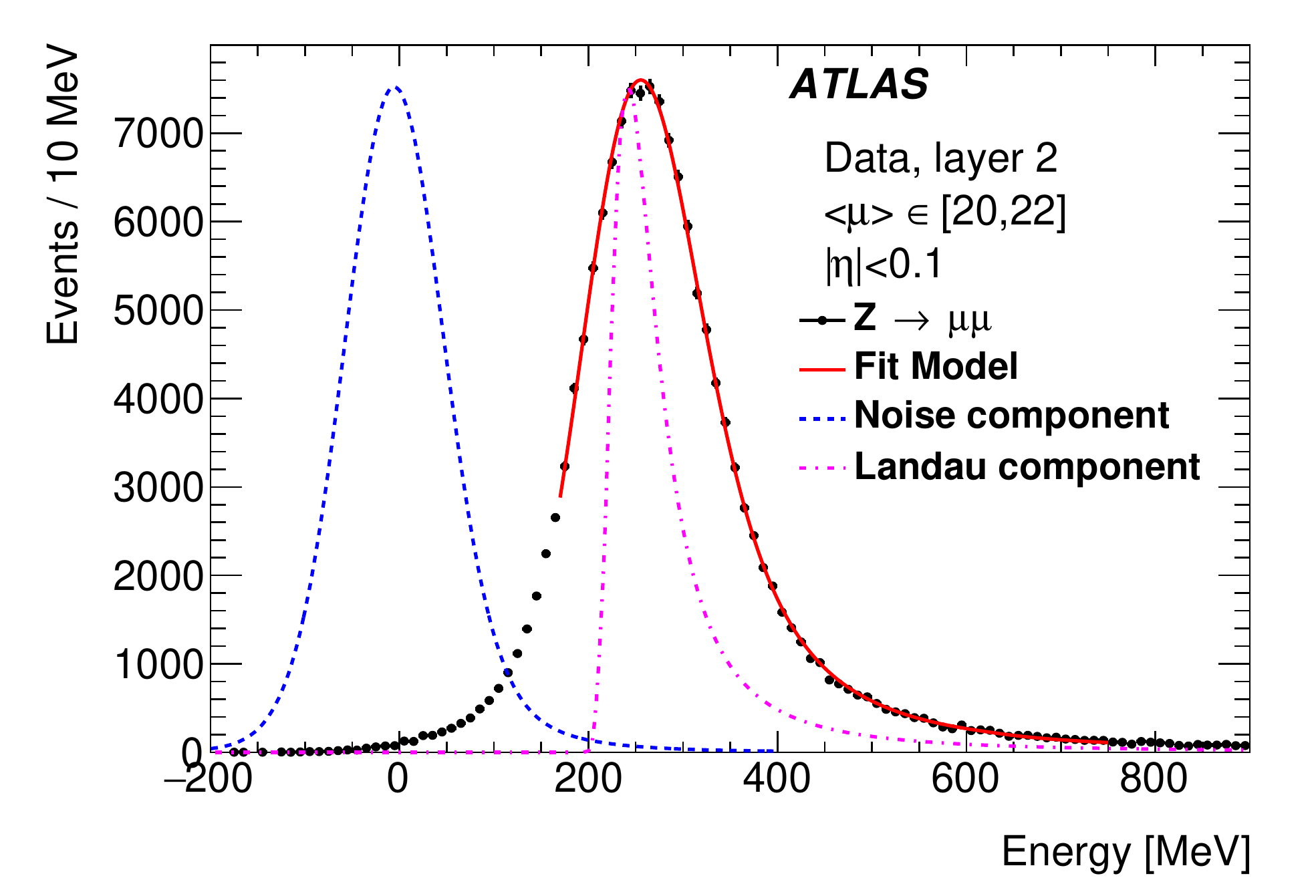} &
            \includegraphics[width=0.53\figwidth]{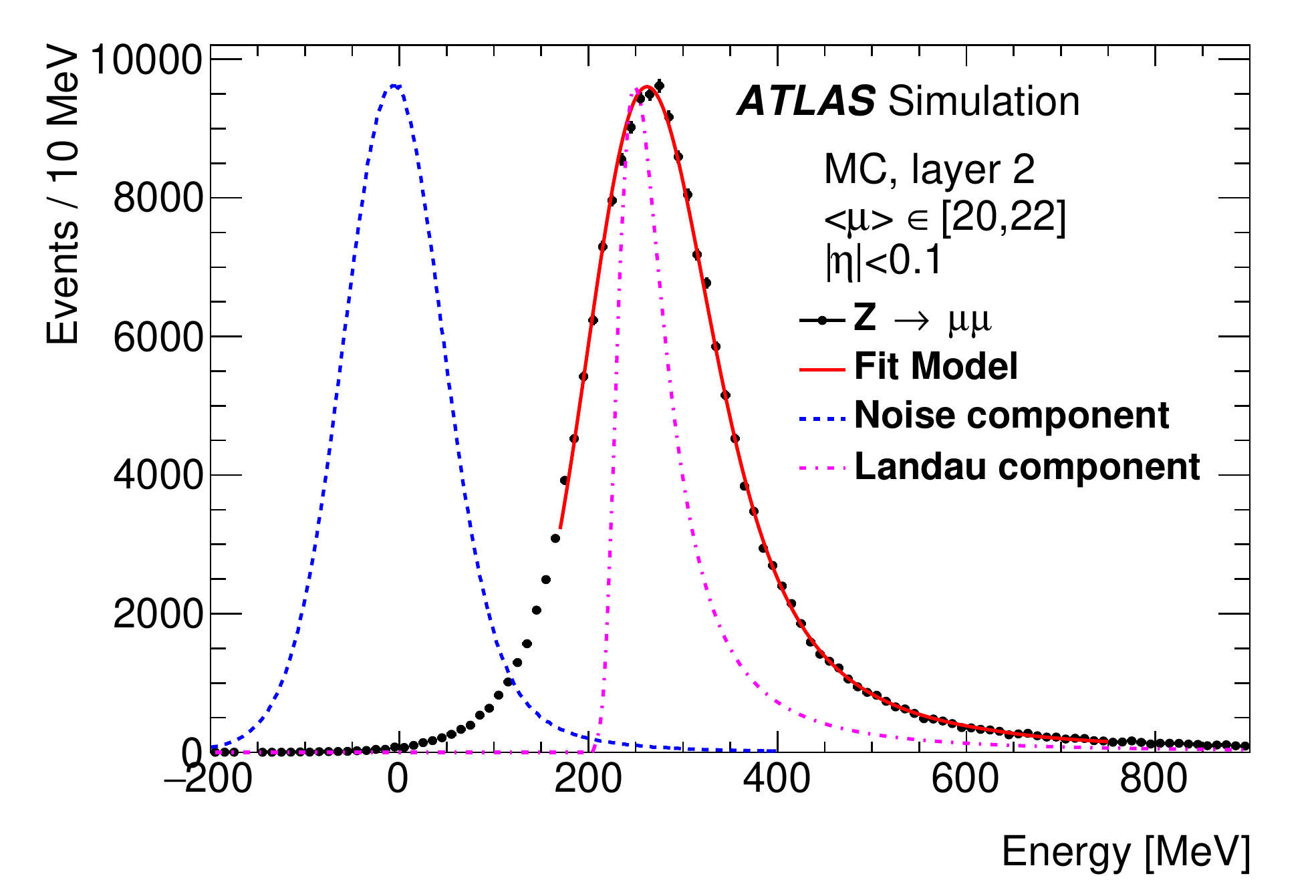} \\
            \end{tabular}
               }
        \caption{Muon energy distributions for two $|\eta|$ regions in data and simulation for the first and second calorimeter layers. The
 fit of the muon data to the convolution of the noise distribution and a Landau function is shown together with the individual
 components: the noise distribution and the Landau function. The distributions are shown for an average number
of interactions per bunch crossing, $\left<\mu\right>$, in the range from 20 to 22.}
        \label{fig:muon_energy}
\end{figure}

In the fit method, the noise distribution is determined from data and MC samples separately to avoid a dependency on a possible pile-up noise mismodelling in the simulation. Events triggered on random LHC proton bunch crossings, 
with a trigger rate proportional to the instantaneous luminosity (``zero-bias events''), are used to estimate the noise distribution in data. The noise distribution is determined in intervals of $|\eta|$ and of $\left<\mu\right>$, where $\left<\mu\right>$ is the average number of pile-up interactions per bunch crossing. Figure~\ref{fig:muon_energy} shows examples of the muon energy deposits in data and MC samples. It also shows the Landau distribution, the noise distribution and their convolution.

In the truncated-mean method, different choices for the window are investigated: ranges of $\pm$2 and $\pm$1.5 times the RMS of the distribution around
the initial mean computed in a wide range,
or the smallest range containing 90\% of the energy distribution. The average of the results obtained with these choices is used as the estimate
of $\alpha_{1/2}$.

To further reduce residual pile-up dependencies of the extracted MPV values, for
both the fit and truncated-mean methods the analysis is performed as
a function of the average number of interactions per bunch crossing. The result is extrapolated to a zero pile-up
value to measure the intrinsic energy scale of each calorimeter layer for a pure signal. This extrapolation
is performed using a first-order polynomial fit, which is found to describe data and MC results well.
The fit is performed in the range from 12 to 30 interactions per bunch crossing to avoid low-statistics bins 
with a large range of the number of interactions per bunch crossing.
The method is validated by comparing the MC extrapolated results with the ones obtained in a MC sample without
any pile-up.
The final result is given by the average of the two signal extraction methods, fit and truncated mean.

\begin{figure}[htb]
          \centering
          \mbox{
            \begin{tabular}{cc}
            \includegraphics[width=0.53\figwidth]{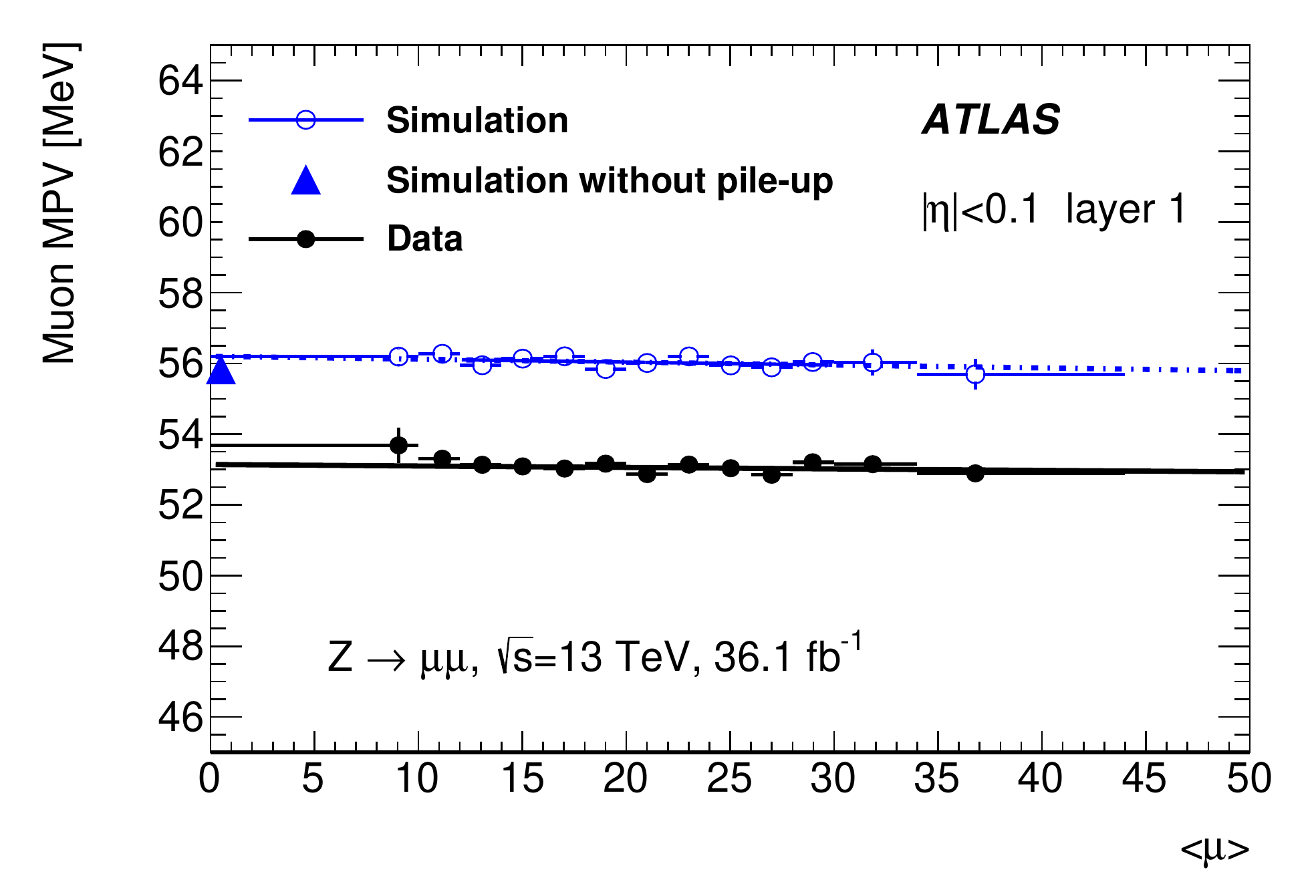} &
            \includegraphics[width=0.53\figwidth]{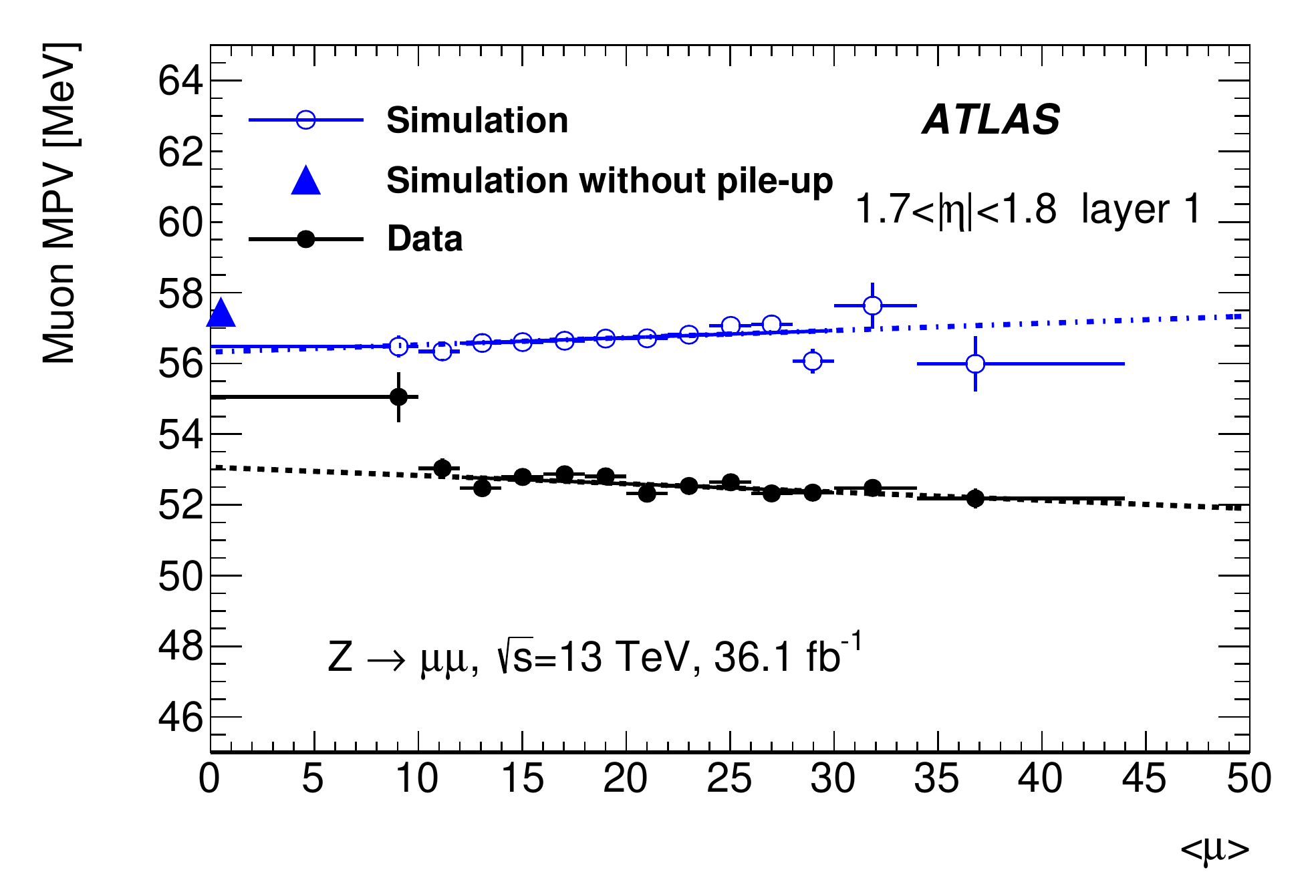} \\
            \includegraphics[width=0.53\figwidth]{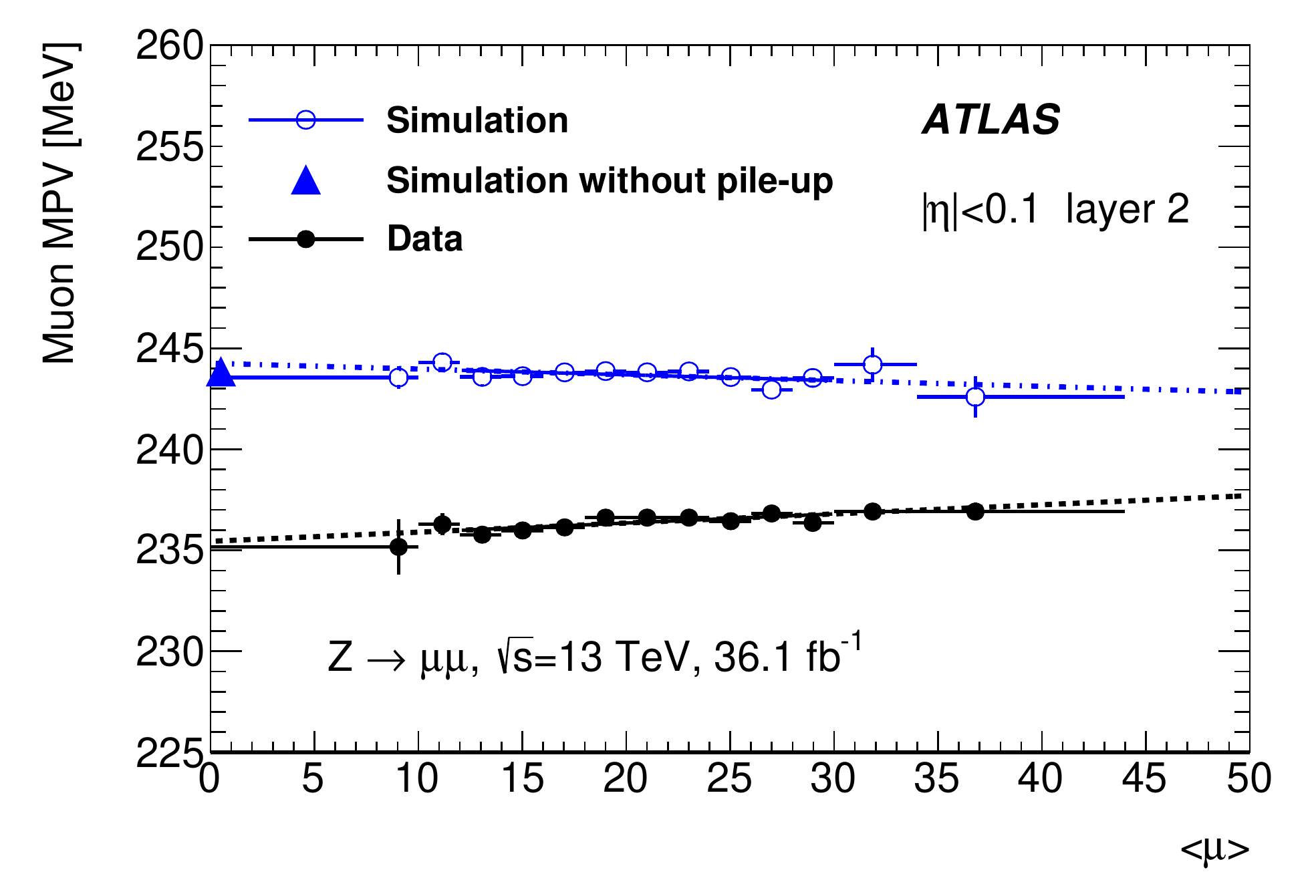} &
            \includegraphics[width=0.53\figwidth]{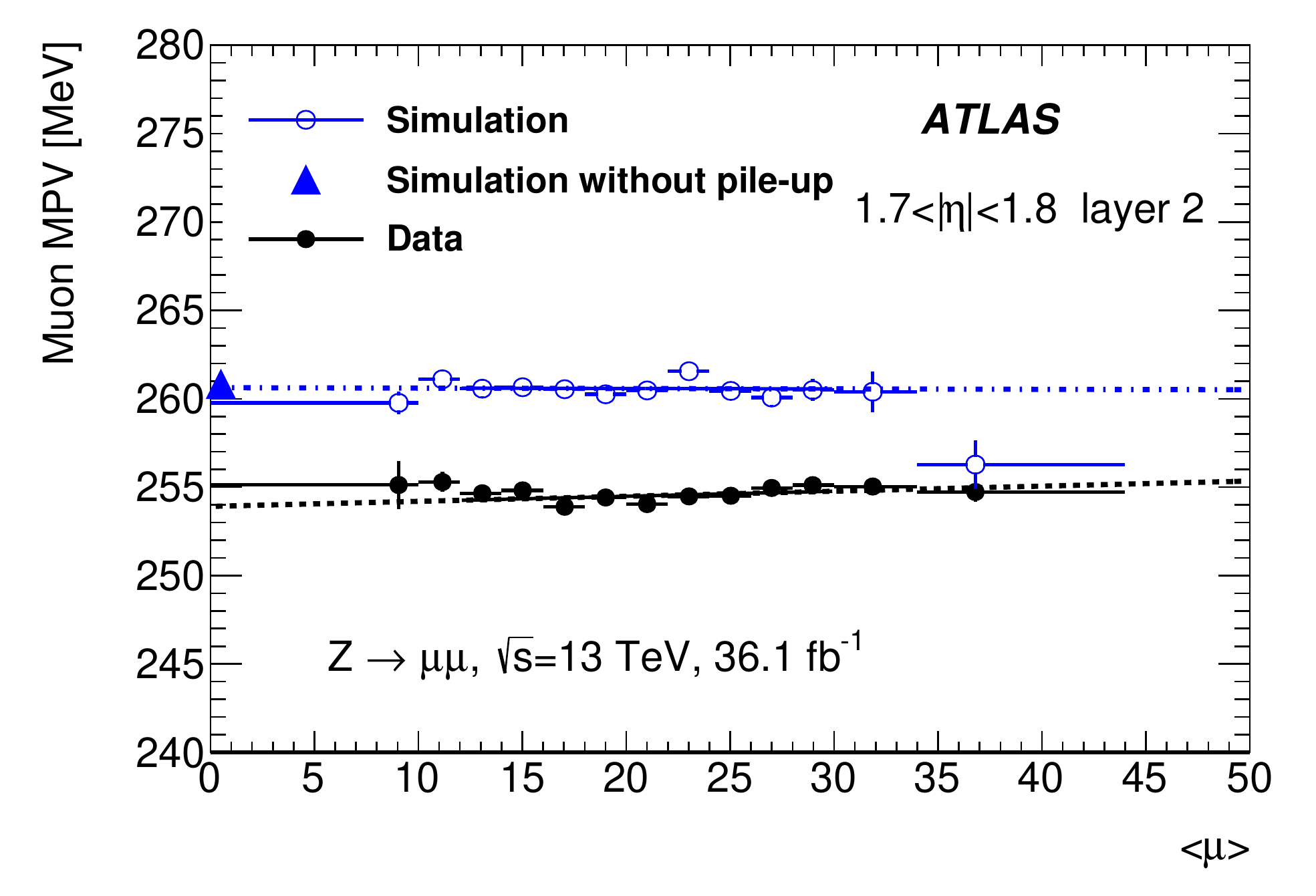} \\
            \end{tabular}
               }
\caption{Distribution of the fitted MPV of the muon energy deposit in two $|\eta|$ intervals, for the first and second calorimeter layers, as
a function of the average number of pile-up interactions per bunch crossing $\left<\mu\right>$.
The values obtained in data and MC samples are shown. 
The linear fits which are used to extrapolate the MPV value to zero pile-up are displayed. 
The solid part of the lines show the range used in the fit while the dashed part of the lines show the extrapolation of the linear fit. The MPV extracted from a MC sample without
pile-up is also shown.}
        \label{fig:muon_vs_muon}
\end{figure}

Figure~\ref{fig:muon_vs_muon} shows examples of the fitted MPV of the deposited muon energy as a function of the number of
interactions per bunch crossing. 
The accurate noise modelling, performed separately for data and simulation, allows the extraction of the MPV of the muon
energy deposit with only a small dependence with pile-up. 
The small slope of the fitted line limits the impact of the extrapolation from the average amount of pile-up in data 
to the zero pile-up point to the percent level.

The following effects are investigated to estimate the uncertainty in the  $\alpha_{1/2}$ value measured with muons
from the average of the results of the fit and truncated mean methods:

\begin{itemize}
\item Accuracy of the method to measure the genuine muon energy loss at zero pile-up: the uncertainty is taken from the
difference between the result from the pile-up extrapolation in the MC sample with pile-up and the
value observed in a MC sample without pile-up. It is typically 0.2\% to 0.5\% depending on $|\eta|$,
up to 1.5\% in some $|\eta|$ intervals in the endcap.
\item Modelling of the energy loss outside the cells used for the measurement: only three (two) cells are used
in the first (second) layer to minimize the noise. For muon trajectories close to the boundaries in $\phi$ ($\eta$) between
the first (second) layer cells a significant fraction of the muon energy deposit can be outside the used cells.
To assess the uncertainty from the modelling of these effects in the simulation, the analysis is repeated using
only muons crossing the centre of the first (second) layer within 0.04 (0.008) in the $\phi$ ($\eta$) direction and
the change induced by these requirements is taken as the uncertainty. The uncertainty varies from 0.5\% to 1\%.
\item Choice of the cell in $\phi$ for the second layer: the analysis is repeated using as the second cell in layer two 
the neighbour closer to the extrapolated muon trajectory instead of the neighbour with the higher energy. 
The difference between the results of these two choices, typically 0.2\%, is taken as the uncertainty.
\item For the truncated-mean method, the results obtained with the different ranges for the truncated-mean computation
are compared. The maximum deviation of these results from their average is taken as the uncertainty. 
The change in the result when varying the upper energy limit used to compute the initial mean is also taken into account in the uncertainty. The resulting uncertainty is 0.5\%.
\item Half of the difference between the fit and truncated-mean methods is taken as an uncertainty in the result.
This leads to an uncertainty varying from 0.5\% to 1\% depending on $|\eta|$.
\end{itemize}

\begin{figure}[htb]
 \centering
 \includegraphics[width=0.55\textwidth]{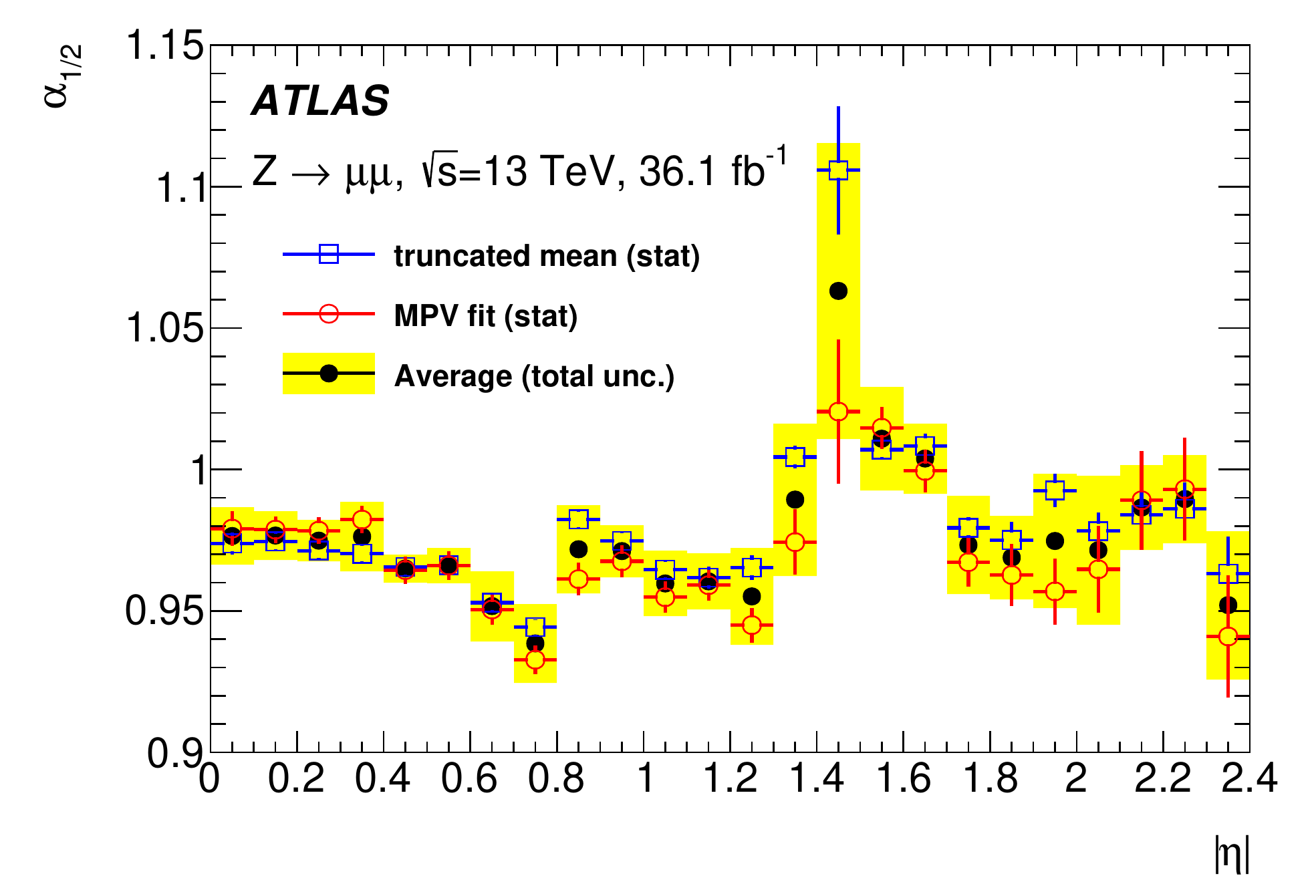}
 \caption{Ratio $\alpha_{1/2} = ( \left<E_1\right>^{\mathrm{data}}/\left<E_1\right>^{\mathrm{MC}}) / ( \left<E_2\right>^{\mathrm{data}} / \left<E_2\right>^{\mathrm{MC}})$ as a function of $|\eta|$, as obtained
from the study of the muon energy deposits in the first two layers of the calorimeters. The results from the
two methods are shown with their statistical uncertainties. The final average measurement is shown with its
total uncertainty including the statistical and systematic uncertainties.}
 \label{fig:muon_e1e2_result}
\end{figure}

Figure~\ref{fig:muon_e1e2_result} shows the results for $\alpha_{1/2}$ and the comparison of the two methods. The
average result is shown with its total uncertainty defined as the sum in quadrature of the statistical
uncertainty and all the systematic uncertainties described above. The total systematic uncertainty is estimated to be 
correlated within $|\eta|$ regions corresponding to the intervals 0--0.6, 0.6--1.4, 1.4--1.5, 1.5--2.4 and 2.4--2.5,
and uncorrelated between two different intervals. 
In the last $|\eta|$ range, no measurement with muons is performed, and a large uncertainty of $\pm20\%$ in the
layer calibration is assigned, derived from a comparison between data and simulation of the ratio $E_1/E_2$ of electron showers.
Despite the high level
of pile-up in the data, the accuracy of the measurement with muons is typically 0.7\% to 1.5\% (1.5\% to 2.5\%)
depending on $\eta$ in the barrel (endcap) calorimeter, for $|\eta|<2.4$, except in the transition region between the
barrel and endcap calorimeters.

The features as a function of $|\eta|$ observed for  $\alpha_{1/2}$ are similar to the ones observed in the Run~1 calibration performed
with muons~\cite{PERF-2013-05}. 
A change in the relative energy scales of the two layers, at a level of less than 1.5\%,
can be expected from the re-optimization of the pulse reconstruction
performed for Run~2 data to minimize the expected pile-up noise.
Within their respective uncertainties, the Run~1 result and this result are in agreement with this expectation.

In addition to the systematic uncertainties specific to the measurement of energy deposits from muons in the calorimeter layers,
the interpretation of this measurement as an estimate of the relative energy scale of the two layers relies
on a proper modelling in the simulation of the ionization current induced by muons. This is subject
to the following sources of uncertainty: uncertainty in the exact path length traversed by the muons, related to the
uncertainty in the geometry of the readout cells; uncertainty in the effect of reduced electric field at the
transition between the different calorimeter layers; uncertainty in the modelling of the conversion of deposited
energy to ionization current due to variations in the electric field following the accordion structure of
the calorimeter, and uncertainty in the cross-talk between different calorimeter cells. These sources of uncertainty affect
muon energy deposits and electron/photon showers differently. The values of these uncertainties are exactly the same as estimated in Ref.~\cite{PERF-2013-05}. They induce an uncertainty varying from 1\% to 1.5\% depending on $|\eta|$ in the relative calibration of the first and second calorimeter layers.
Uncertainties related to possible non-linearities of the energy response for the different calorimeter layers are discussed
in Section~\ref{sec:non_linearity_uncertainty}.

The relative calibration of the first two layers of the calorimeter can also be probed using $Z\rightarrow ee$ decays 
by investigating the variation of the mean of the dielectron invariant mass as a function of the
ratio of the energies of the electron or positron candidates in the first two layers. Good agreement with the results
obtained with muons is observed except in the $|\eta|$ range 1.2 to 1.8. In this region, the results of the method
based on $Z\rightarrow ee$ are very sensitive to the interval used to compute the average invariant mass. Better
agreement with the muon-based results is seen when a narrow mass range around the $Z$ boson mass is used. This points
to differences between data and simulation in the modelling of the tails of the electron energy resolution. The impact of the mass range variation on the energy calibration is studied in Section~\ref{sec:alpha}. Similar results are found if
the ratio of the track momentum measured in the ID to the energy measured
in the calorimeter is used instead of the invariant mass to probe the energy calibration.

\subsection{Presampler energy scale}
\label{sec:E0}

The presampler energy scale $\alpha_{\mathrm{ PS}}$ is determined from the ratio of the presampler energies in data and simulation.
The measured energy in the presampler for electrons from $Z$ boson decays is sensitive to both $\alpha_{\mathrm{ PS}}$
and the amount of material in front of the presampler.
In order to be sensitive only to $\alpha_{\mathrm{PS}}$, the procedure to measure  $\alpha_{\mathrm{ PS}}$~\cite{PERF-2013-05} exploits the correlation between the shower development and the amount of material in front of the presampler; more precisely, several simulations with additional passive material upstream of the presampler are considered, and correlation factors between the presampler energy deposit ($E_0$) and the ratio of the energies deposited in the first two layers ($E_{1/2}$) are extracted. 
The relative calibration of the first two layers, which is described in Section~\ref{sec:E1E2}, is applied.
To minimize the impact on $E_{1/2}$ of any mismodelling of the material between the presampler and the calorimeter, 
an additional correction is applied. This last correction is extracted
from a sample of unconverted photons with small energy deposit in the presampler
to be insensitive to the material in front of the calorimeter.
The presampler energy scale is extracted as

\begin{equation*}
\alpha_{\mathrm{PS}} = \frac{E_0^{\mathrm{data}}(\eta)}{E_0^{\mathrm{MC}}(\eta)} \times \frac{1}{1 + A(\eta) \left( \frac{E_{1/2}^{\mathrm{data}}(\eta)}{E_{1/2}^{\mathrm{MC}}(\eta) b_{1/2}(\eta)} -1\right)}.
\end{equation*}

\begin{itemize}
\item $E_0^{\mathrm{data}}(\eta)$ and $E_0^{\mathrm{MC}}(\eta)$ are the average energies deposited in the presampler by the electrons from $Z$ decays in data and simulation. 
\item $b_{1/2}(\eta)$ is the ratio of $E_{1/2}$ in data and simulation for unconverted photons with small energy deposit in the presampler. It
is estimated using photons from radiative $Z$ boson decays at low \ET and inclusive photons at high \ET. The average value of these two samples is used.
\item $E_{1/2}^{\mathrm{data}}(\eta)$ and $E_{1/2}^{\mathrm{MC}}(\eta)$ are the average values of the ratio of the 
energy deposited in the first layer to
the energy deposited in the second layer for electrons from $Z$ decays in data and simulation, respectively.
After the correction with $b_{1/2}(\eta)$, this ratio is directly proportional to the amount of material in front of the presampler.

\item $A(\eta)$ represents the correlation between the changes in $E_{1/2}$ and $E_0$ when varying the material in front of the presampler.
This correlation is estimated using simulations with different amounts of material (quantity and location in radius) added in front of the presampler.
It varies between 2.5 and 1.5 for different values of $|\eta|$.
\end{itemize}
This procedure is validated using the simulation.

The measurement is performed in intervals of size 0.05 in  $|\eta|$, excluding the transition
region between the barrel and endcap calorimeters ($1.37<|\eta|<1.52$). Within a presampler module of $\Delta\eta$-size 0.2 in the barrel or 0.3 in the endcap, no significant energy scale difference is expected, so the measurements are averaged in $|\eta|$ over each module.

Uncertainties in the measurements of  $\alpha_{\mathrm{PS}}$ include the statistical uncertainties
of the various input quantities in the data and simulation. The residual variations of the measured presampler scale within
a presampler module is also taken as an uncertainty, uncorrelated between the different modules.
In the last module of the barrel, the $b_{1/2}$ correction
exhibits a significant deviation from unity for $|\eta|>1.3$. The reason of this deviation is not understood. 
For this last module, an uncertainty is obtained by comparing the $b_{1/2}$ correction averaged in the neighbouring lower $|\eta|$ interval with the value observed in this module.
Finally, the choice of $E_0$ interval used in the computation of $b_{1/2}$ is studied. From simulation studies, an upper bound in the range from 0.5 to 1.2~\GeV\ reduces the impact of uncertainties in the material in front of the presampler on $b_{1/2}$.
A variation of the result
in the data, not expected from simulation, is observed when the upper bound is changed from 0.5 to 1.2~\GeV.
It is taken as a systematic uncertainty, fully correlated across the whole barrel presampler. 

\begin{figure}[htb]
 \centering
 \includegraphics[width=0.55\textwidth]{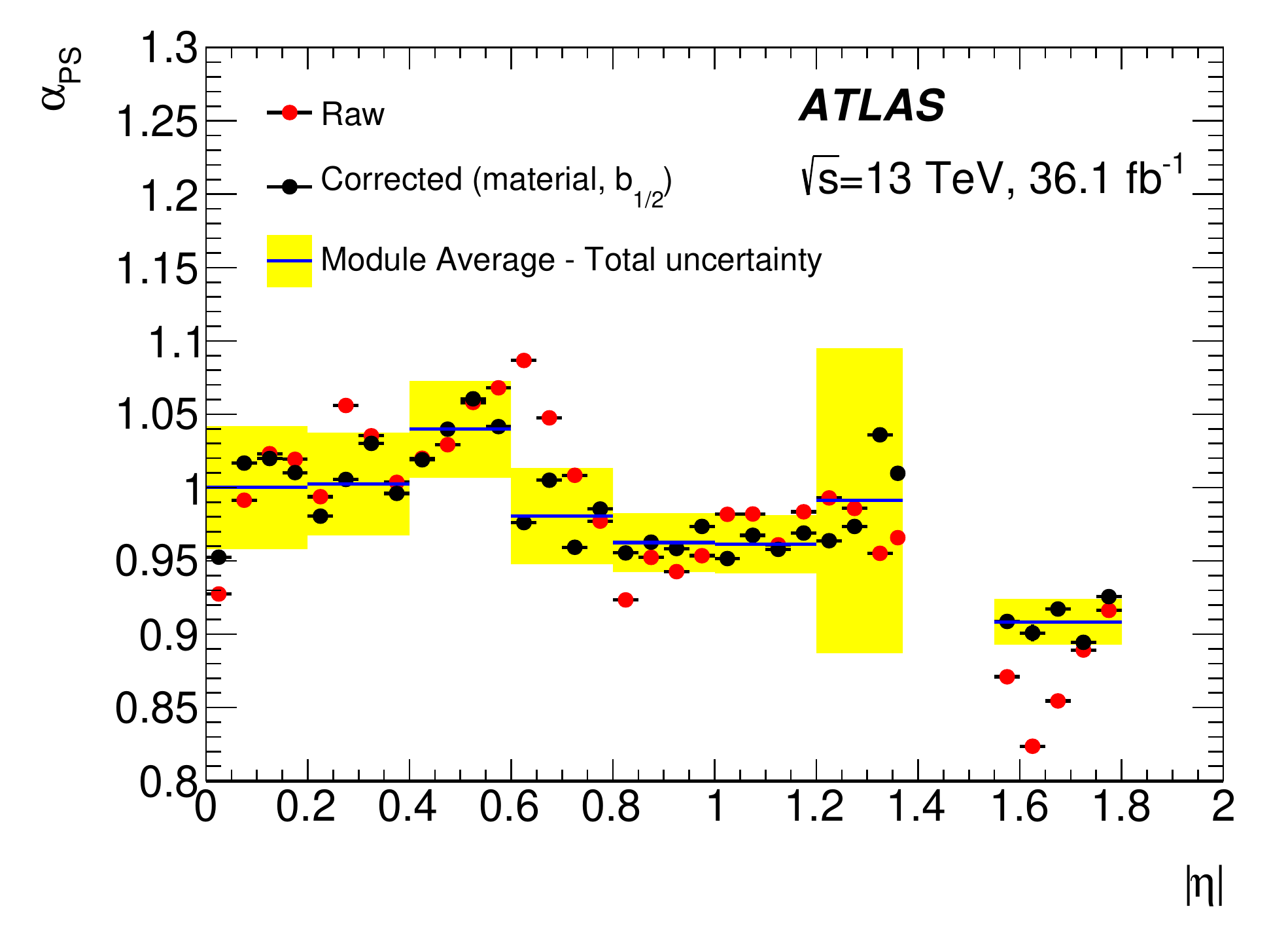}
 \caption{Measurement of the presampler energy scale ratio between data and simulation. The red points show
 the measurement before the material and $b_{1/2}(\eta)$ corrections. The black points show the measurements after these
 corrections are applied. The values averaged per presampler module in $|\eta|$ are shown together with
 the total uncertainties, represented by the shaded areas.}
 \label{fig:ps_scale}
\end{figure}

Figure~\ref{fig:ps_scale} shows the result for  $\alpha_{\mathrm{PS}}$ as a function of $|\eta|$. 
The uncertainty in  $\alpha_{\mathrm{PS}}$
varies between 3\% and 1.5\% depending on $|\eta|$.

\subsection{Pile-up energy shifts}
\label{sec:pile-up_correction}

After bipolar shaping, the average energy induced by pile-up interactions should be zero in the ideal situation
of bunch trains with an infinite number of bunches and with the same luminosity in each pair of colliding bunches. In practice, bunch-to-bunch luminosity
variations and the finite bunch-train length can create
significant energy shifts which depend on the position inside the bunch train and on the luminosity.
For most of the 2016 data, the bunch trains were made of 2 sub-trains
of 48 bunches, with a bunch spacing of 25~ns between the bunches and of 225~ns between the two sub-trains.
To mitigate this effect on the estimation of the cell energies, the average expected pile-up energy shift is subtracted
cell-by-cell. The average is computed as a function of the bunch position inside the full LHC ring, taking
into account the instantaneous luminosity per bunch, the expected pulse shape as a function of the time,
the optimal filtering coefficients used to estimate the amplitude of the signal and
a normalization factor derived from data with single colliding bunches.

Summing the cell-level contributions over an area equal to the size of an electron or photon cluster, the correction can 
reach 500~\MeV\ of transverse energy, about 75~ns after the beginning of a bunch train for 
an average of 20 interactions per bunch crossing. After the correction, residual effects up to around 
30~\MeV\ are observed in zero-bias events. They arise mostly from
inaccuracies in the predicted pulse shape. For instance in the presampler layer, the predicted pulse shape assumes
a drift time corresponding to a high-voltage value of 2000~V while in the 2016 data a significantly lower high voltage
of 1200~V was applied to reduce sporadic noise in the presampler.

To further reduce the impact of pile-up-induced energy shifts for electromagnetic clusters, an additional
correction is applied separately for each calorimeter layer
as a function of the average number of interactions per bunch crossing and as a function of $\eta$. The parameters
of this cluster-level correction are derived from random clusters in zero-bias data.

After this second correction, the residual energy shift from pile-up is less than 10~\MeV\ in transverse
energy for the data collected in 2015 and 2016.

\subsection{Improvements in the uniformity of the energy response}

After all corrections described above are applied to the
electron or photon candidates in data separately
for each calorimeter layer, the energy is computed using the
regression algorithm described in Section~\ref{sec:mva}.
Corrections for variations in the energy response as a function of the impact point of the
shower in the calorimeter affecting only the data are derived and applied to the energy of the electron or photon.
Two effects are considered and corrected:
\begin{itemize}
\item Energy loss between the barrel calorimeter modules: the barrel calorimeter is made of 16 modules of size 0.4 each
in $\Delta \phi$. The gap between absorbers increases slightly at the boundaries between modules, which leads to a reduced energy response. This effect varies as a function of $\phi$ since gravity causes the gaps to be smaller at the bottom of the calorimeter and
larger at the top. A correction of this variation is parameterized using the ratio $E/p$ of the calorimeter
energy to the track momentum as a function of $\phi$. This correction is $\lessapprox$~2\%. It is very similar
to the effect observed with the Run~1 data~\cite{PERF-2013-05}. 
\item  Effect of high-voltage inhomogeneities: in a small number of sectors (of size $0.2 \times 0.2$ in 
$\Delta\eta \times \Delta \phi$)
of the calorimeter, the applied high voltage is set to a non-nominal value due to short circuits occurring in specific LAr gaps.
The value of the high voltage is used to derive a correction applied in the cell-level calibration. Residual effects
can arise for cases where large currents are drawn. In these cases, the correction is not computed accurately.
The $\eta$--$\phi$ profiles of $E/p$ in 2015 and 2016 data are used to derive empirical corrections in the regions which are known
to be operated at non-nominal HV values. The values of the corrections are typically 1\% to 7\% and affect 2\% of the
$|\eta|<2.5$ calorimeter acceptance. Most of these corrections are similar to the ones computed in Ref.~\cite{PERF-2013-05}
with the exception of a few cases where the high-voltage setting was changed between Run~1 and Run~2.
\end{itemize}
These two corrections are validated by checking that the dielectron invariant mass in $Z\rightarrow ee$ events
is uniform as a function of $\phi$ around the $\eta$--$\phi$ regions where these corrections are applied.

\section{Data/MC energy scale and resolution measurements with $Z\rightarrow ee$ decays}
\label{sec:alpha}
\subsection{Description of the methods}
\label{sec:zee_method}

 The difference in energy scale between data and MC simulation, after all the corrections described in Section~\ref{sec:data_correction}
have been applied to the data, is defined as $\alpha_i$, where $i$ corresponds to different regions in $\eta$.
Similarly the difference in energy resolution is assumed to be an additional constant term in the energy resolution, $c_i$,
depending on $\eta$:

\begin{equation*}
  E^\text{data}=E^\text{MC}\left(1 + \alpha_i\right),  \qquad \left(\frac{\sigma_E}{E}\right)^\text{data} = \left(\frac{\sigma_E}{E}\right)^\text{MC}\oplus c_i,
\end{equation*}
where the symbol $\oplus$ denotes a sum in quadrature.

For samples of $Z \rightarrow ee$ decays, with two electrons in regions $i$ and $j$ in $\eta$, the
difference in average dielectron invariant mass is given at first order by $m_{ij}^{\mathrm{data}} = m_{ij}^{\mathrm{MC}} (1 + \alpha_{ij})$
with $ \alpha_{ij}= (\alpha_i+\alpha_j)/2$. The difference in mass resolution
is given by $({\sigma_m}/m)_{ij}^{\mathrm{data}} = ({\sigma_m}/m)_{ij}^{\mathrm{MC}} \oplus c_{ij}$,
with $c_{ij} = (c_i \oplus c_j)/2$.

To extract the values of $\alpha_{ij}$ and $c_{ij}$, the shapes of the invariant mass distributions in data are compared
with histograms of the invariant mass created from the simulation separately for each $(i,j)$ region.
In the simulation distributions the mass scale is shifted by $\alpha_{ij}$ and an extra resolution contribution 
of $c_{ij}$ is applied. The best estimates of  $\alpha_{ij}$ and  $c_{ij}$ are found by minimizing
the  $\chi^2$ of the difference between data and simulation templates.
The measurements are performed using only $(i,j)$ regions
which have at least 10 events and for which the kinematic requirement on the $\Delta \eta$ between the
electrons does not significantly bias the $Z$ mass peak position:
the minimum invariant mass implied by
the $\Delta\eta$ and \ET requirements must not exceed 70~\GeV\ for a back-to-back configuration in $\phi$.
The $\alpha_i$ and  $c_i$ parameters are estimated from the $\alpha_{ij}$ and $c_{ij}$ values 
by a $\chi^2$ minimization of the overconstrained set of equations. The procedure is validated
using pseudo-data samples generated from the simulation samples. From these studies, the residual bias of the method in the estimate of $\alpha_i$ and  $c_i$ parameters
is computed, comparing the extracted values with the values used to generate the pseudo-data samples.
This bias, which is assigned as an uncertainty, is typically (0.001--0.01)\% for $\alpha_i$ and (0.01--0.03)\% for $c_i$, depending on $|\eta|$.

Another method to derive the values of $\alpha_i$ and  $c_i$ is used as a cross-check of the results.
In this second method, both the
data and MC invariant mass distributions are fitted in each $i$-$j$ bin by an analytic function.
A sum of three Gaussian functions provides accurate modelling of the invariant
mass distribution. The parameters describing these functions are fixed to the ones fitted in the simulation
sample. When fitting the data, additional parameters corresponding to an overall energy-scale shift and a resolution
correction per $\eta$ region are added. These $\alpha_i$ and  $c_i$ parameters are then extracted from a simultaneous
fit of all $i$-$j$ regions.
The procedure is optimized and validated using
studies based on pseudo-data samples. The residual bias of the method is smaller than $0.01\%$ in the energy scale
and 0.1\% in $c_i$, except in the transition region between the barrel and endcap calorimeters, where slightly larger
effects are observed.

\subsection{Systematic uncertainties}
\label{sec:zee_uncertainties}

Several sources of uncertainty affecting the comparison of the
dielectron invariant mass distribution in $Z\rightarrow ee$ events between data and simulation are
investigated and their effects on the extraction of $\alpha_i$ and $c_i$ are estimated.

\begin{itemize}
\item Accuracy of the method: the residual bias of the main method, estimated using pseudo-data samples, described in Section~\ref{sec:zee_method},
is assigned as a systematic uncertainty.
\item Method comparison: the difference between the results of the two methods, discussed in Section~\ref{sec:zee_method}, 
is assigned as an uncertainty. For instance, the two methods have a different sensitivity to possible mismodelling of 
non-Gaussian tails in the energy resolution. The difference between the results of the two methods when applied to data can thus be larger than expected from the accuracy of the methods estimated using
pseudo-data samples. In addition, for the $c_i$ measurement, different implementations of the extraction of the $c_i$
parameters from the measured $c_{ij}$ values are compared.
\item Mass range: the results are sensitive to the mass range used to perform the comparison between data and simulation if
the non-Gaussian tails of the energy resolution are not accurately modelled.
The mass range is changed from the nominal 80--100~\GeV\ to 87--94.5~\GeV; the difference is assigned
as a systematic uncertainty.
\item The selection used to remove $i$-$j$ regions with a biased mass distribution is changed by varying the requirement
 on the minimum invariant mass implied by the $\Delta \eta$ selection in a given $i$-$j$ region.
\item Background with prompt electrons: the small contribution of backgrounds from $Z\rightarrow \tau\tau$,
diboson pair production and top-quark production, leading to a dielectron final state with both electrons originating from
$\tau$-lepton or vector-boson decays, is neglected in the extraction of the parameters $\alpha_i$ and $c_i$. 
The procedure is repeated with the contributions from these backgrounds, as estimated from MC simulations, included in 
the mass template distribution.
The differences between the results are assigned as systematic uncertainties in $\alpha_i$ and $c_i$.
\item Electron isolation: the requirement on the electron isolation strongly rejects the
backgrounds where at least one electron does not originate from a vector-boson or $\tau$-lepton decay, but from
semileptonic heavy-flavour decay, from conversions of photons produced in jets or from hadrons. To estimate
the residual effect of these backgrounds on the result, the extraction of $\alpha_i$ and $c_i$ is
repeated without the isolation selection and the differences are assigned as systematic uncertainties.
\item Electron identification: the selection uses Medium quality electrons. Small correlations between
the electron energy response and the quality of the electron identification are expected, since the latter uses as input
the lateral shower development in the calorimeter. If these correlations are not properly modelled in the simulation, the
data-to-MC energy scale and resolution corrections can depend on the identification requirement.
In order to make the corrections applicable to measurements using electron selections that are different from those 
used in this paper, additional systematic uncertainties are estimated by comparing the results for $\alpha_i$ and $c_i$ 
obtained using the Tight identification requirement instead of the Medium quality requirement.
\item Electron bremsstrahlung probability: electrons can lose a significant fraction of their energy by
bremsstrahlung before reaching the calorimeter. To determine to what extent the measured $\alpha_i$ and $c_i$ parameters
are intrinsic to the calorimeter response and to what extent they are sensitive to the modelling of energy loss before the 
calorimeter, a requirement on the fraction of electron bremsstrahlung is applied, using the change in track curvature 
between the perigee and the last measurement before the calorimeter. The difference in $\alpha_i$ and $c_i$ values 
obtained with or without this additional requirement applied is assigned as an uncertainty.
\item Electron reconstruction, trigger, identification and isolation efficiencies: the MC simulation is corrected for
the difference in efficiencies between data and simulation~\cite{ATLAS-PERF-2017-01}. These corrections, which depend on 
$\ET$ and $\eta$, can slightly change the shape of the invariant mass distribution predicted by the MC simulation. 
The corrections are varied within their uncertainties and the resulting uncertainty in $\alpha_i$ and $c_i$ is estimated.
\end{itemize}

All the listed uncertainties are computed separately in each $\eta$ interval.
The typical values in different $\eta$ ranges are given in Table~\ref{table:alpha_systematics}.
The table shows a wide range of uncertainties for the interval $1.2<|\eta|<1.8$. Inside this interval, the uncertainties are largest for the region around $|\eta|=1.5$.
For $|\eta|>2.4$, near the end of the acceptance, the uncertainties are significantly larger than for the other regions. 

The total systematic uncertainty in $\alpha_i$ and $c_i$ is computed adding in quadrature all the effects described above.
This procedure may lead to slightly pessimistic uncertainties because some of the variations discussed above can 
double-count the same underlying source of uncertainty and also because the results must remain valid in a variety of 
final states and with different event selections.
The systematic uncertainty in $\alpha_i$ varies from $\approx$~0.03\% in the central part of the barrel calorimeter,
to $\approx$~0.1\% in most of the endcap calorimeter and reaches a few per mille in the transition region
between the barrel and endcap calorimeters. The uncertainty in $c_i$
is typically 0.1\% in most of the barrel calorimeter, 0.3\% in the endcap and as large as 0.6\% in the
transition region. The statistical uncertainty from the size of the $Z\rightarrow ee$ sample
in the 2016 dataset is significantly smaller than the systematic uncertainty.

Uncertainties from the modelling of the $Z$ boson production and decay, including
the modelling of final-state QED radiation from the charged leptons, were investigated
in Ref.~\cite{PERF-2013-05} and found to be negligible compared with the total uncertainty
quoted above.

\begin{table}
\begin{center}
\caption{Ranges of systematic uncertainty in $\alpha_i$ and $c_i$ for different $\eta$ ranges.}
\resizebox{\textwidth}{!}{
\begin{tabular}{lcccccc}
\toprule
                    & \multicolumn{3}{c}{Uncertainty in $\alpha_i$ $\times 10^3$ } & \multicolumn{3}{c}{Uncertainty in $c_i$ $\times 10^3$} \\
\cmidrule(l){2-7}                    
\hline
$|\eta|$ range      &  0--1.2  & 1.2--1.8   & 1.8--2.4  &  0--1.2  & 1.2--1.8  & 1.8--2.4 \\
\midrule
Uncertainty source                  & & & & & & \\
Method accuracy                     & (0.01--0.04)  & (0.04--0.10) & (0.02--0.08) &  (0.1--0.7) & (0.2--0.4) & (0.1--0.2)  \\
Method comparison                   & (0.1--0.3)    & (0.3--1.2) & (0.1--0.4) &  (0.1--0.5) & (0.7--2.0) & (0.2--0.5)  \\
Mass range                          & (0.1--0.5)    &  (0.2--4.0)   & (0.2--1.0)    &  (0.2--0.8) & (1.0--3.5) & 1.0         \\
Region selection                    & (0.02--0.08)  &  (0.02--0.2) & (0.02--0.2)  &  (0--0.1)    & 0.1        & (0.2--1.0)  \\
Bkg. with prompt electrons          & (0--0.05)     &  (0--0.1)     &  (0--0.5)    &  (0.1--0.4) & 0.2        & (0.1--0.2)  \\
Electron isolation requirement      & (0--0.02)      &  (0.02--5.0) & (0.02--0.20) & (0.1--0.9)  & (0.1--1.5)  & (0.5--1.5)   \\
Electron identification criteria    & (0--0.30)      &  (0.20--2.0) & (0.20--0.70) &   (0--0.5)   &  0.3       &  0.0         \\
Electron bremsstrahlung removal     & (0--0.30)      &  (0.05--0.7) & (0.20--1.0) &  (0.2--0.3) & (0.1--0.8) & (0.2--1.0)  \\
Electron efficiency corrections     &  0.10         &  (0.1--5.0)  & (0.10--0.20) &   (0--0.3)   &  (0.1--3.0) & (0.1--0.2)  \\
\midrule
Total uncertainty                   &  (0.2--0.7)   & (0.5--10)     &  (0.6--2.0) &  (0.3--1.2) &  (1.0--6.0) & (2.0--3.0) \\
\bottomrule
\end{tabular}
}
\label{table:alpha_systematics}
\end{center}
\end{table}

\subsection{Results}

The extraction of the energy scale correction is performed in 68 intervals in $\eta$. These intervals cover a
range of 0.1 in the barrel calorimeter and are usually a bit smaller in the endcap calorimeter. 
The resolution corrections $c_i$ are computed
in 24 intervals. In each of these $\eta$ regions, $c_i$ corresponds to the effective additional constant term for the data
after the fine-grained $\eta$-dependent energy scale corrections are applied.

\begin{figure}[htb]
          \centering
          \subfloat[]{\includegraphics[width=0.7\figwidth]{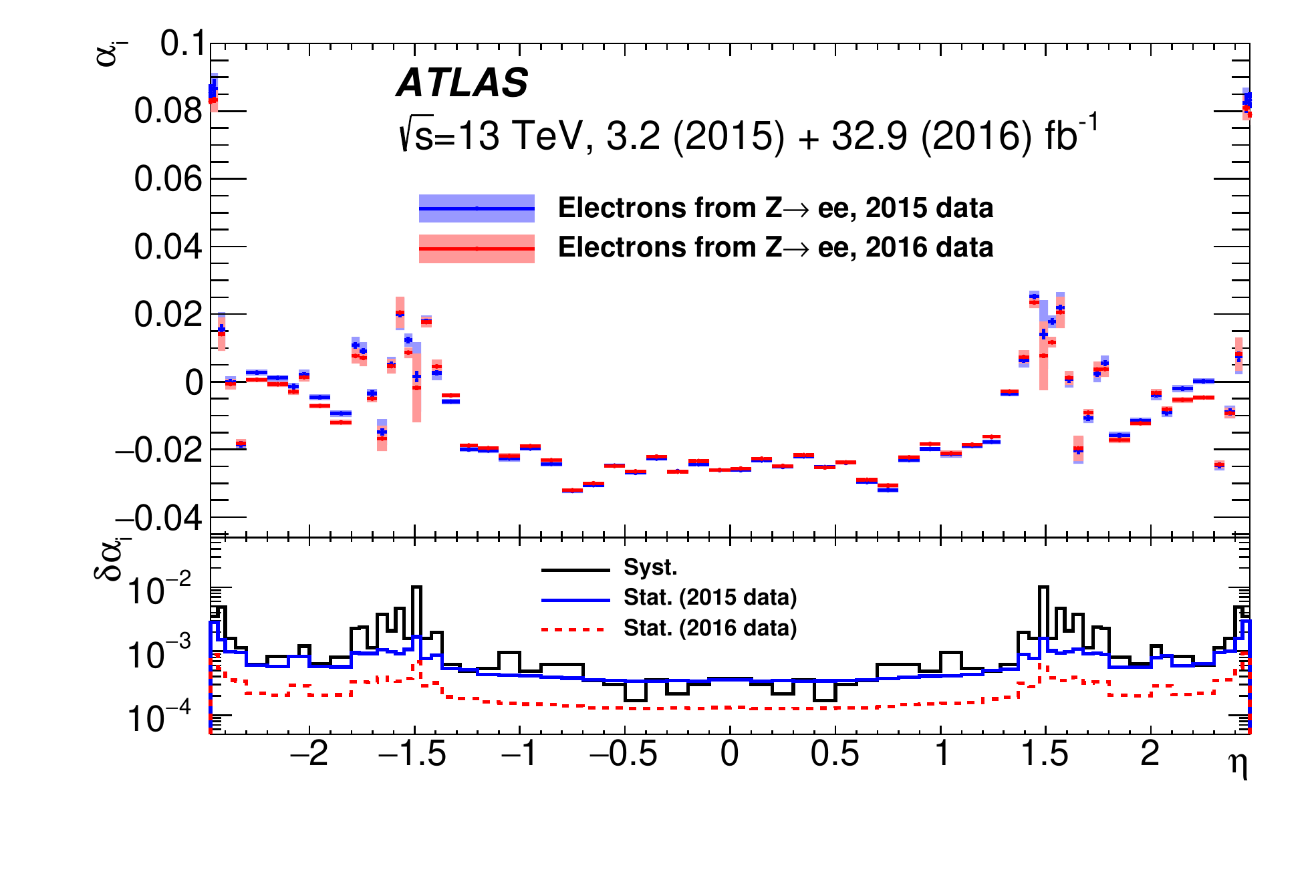}} \\
          \subfloat[]{\includegraphics[width=0.7\figwidth]{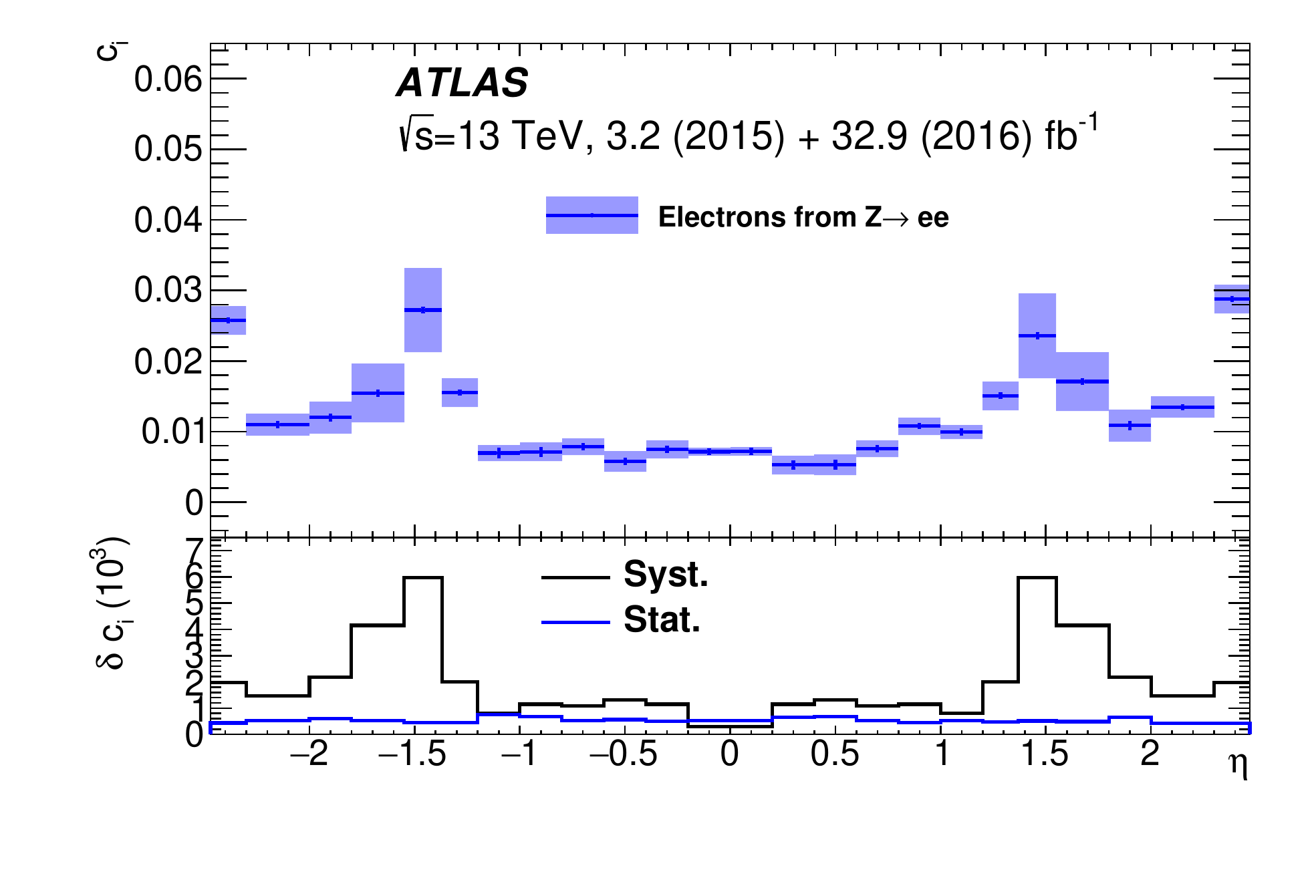}}
        \caption{Results of the data-to-MC calibration from $Z\rightarrow ee$ events for (a) the energy scale
 corrections ($\alpha_i$) and (b) the energy resolution corrections ($c_i$)
 as a function of $\eta$. The systematic and statistical uncertainties are shown separately in the bottom panels.}
        \label{fig:zee_calibration}
\end{figure}

Figure~\ref{fig:zee_calibration} shows the results for $\alpha_i$ and $c_i$ from the 2015 and 2016 datasets.
The energy scale correction factors are derived separately for the 2015 and 2016 datasets to
take into account the difference in instantaneous luminosity between the two samples which is detailed
in Section~\ref{sec:luminosity_scale}.
As the resolution corrections are consistent between the two years, they are derived from the combined dataset, after the energy scale correction has been applied.

The additional constant term of the energy resolution present in the data is typically less than 1\% in most of the
barrel calorimeter. It is between 1\% and 2\% in the endcap, with slightly larger values in the transition
region between the barrel and endcap calorimeters and in the outer $|\eta|$ range of the endcap.

No parameterization of the $\alpha_i$ as a function of $\phi$ is performed.
The calorimeter uniformity in $\phi$ is typically at the 0.5--1\% level
and the residual variations of the energy response with $\phi$ contribute at this level to the additional constant term. 
These variations are a bit larger in the endcap calorimeter because of the larger variation of the calorimeter
gaps under the influence of gravity as a function of $\phi$.

\begin{figure}[htb]
         \centering
          \includegraphics[width=0.7\figwidth]{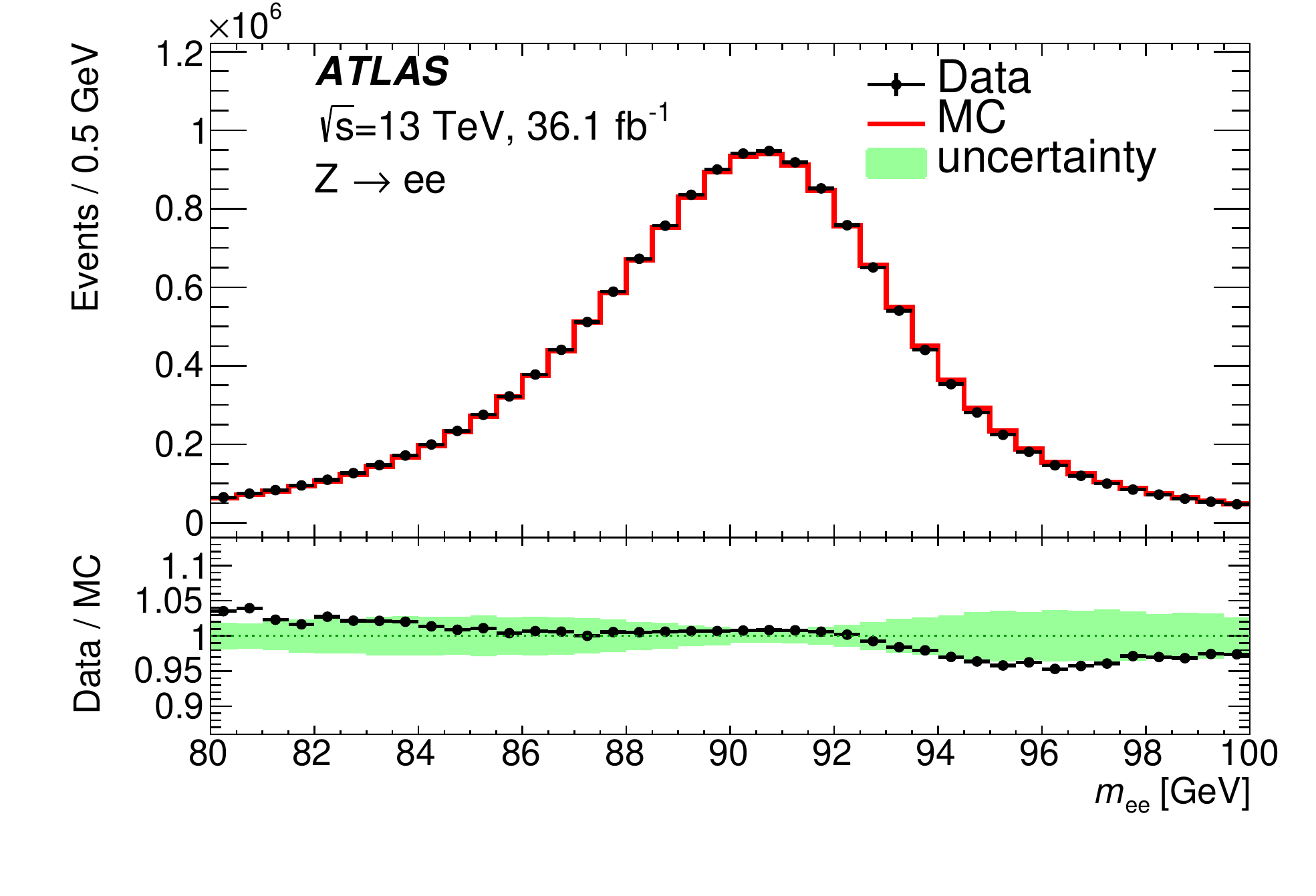}
         \caption{Comparison of the invariant mass distribution of the two electrons in the selected $Z\rightarrow ee$ candidates,
         after the calibration and resolution corrections are applied.
          The total number of events in the simulation is normalized to the data. The ratio is shown in the bottom plot.
        The uncertainty band of the bottom plot represents to the impact of the uncertainties in the calibration and
        resolution correction factors. }
      \label{fig:zee_mass_ratio}
\end{figure}

Figure~\ref{fig:zee_mass_ratio} shows a comparison of the invariant mass distribution for $Z\rightarrow ee$
candidates between data and simulation after the energy scale correction has been applied to the data and the
simulation corrected for the difference in energy resolution between data and simulation.
No background contamination is taken into account in this comparison. The non-$(Z \rightarrow ee)$ background
is smaller than $\approx$~1\% over the full shown mass range.
The uncertainties in the ratio of the data and simulation distributions are computed varying
the  $\alpha_i$ and $c_i$ correction factors within their uncertainties.
These uncertainties are estimated as discussed in Section~\ref{sec:zee_uncertainties} and
take into account changes in the selections applied to $Z\rightarrow ee$ candidates and variations
in the mass window used to extract the calibration.
The decrease in the ratio
near a mass of 96~\GeV\ is most likely related to imperfect modelling of the tails of the energy resolution
by the simulation, which affects the extraction of the energy scale and resolution correction factors.
This variation is covered by the estimated uncertainties in the
correction factors.
Within these uncertainties, the data and simulation are in fair agreement.

\subsection{Stability of the energy scale, comparison of the 2015 and 2016 data}
\label{sec:luminosity_scale}

\begin{figure}[htb]
  \centering
         \includegraphics[width=0.7\figwidth]{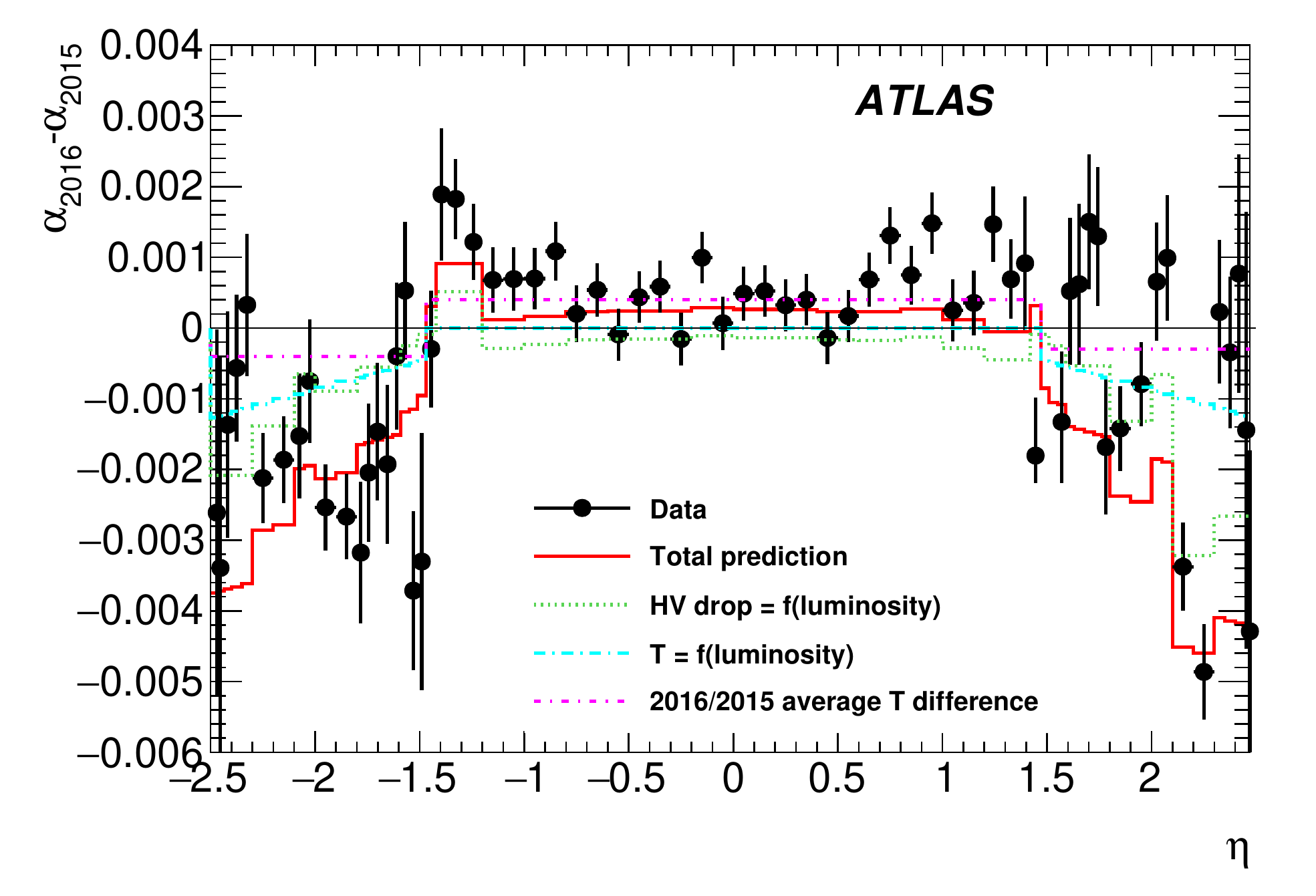}
         \caption{Comparison between the energy scale corrections derived from $Z\rightarrow ee$ events in 2015 and 2016
 as a function of $\eta$. The difference of the energy scales measured in the data are compared with predictions taking into account the luminosity-induced high-voltage
 reduction and LAr temperature changes as well as the small overall difference in LAr temperature between 2015 and 2016.}
      \label{fig:delta_alpha}
\end{figure}

Figure~\ref{fig:delta_alpha} shows a comparison of the energy scale corrections extracted from the 2015 and 2016 data.
Small differences up to a few per mille are observed, mostly in the endcap calorimeter. These effects can qualitatively
be explained by the difference in instantaneous luminosity between the two years: the average instantaneous luminosity is
around \SI{0.3E34}{\centi\meter^{-2}\second^{-1}} in 2015 and  \SI{E34}{\centi\meter^{-2}\second^{-1}} in 2016.
The following effects are expected to create small variations of the calorimeter response as a function of the luminosity:
\begin{itemize}
\item The large amount of deposited energy increases the temperature of the calorimeter, creating a small drop in the energy response of about
 --2\%/K~\cite{ATL-LARG-95-029}.
The LAr temperature is measured with probes inside the cryostat at the inner and outer radius of the endcap entrance face.
The measured temperature increase is $\SI{0.07}{\kelvin}$ at the inner radius ($|\eta|=2.65$) when collisions occur at high luminosity and $\SI{0.02}{\kelvin}$ at the outer radius ($|\eta|=1.4$).
The effect on the energy response is estimated assuming a linear temperature variation as a function of $\eta$
in the endcap.
In addition, there is a small change in the LAr temperature of the different cryostats in the absence of collisions between 2015 and 2016.
\item The large amount of deposited energy in the liquid-argon gap creates a current in the HV lines. The current $I$ induced on the HV line
is equal to the total ionization current, from the drift of electrons and ions, which is created by the steady flux of deposited energy in the calorimeter.
Since there is a significant resistance, $R$, between the power supply where the voltage is set to a constant value and
the LAr gap, the voltage effectively applied to the gap is reduced by $R \times I$.
The pattern of resistances across the LAr electrodes~\cite{AUBERT2005558} is quite complex, but the dominant contribution to the resistance is
due to the filter-box resistance in the high-voltage feedthrough~\cite{1748-0221-2-10-T10002}.
The HV drop can thus be estimated
from the current drawn by the power supply and the value of the filter-box resistance, which
is \SI{100}{\kilo\ohm} for the EM calorimeter ($|\eta| <2.5$).
The change in HV induces a change in the calorimeter response because the drift velocity of ionization electrons varies
approximately with the power 0.3 of the electric field in the gap~\cite{PhysRev.166.871,Yoshino:1976zz,LARG-2009-02} 
and the amplitude of the shaped calorimeter signal is proportional to the drift velocity.
\end{itemize}
The predictions for the energy scale difference are included in Figure~\ref{fig:delta_alpha}. The changes observed in the
data in the endcap are qualitatively reproduced although the difference seen in the data is somewhat smaller
than expected for the highest $|\eta|$ values.

\begin{figure}[htb]
  \centering
  \subfloat[\label{fig:mee_mu1}]{\includegraphics[width=0.6\figwidth]{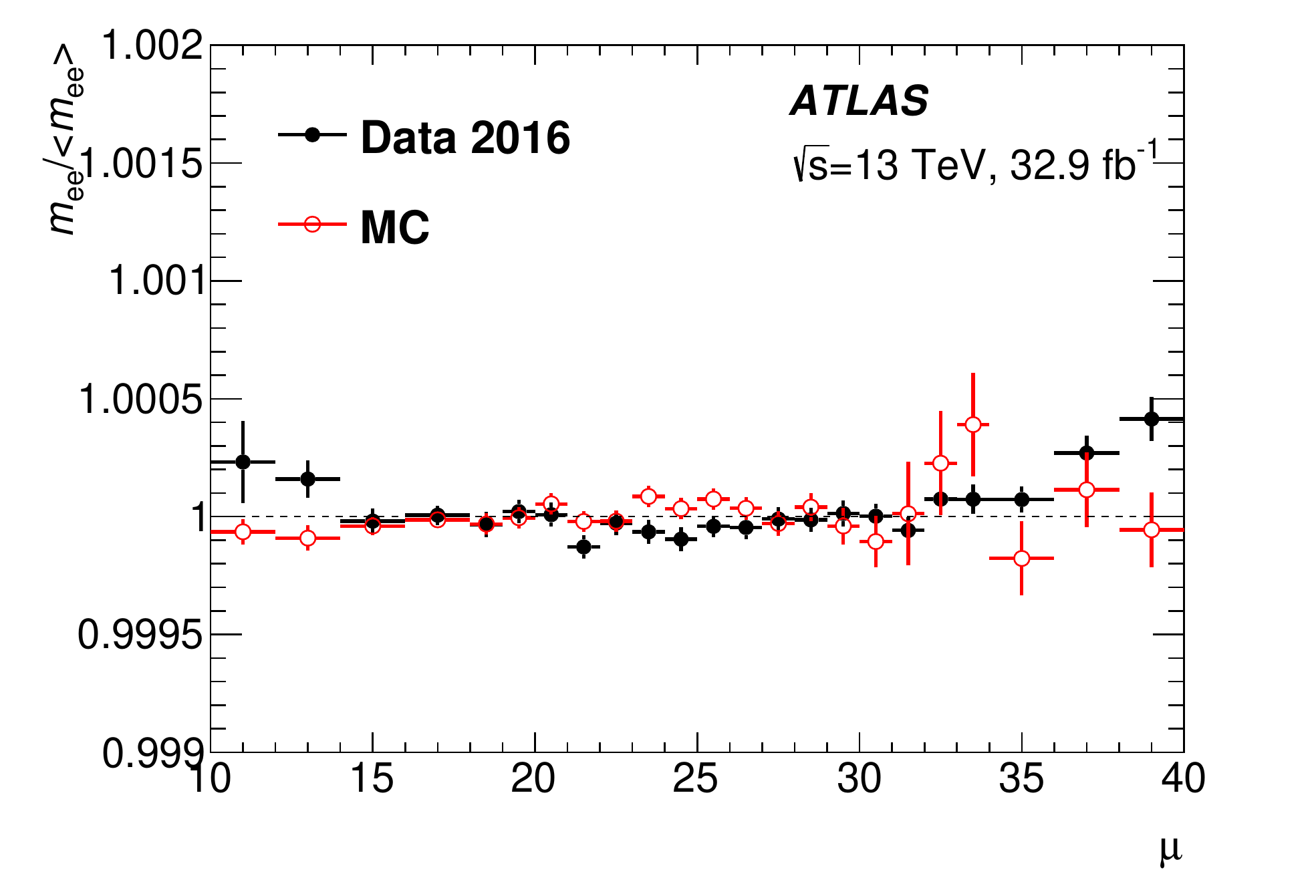}}
  \subfloat[\label{fig:mee_mu2}]{\includegraphics[width=0.6\figwidth]{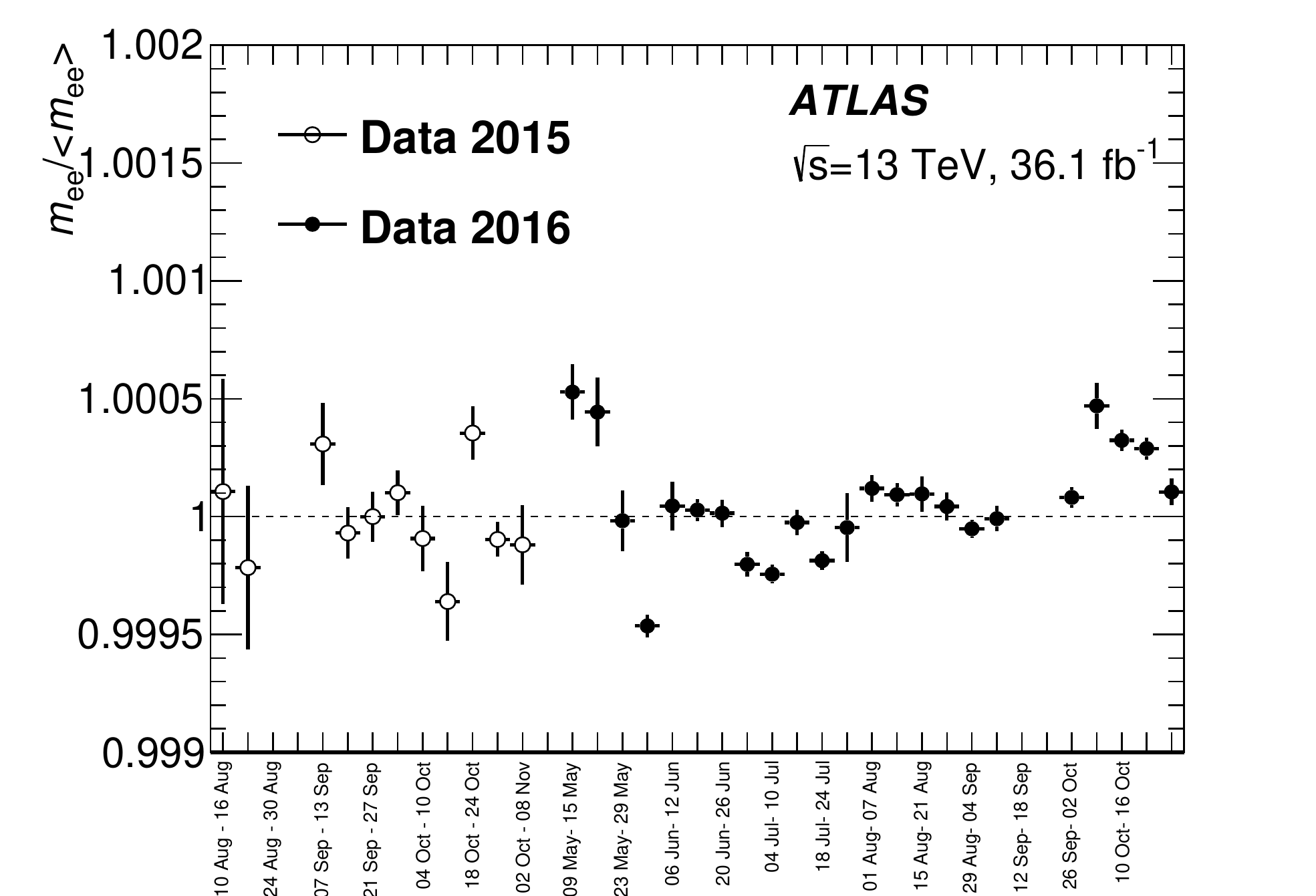}}
  \caption{Relative variation of the peak position of the reconstructed dielectron mass distribution in $Z\rightarrow ee$ events \protect\subref{fig:mee_mu1}
  as a function of the average number of interactions per bunch crossing for the 2016 data and \protect\subref{fig:mee_mu2} as a function of time over the full 2015 and 2016 data-taking periods.}
      \label{fig:mee_mu}
\end{figure}

Figure~\ref{fig:mee_mu1} shows the variation of the reconstructed peak position of the dielectron mass distribution as a function
of the average number of interactions per bunch crossing for the data collected in 2016.
When integrated over the full $|\eta|$ range, the variation of the energy scale with the number of interactions
per bunch crossing is well below the 0.1\% level in the data. No effect is visible in the simulation either.
Figure~\ref{fig:mee_mu2} shows the stability of the energy scale as a function of time, probed with $Z\rightarrow ee$
events. The stability is significantly better than 0.1\%.

\section{Systematic uncertainties in the energy scale and resolution}
\label{sec:systematics}
Several systematic uncertainties impact the measurement of the energy of
electrons or photons  (converted or unconverted) in a way that depends on their transverse energy and pseudorapidity.
After the $Z$-based calibration, which  fixes the energy scale and its uncertainty
for electrons with transverse energy close to the average of those produced in $Z$ decays,
the relative uncertainty for any given electron or photon with transverse energy $\ET$ and pseudorapidity $\eta$
can be written as:
\begin{equation*}
\delta E_i^{e,\gamma}(\ET,\eta) = \Delta E_i^{e,\gamma}(\ET,\eta) - \Delta E_i^{e}\left(\left<\ET^{e(Z\rightarrow ee)}\right>,\eta\right),
\label{equation:energy_syst}
\end{equation*}
where $\Delta E_i^{e,\gamma}(\ET,\eta)$ is, for a given uncertainty variation $i$, its relative impact on the energy
as a function of $\eta$ and $\ET$ before the application of the $Z$-based calibration and $\left<\ET^{e(Z\rightarrow ee)}\right> \approx~40$~\GeV\ is
the average transverse energy for electrons produced in $Z$ boson decays.
The $Z$-based calibration absorbs the effect for electrons with $\ET = \left<\ET^{e(Z\rightarrow ee)}\right>$ and leaves the
residual uncertainty $\delta E_i^{e,\gamma}(\ET,\eta)$.

For a given uncertainty variation $i$, $\delta E_i^{e,\gamma}(\ET,\eta)$ can change sign as a function of \ET. This
is often the case for electrons where $\delta E_i^{e,\gamma}\left(\left<\ET^{e(Z\rightarrow ee)}\right>,\eta\right)$ is zero.
In addition, for most of the considered uncertainty variations, their impact on photon energy is computed separately
for reconstructed converted and unconverted photons. The converted photons have a shower development more similar to 
that of electrons and therefore usually smaller energy scale systematic uncertainties than unconverted photons.
The different uncertainties affecting the energy scale of electrons and photons are described in this section.

\subsection{Uncertainties related to pile-up}

After correction, the energy shift induced by pile-up is estimated to be less than $\pm\SI{10}{\MeV}$ in 
transverse energy (see Section~\ref{sec:pile-up_correction}).
The energy scale uncertainty after the $Z$-based calibration is thus
$\delta E_i^{e,\gamma}(\ET,\eta) = \SI{10}{\MeV}/\ET - \SI{10}{\MeV}/\left<\ET^{e(Z\rightarrow ee)}\right>$.
For electrons or photons with \ET ~=~10~\GeV,
the uncertainty is $\approx$~0.075\% and it is $\approx$~0.02\% for $\ET > 100$~\GeV.

\subsection{Impact of the layer calibration uncertainties}

The uncertainties in the calibration of the first two layers of the calorimeter and of the presampler
are discussed in Sections~\ref{sec:E1E2} and ~\ref{sec:E0}.
The impact of these uncertainties
depends on the reconstructed particle energy since the fraction
of energy observed in the presampler and in the different calorimeter layers is
a function of the energy and of the particle type. Typically, the fraction of energy deposited
in the presampler and in the first layer increases when the energy decreases and these
fractions are higher for electrons and converted photons than for unconverted photons.
The effect of the uncertainties in the layer calibration is propagated to the energy measurement
using parameterizations of these fractions as a function of \ET and $\eta$.
In the low-$|\eta|$ region of the barrel calorimeter, the impact of the uncertainties affecting the calibration
of the different calorimeter layers and of the presampler is for instance $\approx \pm 0.2\%$ on
the electron energy scale in the range $10 < \ET < 200$~\GeV.

Since only a small energy fraction is deposited in the third layer of the EM calorimeter,
uncertainties in the relative calibration of this layer have a negligible impact on the total
calibration uncertainty.

\subsection{Impact of the $E_4$ scintillator calibration}

In the region $1.4<|\eta|<1.6$, the signals of the $E_4$ scintillators are used as input to the energy measurement,
as discussed in Section~\ref{sec:mva}.
The accuracy of the calibration of the energy deposited in these scintillators varies from 4\% to 6\% depending on
$|\eta|$.  These uncertainties are based on the comparison of the energy deposited in $E_4$
between data and simulation for electrons from $Z \rightarrow ee$
events and the monitoring accuracy of the time-dependence of the reconstructed signal of the scintillator.
The impact on the total electron energy
is found to be typically 0.3 times the uncertainty in the $E_4$ scintillator calibration, where the factor 0.3
reflects the typical weight of the $E_4$ cell information in the calibration regression algorithm.

\subsection{Uncertainties due to the material in front of the calorimeter}
\label{sec:material_uncertainty}

The material in front of the calorimeter was studied in Ref.~\cite{PERF-2013-05} using data collected in 2012.
The impact of the material on the energy response depends on the radial location of the material.
Different uncertainty variations are thus considered for material in different regions in front of the calorimeter.

\begin{itemize}
\item Material inside the active area of the ID. From measurements performed during the detector
construction~\cite{PERF-2007-01}, the material integral is known with a $\pm 5\%$ accuracy in four independent $|\eta|$ regions.
In addition, uncertainties in the description of the material in the new innermost pixel layer and in the
modified layout of the pixel detector services at low radius are added for the 2015--2016 data. These
uncertainties (expressed in units of radiation lengths) range from 0.01 or less for $|\eta|<1.5$ to 0.05 at $|\eta|=2.0$ and 0.2
at $|\eta|=2.3$. 
These uncertainties include the impact of missing some detector components in the description of the new innermost pixel layer 
and uncertainties in the description of the modified services for the detector description used in this paper,
which corresponds to the \textit{original} geometry model described in Ref.~\cite{PERF-2015-07}.
\item Material between the end of the active area of the ID and the presampler (or the calorimeter for $|\eta|>1.8$)
and material between the presampler and the calorimeter (for $|\eta|<1.8$). The uncertainties in the amount of material in these
regions are the same as the ones derived from the Run~1 studies, since the detector layout is unchanged.
The amount of material in these regions was constrained by the longitudinal development of electron- and photon-induced
showers in Run~1 data. The uncertainties
include the longitudinal shower shape modelling uncertainties after calibration of the presampler and the first two calorimeter layers,
in addition to the uncertainties in the GEANT4 simulation.
The latter is estimated by varying the associated GEANT4 options to test refinements
in the theoretical description of bremsstrahlung and photon conversion cross sections,
as well as alternative electron multiple-scattering models.
The total uncertainty in the amount of material between the end of the active ID area and the presampler
is typically 0.03 to 0.1 radiation lengths for $|\eta|<1.4$, up to 0.7 radiation lengths at $|\eta|=1.5$ and 0.1 to 0.3 radiation lengths
for $1.5<|\eta|<1.8$. The uncertainty in the amount of material between the presampler and the first calorimeter layer
is typically 0.04 to 0.1 radiation lengths in the full range $|\eta|<1.8$.
Finally, the uncertainty in the material in front of the calorimeter for the region $1.8<|\eta|<2.5$ is about 0.1 to 0.15 radiation
lengths.
\end{itemize}

In the low-$|\eta|$ region of the barrel calorimeter, the total uncertainties related to the description of
material in front of the calorimeter give an uncertainty in the energy scale of $\pm$0.3\% for \ET=10~\GeV\ electrons.
This uncertainty increases to $\pm$0.5\% at $|\eta|=2.3$.

\subsection{Non-linearity of the cell energy measurement}
\label{sec:non_linearity_uncertainty}

Non-linearity in the cell energy measurement induces a dependence of the energy response on the energy of the particle.
The linearity of the readout electronics is better than 0.1\%~\cite{1748-0221-5-09-P09003} in each of the three gains
used to digitize the calorimeter signals in the ranges where they are used to collect data.
However, the relative calibration of the different readout gains is less well known.
To study the accuracy of this relative calibration, data recorded
under special conditions in 2015 and 2017 are used, corresponding to an integrated luminosity of 12~pb$^{-1}$ in 2015 and 160~pb$^{-1}$ in 2017.
For these data, the threshold to switch from high gain (HG) to medium gain (MG)
readout for the cells in the second layer was significantly lowered, by a factor 5. With this special configuration,
almost all electrons
from $Z$ boson decays have at least the highest-energy cell in layer two recorded in the MG readout. In the
standard configuration, the HG readout is almost always used, at least in the barrel, where the transition between
the two gains is typically at an energy of $\approx$~25~\GeV\ for the cells in the second layer at low $|\eta|$.

The reconstructed dielectron invariant mass distribution in these data is compared with the one in data recorded with the standard
gain transition configuration taken around the same time.
To properly calibrate the ADC-to-current conversion function for low numbers of ADC counts in the MG range in the
special configuration, a non-linear ADC-to-current conversion is used, derived from dedicated pulser calibration runs.
Uncertainties at the 0.05\% level in this conversion can arise from non-linearity of the calibration system for this situation.

\begin{figure}[htb]
  \centering
  \includegraphics[width=0.50\textwidth]{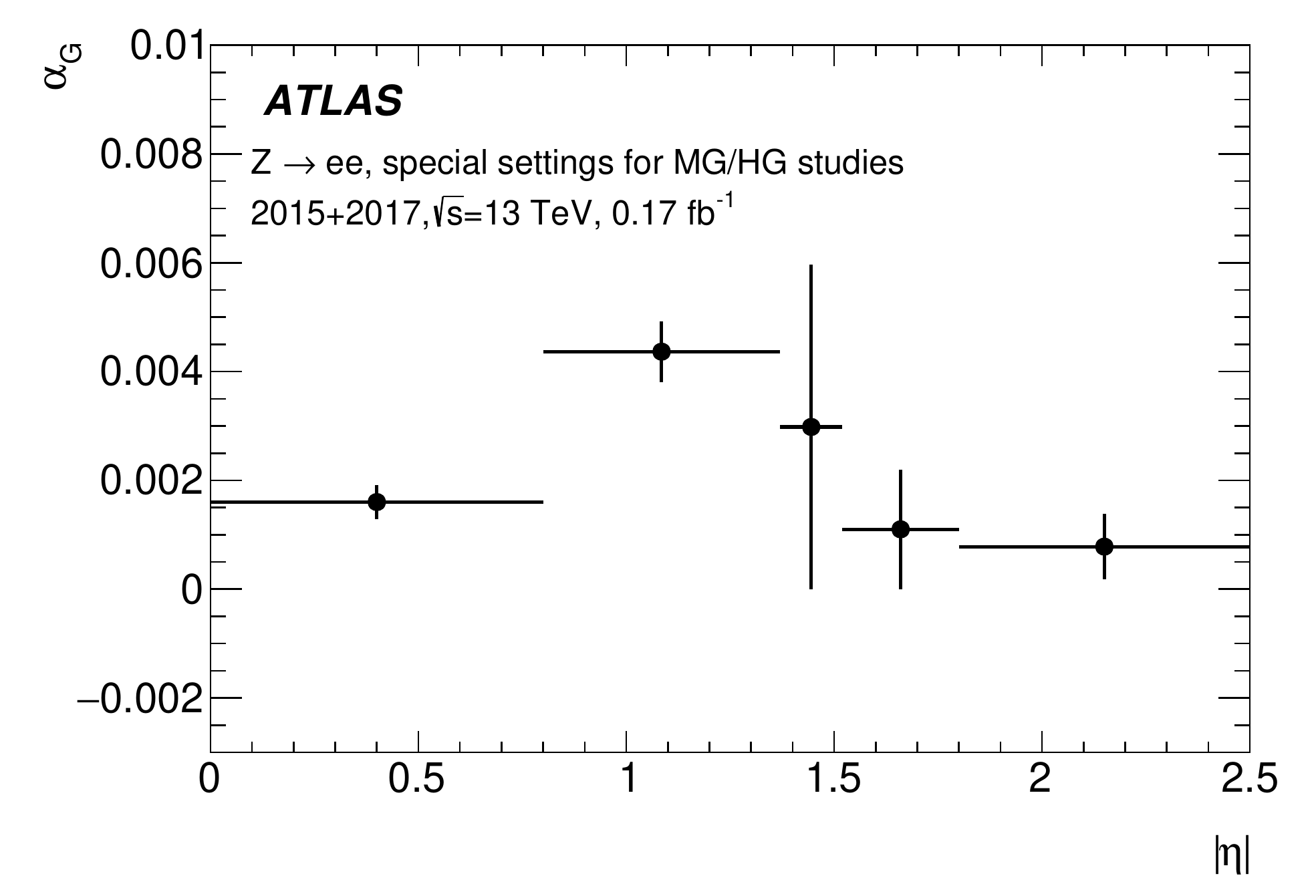}
  \caption{Difference of energy scales, $\alpha_\mathrm{G}$, extracted from $Z\rightarrow ee$ events, as a function of $|\eta|$ between data recorded
with the standard thresholds for the transition between HG and MG in the readout of the layer-two
cells and data with lowered thresholds. Only statistical uncertainties are shown.}
 \label{fig:alpha_specialRun}
\end{figure}

Figure~\ref{fig:alpha_specialRun} shows the measured values of the energy scale difference between
the two datasets, $\alpha_\mathrm{G}$, as a function of $|\eta|$. If the HG and MG are perfectly intercalibrated,  $\alpha_\mathrm{G}$
will be zero.
A small but significant
difference is observed, especially in the region $0.8<|\eta|<1.37$. Further investigations did not reveal any significant
energy dependence or further $\eta$ dependence of this effect.
The observed difference is assigned as a systematic uncertainty.
This uncertainty is assumed to be a
scale factor between the calibration of the two gains, independent of the cell energy.

From this measurement, the impact of the gain intercalibration uncertainty for data taken under standard conditions can be
written as a function of the particle type and of \ET and $\eta$ as follows:

\begin{equation*}
\frac{\Delta E}{E} = \alpha_\mathrm{G} (\eta) \cdot \frac{1}{\delta_Z(\eta)} \cdot \delta_\mathrm{G}^{e,\gamma}(\eta,\ET)
\label{eqn:special_run_formula}
\end{equation*}

where:
\begin{itemize}
\item  $\alpha_\mathrm{G} (\eta)$  is the measured energy scale difference as a function of $\eta$ from $Z \rightarrow ee$ decays, comparing the
data recorded with lower gain threshold with data recorded in standard conditions. 
\item $\delta_Z(\eta)$ quantifies the fractional change in energy for electrons from $Z$ boson decays between the data with lower
and standard thresholds for a given change in the energy recorded in MG in the second layer.  This sensitivity factor is about 0.3 to 0.4 in the barrel calorimeter and about 0.2 to 0.25 in the endcap calorimeter.
It takes into account the fact that only a fraction of the electron energy is recorded in MG layer-two cells in the data with special settings and
that in the data with normal settings some layer-two cells can be read out in MG, especially in the endcap where
the electron energies are larger.
\item  $\delta_\mathrm{G}^{e,\gamma}(\eta,\ET)$ quantifies, for a given particle, the fractional change in the total energy
for a given change in the energy recorded in MG in the second layer, when the standard settings of the gain threshold are used.
It is estimated using simulated single-particle samples. It is close to 0 up to \ET $\approx$~40 to 60~\GeV, depending on $\eta$
and on the particle type, and then rises to reach an asymptotic value of about 0.8 for \ET above a few hundred \GeV, reflecting the
fraction of electron energy measured with second-layer cells read out in MG.
\end{itemize}

The calibration uncertainty for the low-gain readout is assumed to be the same as for the MG.
The low-gain readout is used in the second layer for electrons or photons with transverse energy above 350 to 500~GeV (100 to 300~GeV) 
depending on $\eta$ in the barrel (endcap).
Studies of a small sample of high transverse momentum $Z$ boson decays, where some of the electrons are recorded in low gain, do
not indicate any significantly larger effect.

The uncertainty in the total energy is typically 0.05\% to 0.1\% depending on $\eta$ for photons of \ET=~60~\GeV. It reaches 0.2\% to 1\%
for very high energy electrons and photons.

The uncertainty in the MG-to-HG relative calibration in the first layer 
has a much smaller  effect than the one in the second layer, 
except in the endcap for $1.8< |\eta| < 2.3$. 
In this region, the relative  calibration of the two gains in the first layer was found 
to be sensitive to the pile-up-dependent optimization  of the optimal filtering coefficients 
with an uncertainty rising from 1\% to 5\%. 
In this region, the highest-energy cell in the first layer of most high-\ET electromagnetic showers is recorded in MG. 
The application of the muon-based layer calibration to electrons or photons therefore leads to an uncertainty in the energy scale for electrons or photons,
which  reaches about 0.8\% for unconverted photons with $|\eta| >2.0$.

In the studies reported in Ref.~\cite{PERF-2013-05}, the gain calibration uncertainty was investigated by splitting the sample of $Z$ boson decays
recorded in standard conditions according to the gain used to measure the highest-energy cell in the second layer.
The uncertainty estimated in this way combined the effect of the genuine intercalibration of the different readout gains with
systematic effects related to the modelling of lateral shower shapes. The latter impacts a selection based on the gain of the 
highest-energy cell since showers with narrow lateral shape are more likely to have a second-layer cell with high
energy deposit and thus are more likely to have this cell recorded in MG. For the results presented in this paper, the
two uncertainties are separated with the relative gain calibration discussed above and with a
separate investigation of the impact of the modelling of lateral
shower shape on the energy response, reported in Section~\ref{sec:lateral_shower_shape}.

\subsection{Modelling of the lateral shower shape}
\label{sec:lateral_shower_shape}

Any energy-dependent mismodelling of the energy response as a function of the lateral shower shape can create differences between
data and simulation
in the energy response relative to the energy response for electrons at $\ET~=~ 40$~\GeV, the average
value of \ET for electrons from $Z$ boson decays.

Two effects are investigated to derive uncertainties related to the modelling of the lateral shower shape:
\begin{itemize}
\item The variation of the electron energy response as a function of the shower width in the $\eta$ direction is studied using $Z$
boson decays.
The measured differences between data and simulation are used to derive uncertainties in the energy response for electrons and photons of any transverse energy.
\item To take into account possible differences between electron and photon showers related to the different interaction probabilities
with the material in front of the calorimeter,
the lateral energy leakage in the calorimeter outside the area of the cluster
is studied directly in data and simulation.
From the differences between data and simulation, an uncertainty in the photon energy calibration is derived.
\end{itemize}

\begin{figure}[htb]
\centering
  \includegraphics[width=0.45\textwidth]{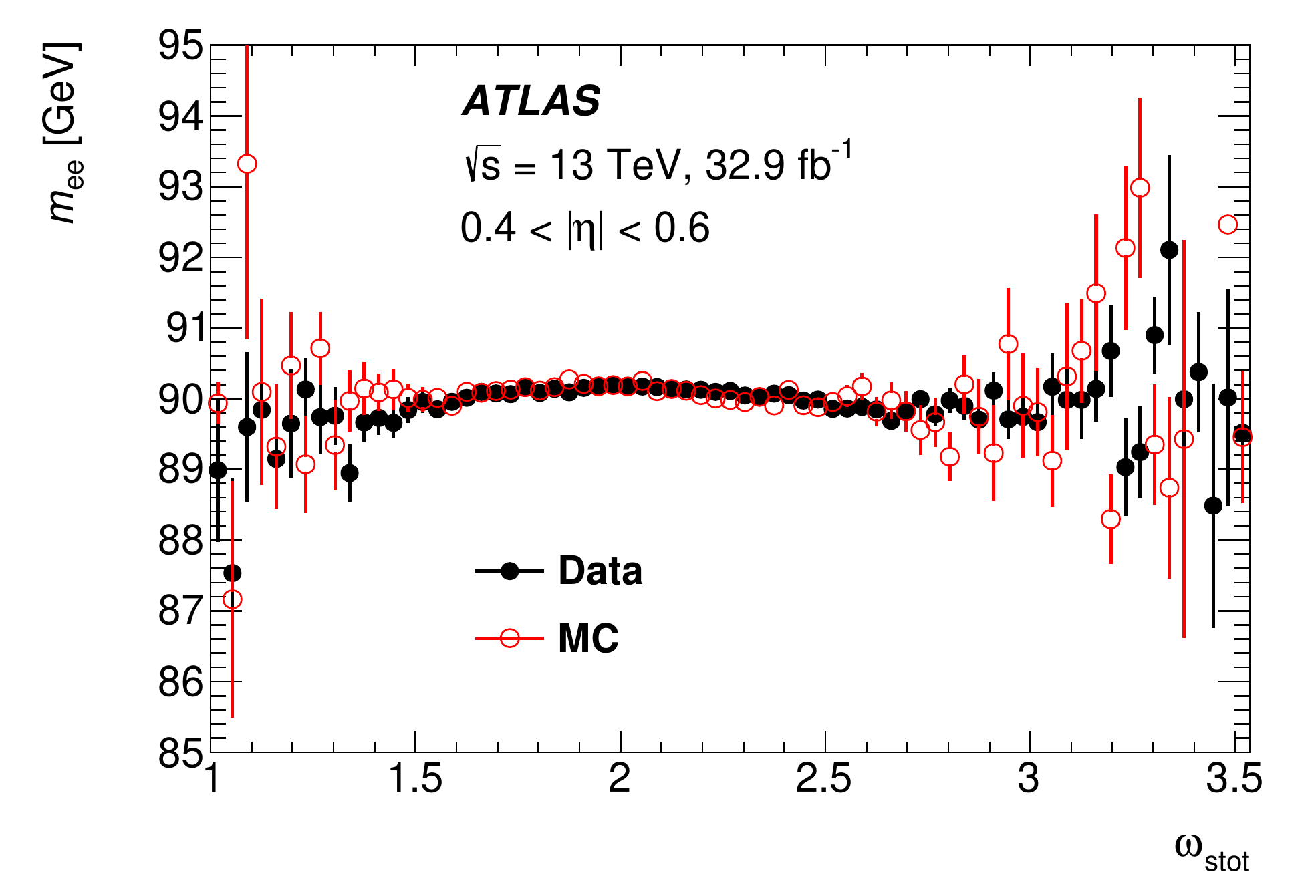} 
  \includegraphics[width=0.45\textwidth]{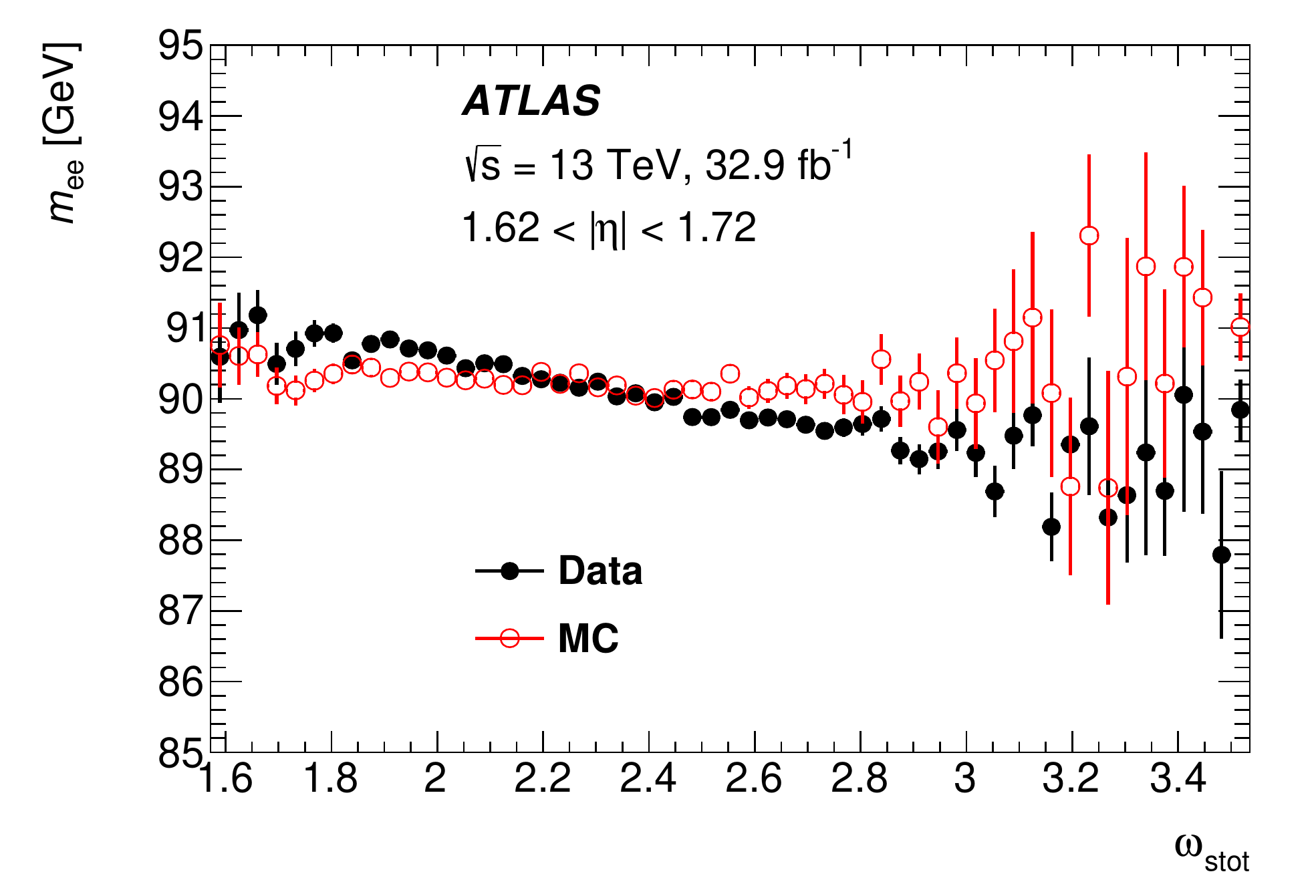}   
  \includegraphics[width=0.45\textwidth]{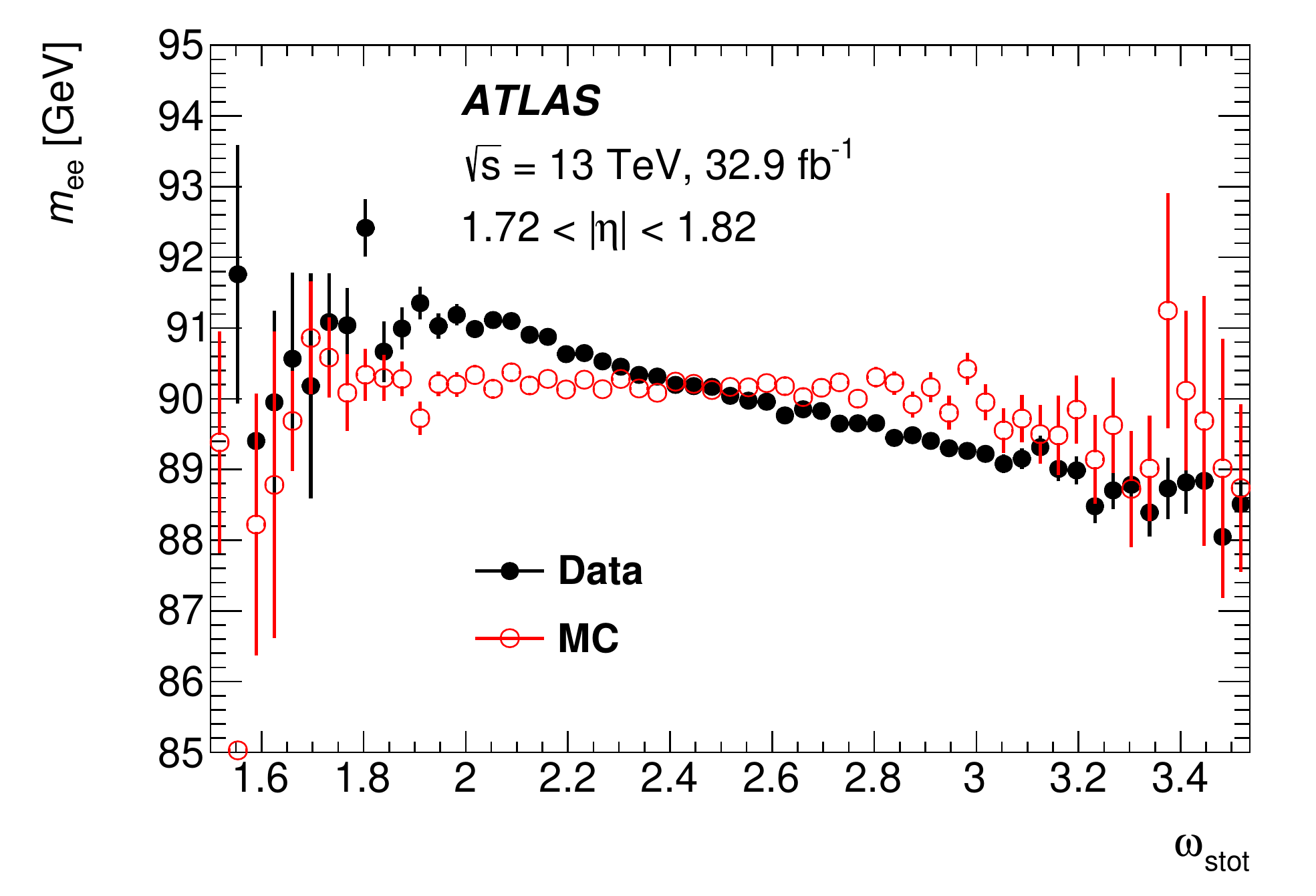} 
  \includegraphics[width=0.45\textwidth]{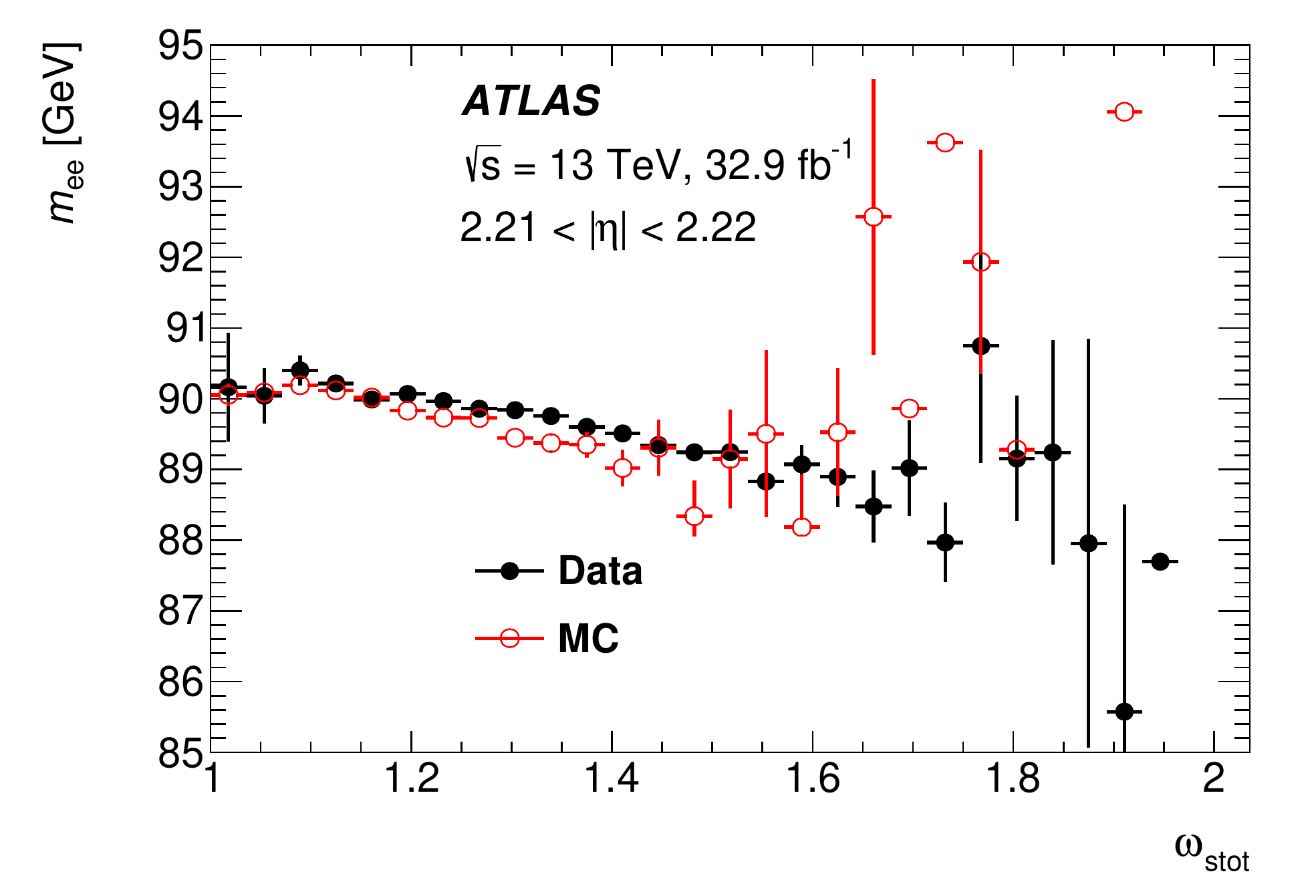}
\caption{The mean of the dielectron mass for $Z$ candidates ($m_{ee}$), in data and simulation, as a function of the lateral shower width ($\wstot$) for different regions in $|\eta|$.}
\label{fig:mZ_wstot}
\end{figure}

To characterize the lateral shower shape in the $\eta$ direction, the measurement of the shower width using first-layer cells ($\wstot$) is used.
The variable $\wstot$ is defined as the RMS of the energy distribution as a function of $\eta$ using all first-layer cells included
in the cluster.
The energy response as a function of the lateral shower shape is investigated by examining the
reconstructed $Z$ mass as a function of $\wstot$, separately for data and simulation.
This is illustrated in Figure~\ref{fig:mZ_wstot}.
In most of the detector, the difference between data and simulation is small albeit not zero. However, in the
region around $|\eta|=1.7$ a large difference between data and simulation is observed. This area is where the
material in front of the calorimeter is the largest. Changes in the amount of material in the simulation
do not, however, reproduce the effect observed in data.
In addition, an overall difference in lateral shower shape, as observed in Refs.~\cite{PERF-2013-04} and~\cite{PERF-2016-01}, can also induce a difference in the energy response, even if the dependence of the energy response on the shower width
is the same in data and simulation.
A systematic uncertainty in the energy response from the dependence of the energy response on the shower width is thus
estimated.

This uncertainty, taking into account the calibration performed with $Z\rightarrow ee$ events, can be estimated as
\begin{equation*}
\begin{aligned}
\frac{\Delta E}{E} (\ET,\eta) =  &
    A\times \Big[ \left< \wstot^{e,\gamma}(\mathrm{data},\ET,\eta) \right> - \left<\wstot^{e}\left(\mathrm{data},\left<\ET^{e(Z\rightarrow ee)}\right>,\eta\right)\right>  \Big]  \\
    &
-  B\times \Big[ \left<\wstot^{e,\gamma}(\mathrm{MC},\ET,\eta)\right> - \left<\wstot^e\left(\mathrm{MC},\left<\ET^{e(Z\rightarrow ee)}\right>,\eta\right)\right>  \Big],
\label{eq:wstotsyst_eq}
\end{aligned}
\end{equation*}

where $A$ ($B$) is the slope of the energy response as a function of $\wstot$ in data (MC simulation)
in a given $\eta$ bin. This slope is extracted from the variation of the dielectron mass as a function of  $\wstot$ in $Z\rightarrow ee$ events.
The variation of $\wstot$ as a function of \ET and $\eta$ for electrons and photons is parameterized
from the $Z\rightarrow ee$ and inclusive photon samples separately for data and MC simulation.
Simulated single-particle MC samples are used to extrapolate the behaviour to the highest energies.

The resulting energy scale uncertainty is significantly smaller than 0.1\% in most of the detector acceptance
except in the region with $|\eta|$ between
1.52 and 1.82 where it is up to 1\% for electrons with $\ET>$~500~\GeV, up to 1.5\% for unconverted photons with  $\ET>$~400~\GeV\
and around 0.5\% for converted photons.

The lateral energy leakage in the calorimeter is estimated as the difference between the energy collected in an area corresponding to $7 \times 11$ second-layer cells
and the energy collected in the cluster size, which is $3 \times 7$ ($5 \times 5$) second-layer cells in the barrel (endcap) calorimeter.
The leakage measured for photons from radiative $Z$ boson decays is compared with the leakage measured for electrons from $Z$
decays. This is done separately for converted and unconverted photons in wide $|\eta|$ regions and for transverse energies below and above 25~\GeV.

\begin{figure}[htb]
  \centering
  \includegraphics[width=0.50\textwidth]{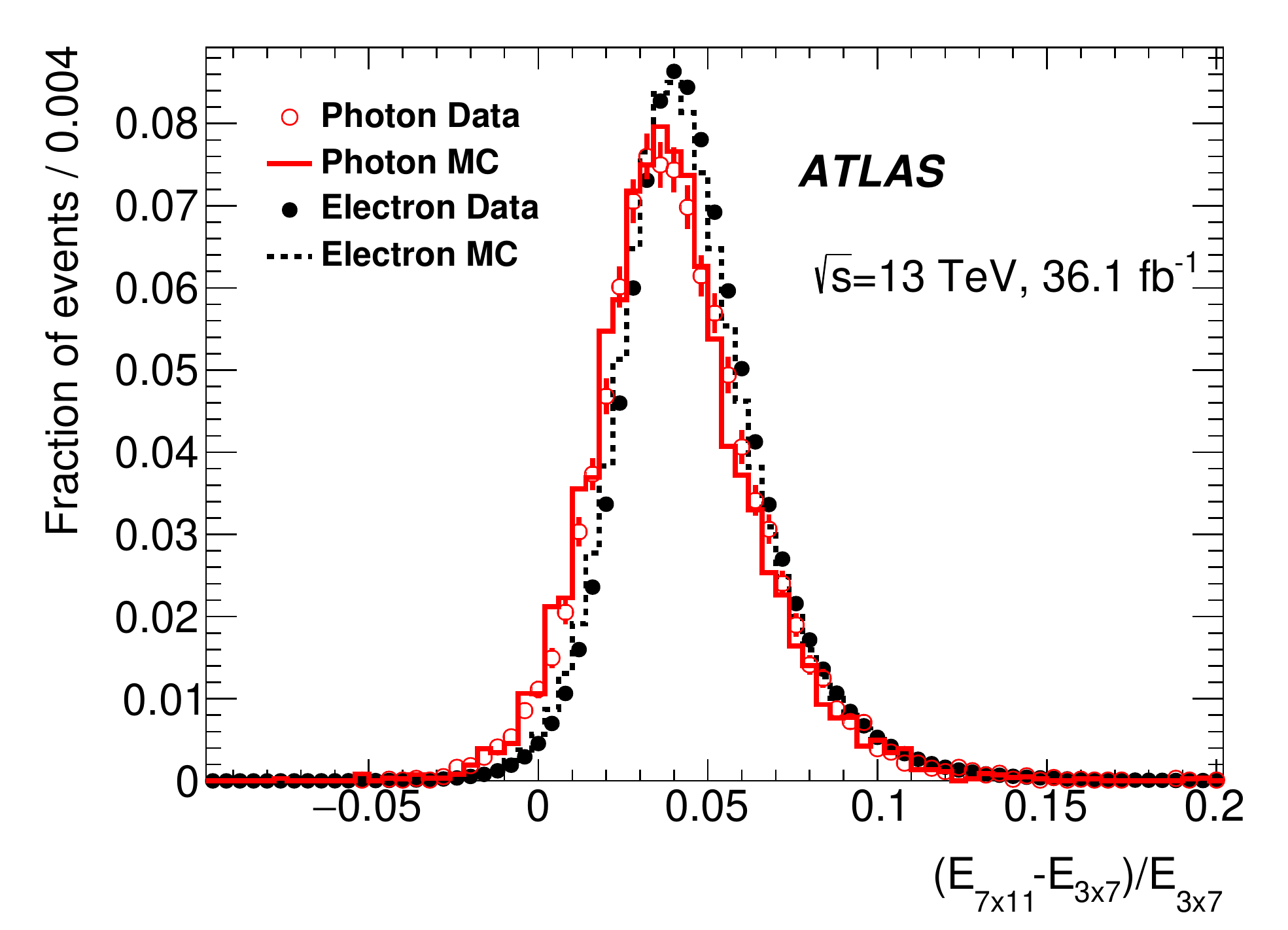}
  \caption{Distributions of the lateral leakage in data and simulation for electron and unconverted photon candidates with \ET~$>$~25~\GeV\ and $|\eta|<0.8$.
 Photons from $Z \rightarrow \ell \ell \gamma$ decays are compared with electrons from $Z\rightarrow ee$.}

 \label{fig:leakage}
\end{figure}

The energy measured in  $7 \times 11$ layer-two cells is corrected for the pile-up-induced energy shifts, in the same way
as for the cluster energy (see Section~\ref{sec:pile-up_correction}).
Figure~\ref{fig:leakage} shows the distribution of the energy leakage for electrons and unconverted photons with \ET~$>$~25~\GeV\
in the $|\eta|$ range 0 to 0.8. The average leakage is larger in the data than in the simulation, which is consistent with a
wider lateral shower shape in data.
The differences between data and simulation are, within statistical uncertainties, mostly consistent between electrons and photons.
To quantify the effect, the double difference of lateral leakage between electrons and photons and between data and simulation is investigated.
The largest deviations of the double difference from zero are observed
for converted photons with $|\eta|<0.8$, with a value of $(0.25\pm0.10_\mathrm{stat})\%$ and for unconverted photons across the full
$\eta$ range where an average value of around $-0.1\%$ is observed.
The impact of the photon conversion reconstruction and the classification between converted and unconverted photon
categories is estimated by applying the procedure discussed in Section~\ref{sec:conversion_reconstruction}.
The measured double differences are taken as additional systematic uncertainties in the photon energy calibration. If the double difference
is consistent with zero within its statistical uncertainty, the statistical uncertainty is taken instead as an estimate
of the systematic uncertainty.

\subsection{Modelling of the photon reconstruction classification}
\label{sec:conversion_reconstruction}

The energy estimating algorithm (Section~\ref{sec:mva1}) is trained separately for reconstructed converted and unconverted photons. Misclassifications in
conversion category arise from inefficiencies in the conversion-finding algorithm and from fake classification
of genuine unconverted photons as converted photons by matching the cluster to pile-up-induced track(s).
For conversions occurring at a radius smaller than 800~mm from the beam line, the typical reconstruction
efficiency, as estimated by the simulation, is 65\% to 85\% depending on $\eta$. The fake rate, i.e.\ the fraction of genuine unconverted photons
reconstructed as converted photon candidates, is typically 1\% to 4\% depending on $\eta$ for the pile-up conditions
of the 2015 and 2016 datasets. The efficiency and fake rate are smaller for $|\eta|>2.0$ where the absence of
transition radiation tracker coverage does not allow reconstruction of photon conversions occurring at large radius.

If the misclassification rate is different between data and simulation, a bias in the photon energy
scale is induced. The efficiency and fake rate are studied using a sample of radiative $Z$ boson decays,
$Z \rightarrow \mu\mu\gamma$~\cite{ATLAS-PERF-2017-02}. The longitudinal shower shape of the photon candidates is used to provide
a statistical discrimination between genuine converted and unconverted photons and to estimate the efficiencies
and fake rate in both data and simulation. The ratio of the efficiencies of the conversion finding algorithm 
in data and simulation is
typically around 0.9. The ratio of fake rates is between 1 and 1.7 depending on $\eta$.
The impact on the photon energy measurement is estimated by reweighting the MC events with the data-to-MC
ratio of efficiencies and fake rates. The uncertainty is taken as the difference between this reweighted MC sample
and the original MC sample. This is done separately for the efficiency and the fake rate, which are treated
as independent uncertainty sources. A change of the conversion-finding efficiency mostly affects the energy
scale of the reconstructed unconverted photon candidates while a change in the fake rate mostly affects
the sample of reconstructed converted photon candidates. For photons with $\ET~=~60$~\GeV, the
uncertainty in the energy scale is about 0.04\% (0.2\% to 0.02\%) in the barrel (endcap) for reconstructed
unconverted photon candidates and about 0.05\% (0.005\%) in the barrel (endcap) for reconstructed converted photon candidates.

\subsection{Summary of systematic uncertainties in the energy scale}

The systematic uncertainties are described by a set of 64 independent uncertainty variations.
A given systematic uncertainty can be described by multiple variations for different regions in $|\eta|$.
The list of these uncertainties is given in Table~\ref{table:systematics_list}.
For simplification, only one uncertainty variation is assigned to the statistical accuracy of the $Z\rightarrow ee$
calibration since this uncertainty is always negligible compared with the other uncertainties.

\begin{table}
\begin{center}
\caption{List of the different independent systematic uncertainties affecting the energy calibration
and their divisions in $|\eta|$ regions between which the uncertainties are not correlated.
Uncertainties with one $|\eta|$ region are fully correlated across the full $\eta$ acceptance.}
\begin{tabular}{llc}
\toprule
\multicolumn{2}{c}{Uncertainty source} & Number of \\
\multicolumn{2}{c}{} & $|\eta|$ regions \\
\midrule
$Z\rightarrow ee$ calibration & Statistical uncertainty & 1 \\
                                & Systematic uncertainty & 1 \\
\midrule
Cell energy non-linearity &  Medium Gain/High Gain layer 2 &  1 \\
                          &  Medium Gain/High Gain layer 1 &  1 \\
                          &  Pile-up shift                  &  1 \\
\midrule
Layer 1/Layer 2 calibration &  $\alpha_{1/2}$ $\mu$ measurement & 5 \\
                            &  $\alpha_{1/2}$ $\mu \rightarrow e$ extrapolation & 2 \\
Presampler calibration      &  Module spread &           8 \\
                            &  Uncertainty for last EMB module & 1 \\
                            &  $b_{1/2}$ correction &    1 \\
Barrel--endcap gap scintillator & Scintillator calibration & 3 \\
($1.4<|\eta|<1.6$) & &  \\
\midrule
ID material                 & Run 1 detector construction & 4 \\
                            & Run 2 inner most pixel layer description & 1 \\
                            & Pixel services description               & 1 \\
Material presampler (PS) to calorimeter  & Run 1 measurement with unconv. photon & 9  \\
($|\eta|<1.8$)                      & Simulation of long. shower shape unconv. photon & 2 \\
Material ID to presampler           & Run 1 measurement with electrons &  9    \\
($|\eta|<1.8$)                      & Simulation of long. shower shape electrons   & 2 \\
Material ID to calorimeter          & Run 1 measurement with electrons &  3  \\
($|\eta|>1.8$)                      & Simulation of long. shower shape electrons   & 1 \\
All material ID to calorimeter      & Variations of GEANT4 physics list & 1 \\
\midrule
Lateral shower shape modelling      & Dependence on shower $\eta$ width & 1 \\
                                    & Lateral leakage for unconv. photons & 1  \\
                                    & Lateral leakage for conv. photons & 1 \\
\midrule
Conversion reconstruction           & Conversion efficiency &  1 \\
                                    & Conversion fake rate  &  1 \\
                                    & Radius dependence of conversion reconstruction & 1 \\
\bottomrule
\end{tabular}
\label{table:systematics_list}
\end{center}
\end{table}

\begin{figure}[htb]
 \centering
          \mbox{
            \begin{tabular}{c}
            \subfloat[\label{fig:syst_elec1}]{\includegraphics[width=0.53\figwidth]{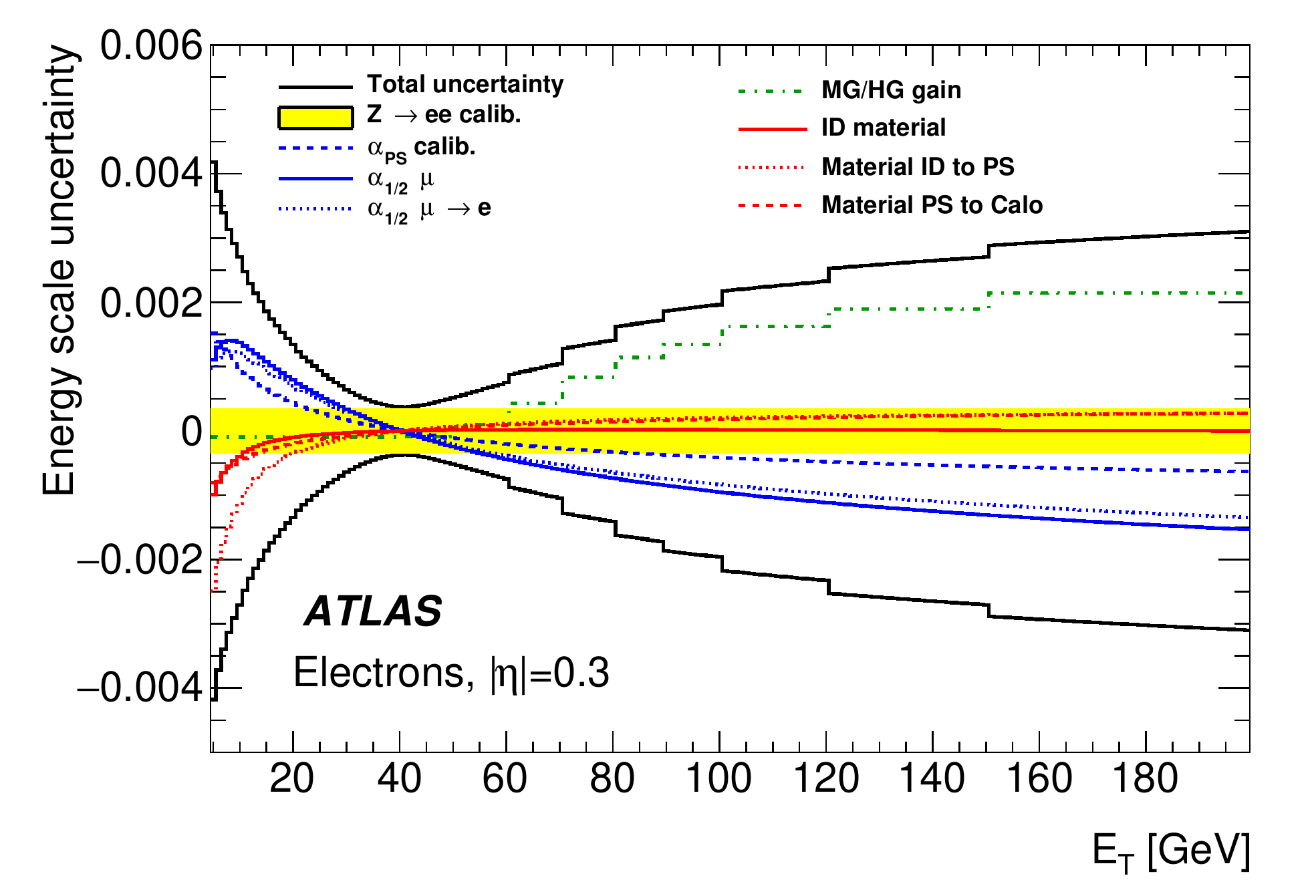}} \\
            \end{tabular}
           }
           \mbox{
            \begin{tabular}{cc}
             \subfloat[\label{fig:syst_unco1}]{\includegraphics[width=0.53\figwidth]{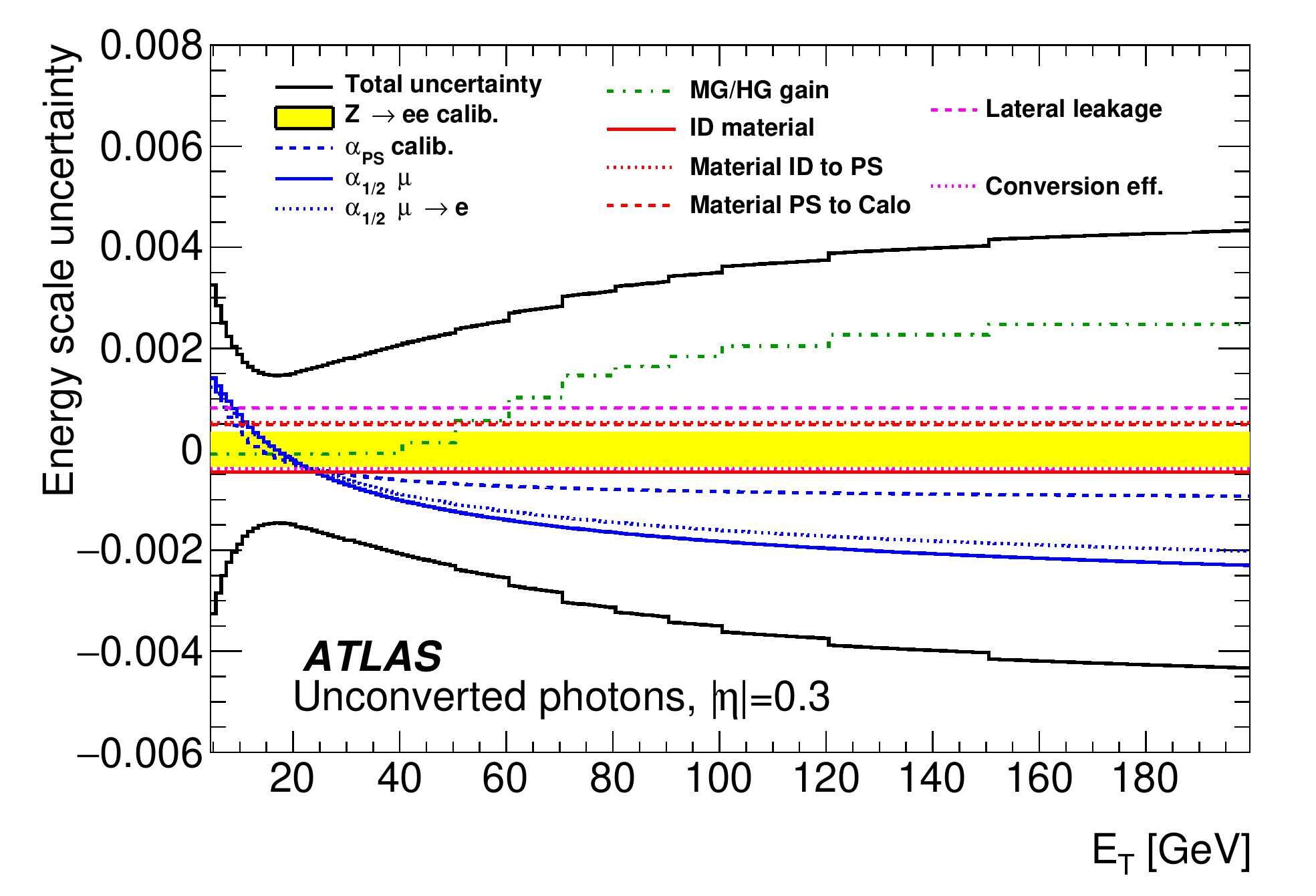}}   &
            \subfloat[\label{fig:syst_conv1}]{\includegraphics[width=0.53\figwidth]{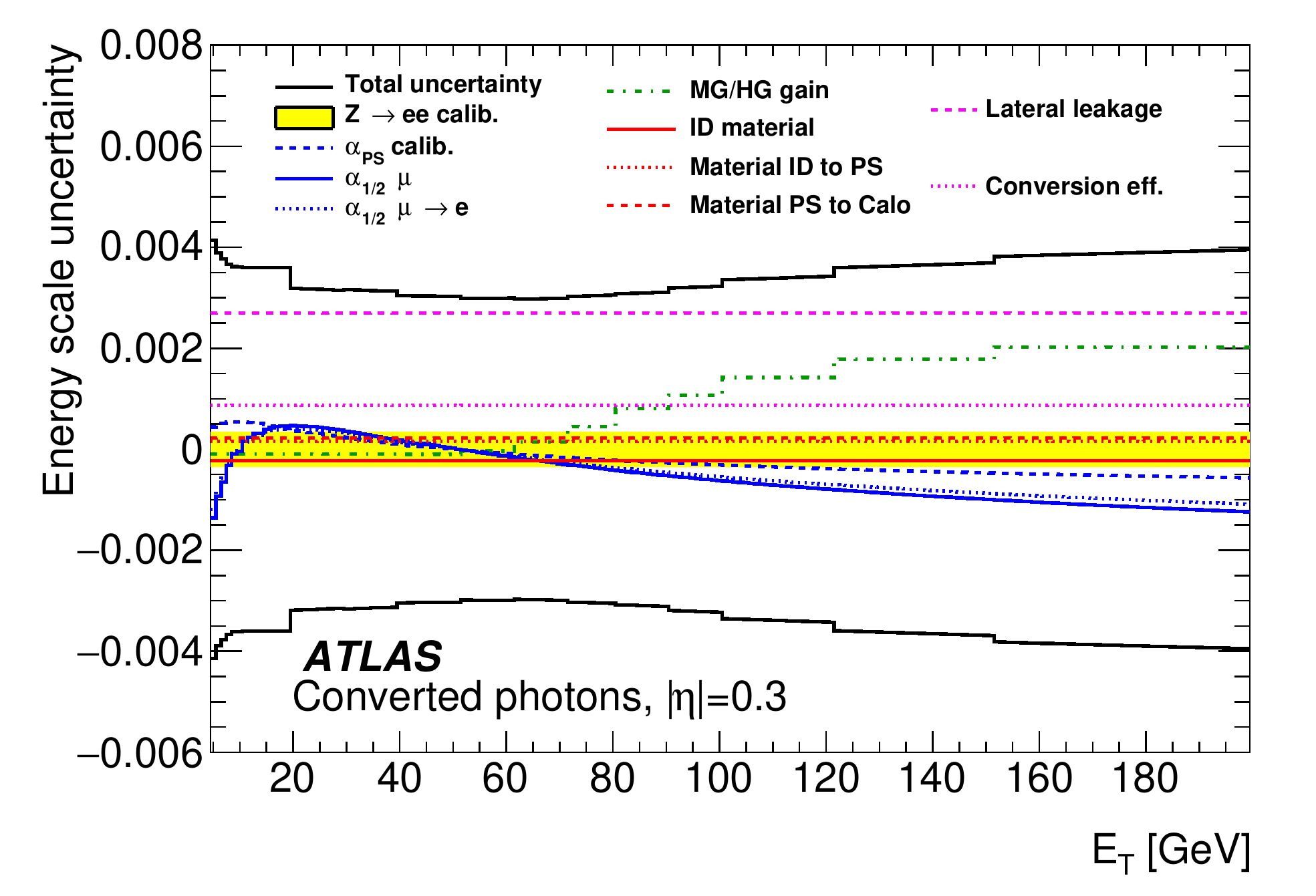}} \\
            \end{tabular}
               }
  \caption{Fractional energy scale calibration uncertainty for (a) electrons, (b) unconverted photons and (c) converted photons,
 as a function of \ET for $|\eta|=0.3$. The total
 uncertainty is shown as well as the main contributions, which are represented by the signed impact of a one-sided
variation of the corresponding uncertainty. Only a one-sided variation for each uncertainty is shown for clarity.}
 \label{fig:sys_elec}
\end{figure}

Figure~\ref{fig:sys_elec} illustrates the impact of the main systematic uncertainties affecting the energy scale of
electrons, unconverted photons and converted photons at $|\eta|=0.3$ as a function of the transverse energy.
For a given uncertainty variation, the effect
on the energy can be positive or negative with a possible change of sign near the average \ET of electrons from $Z$
decays. This is illustrated in this figure by showing the signed uncertainty, i.e.\ the impact of a one-sided variation
of the systematic uncertainties. The opposite-sign variation will give a systematic impact with the opposite sign.
Keeping track of the relative sign across \ET and $\eta$ of the impact of each uncertainty source is important
for properly computing the total uncertainty for a sample covering a range of \ET and $\eta$ values.
At given values of both \ET and $\eta$, the total systematic uncertainty is given by the sum in quadrature of the
uncertainties related to each of the independent uncertainty sources.

\begin{figure}[htb]
          \centering
          \mbox{
            \begin{tabular}{cc}
            \subfloat[\label{fig:syst_summary1}]{\includegraphics[width=0.5\figwidth]{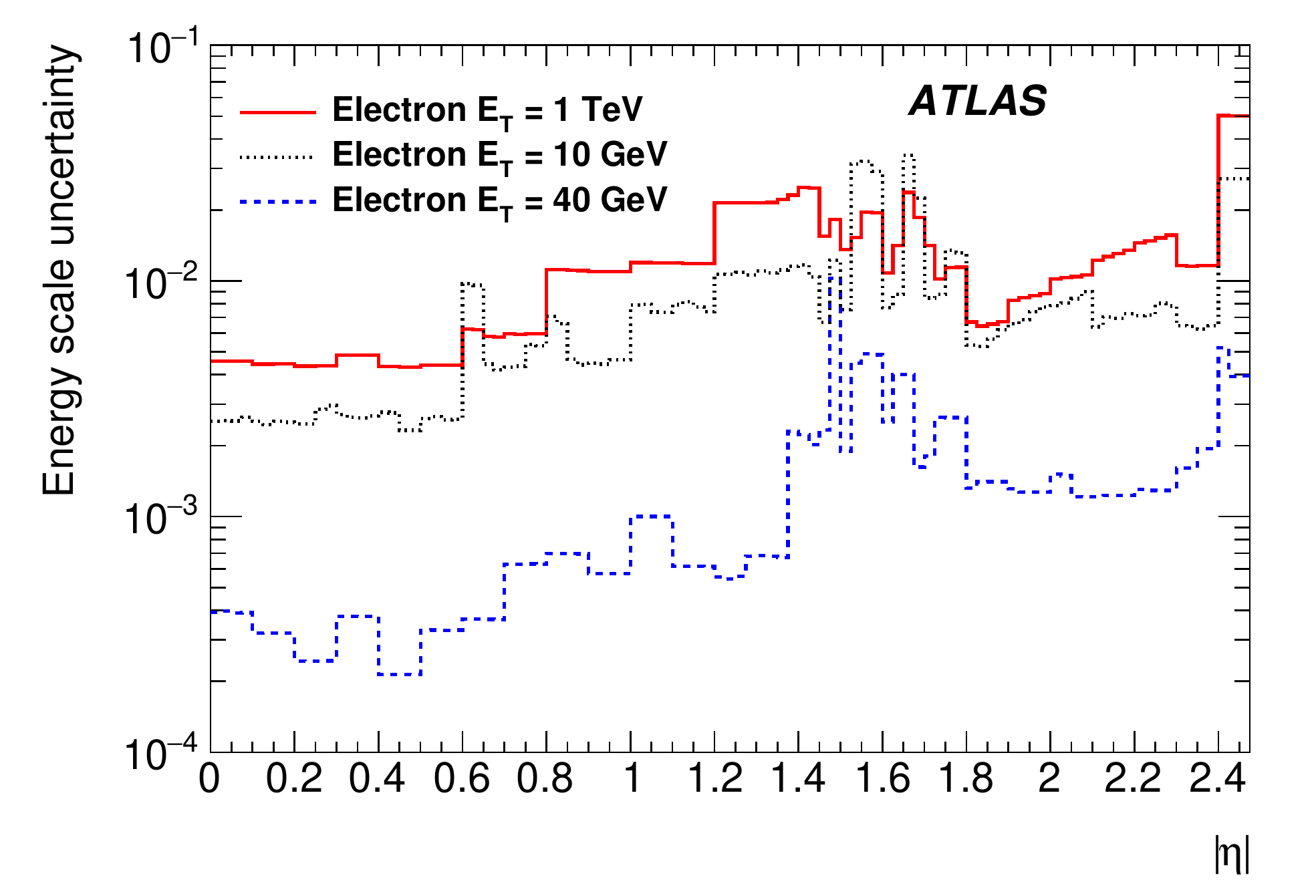}} &
            \subfloat[\label{fig:syst_summary2}]{\includegraphics[width=0.5\figwidth]{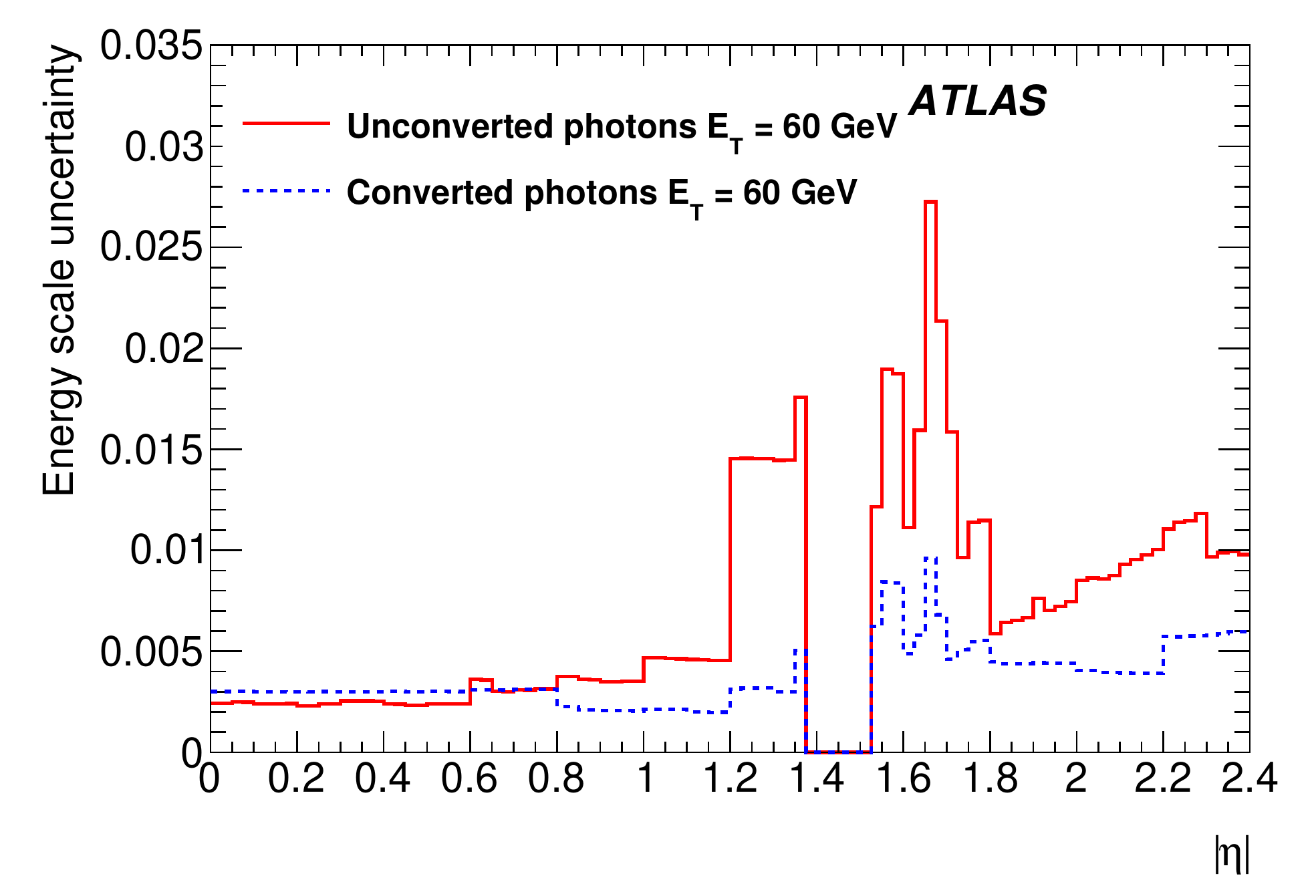}}
            \end{tabular}
               }
        \caption{Total fractional systematic uncertainty in the energy scale as a function of $|\eta|$ for (a) electrons of
 \ET~=~10~\GeV, 40~\GeV\ and 1~\TeV\ and (b) photons of \ET~=~60~\GeV.}
        \label{fig:syst_summary}
\end{figure}

Figure~\ref{fig:syst_summary} summarizes the total uncertainty in the energy scale as a function of $\eta$
for electrons and photons for given values of transverse energy. Uncertainties for converted and unconverted photons
are shown separately.

To illustrate the $\eta$-dependence of the different uncertainties,
Table~\ref{table:photon_systematics_summary} gives
the photon energy scale systematic uncertainties for \ET~=~60~\GeV\ in wide $\eta$ regions corresponding to either the barrel
or the endcap acceptance. A uniform $\eta$ distribution of the photons is assumed.
The typical photon energy scale uncertainty is 0.2\% to 0.3\% averaged over the barrel and  0.45\% to 0.8\% in the endcap.
For this value of \ET, the uncertainties from the relative calibration of the different layers is significantly smaller for
converted photons than for unconverted photons as they have a longitudinal shower development closer to that of \ET~=~40~\GeV\
electrons. The cell energy non-linearity uncertainty is higher for unconverted photons as they have a higher probability to use
medium-gain readout in the second layer, given that they deposit a higher energy fraction in the second calorimeter layer.
The total uncertainties are only partially correlated  between converted and unconverted photons.

\begin{table}
\begin{center}
\caption{Photon energy scale fractional systematic uncertainty for a sample with uniform $\eta$ distribution at \ET~=~60~\GeV, with the
contributions of the different types of uncertainties.}
\adjustbox{max width=\textwidth}{
\begin{tabular}{lcccc}
\toprule
 Systematic category & \multicolumn{4}{c}{Photon energy scale uncertainty $\times 10^3$} \\
 \cmidrule(l){2-5}
                     & \multicolumn{2}{c}{$|\eta| < 1.37$}   &  \multicolumn{2}{c}{$1.52 < |\eta| < 2.37$} \\
\cmidrule(l){2-5}
                     & Unconverted & Converted &  Unconverted & Converted \\
\midrule
$Z\rightarrow ee$ calib.   &                      0.45   &   0.45  &  1.41  &  1.41 \\
Cell energy non-linearity &                         0.88   &   0.10  &  3.89  &  0.38 \\
Layer (presampler, E1/E2, scintillator) calibration & 2.34   &   0.29  &  3.04  &  0.60 \\
ID material &                                       0.96   &   0.82  &  3.71  &  3.89 \\
Other material &                                    1.66   &   0.26  &  3.19  &  1.02 \\
Conversion reconstruction      &                    0.40   &   0.99  &  0.76  &  0.97 \\
Lateral shower shape modelling &                    1.03   &   1.95  &  3.20  &  0.85 \\
\midrule
Total                          &                    3.37   &   2.41  &  7.81  &  4.50  \\
\bottomrule
\end{tabular}
}
\label{table:photon_systematics_summary}
\end{center}
\end{table}

\subsection{Energy resolution uncertainties}

The different contributions to the energy resolution are: the shower and sampling fluctuations in the calorimeter,
the fluctuations in energy loss upstream of the calorimeter, the effect of electronics and pile-up noise and the
impact of residual non-uniformities affecting the measurement of the energy in the data.
The total contributions of the effects of shower and sampling fluctuations, energy loss before the calorimeter
and electronics noise are given in Section~\ref{sec:mva}.
The intrinsic energy resolution, defined as the expected resolution in the absence of upstream material
and with uniform response,
is derived from the energy resolution in the simulation of genuine unconverted photons.
A 10\% relative uncertainty is assumed for this intrinsic energy resolution, based on
test-beam studies~\cite{ctb}.
The impact of uncertainties in the detector material upstream of the calorimeter on the energy resolution
is derived from simulations with additional material as described
in Section~\ref{sec:material_uncertainty}.
The uncertainty in the electronics and pile-up noise modelling
is derived from a comparison of data and simulation for a sample of zero-bias events introduced in Section~\ref{sec:E1E2}.
The noise is typically 350--400~\MeV\ expressed in transverse energy. The noise uncertainty is defined
as the difference in quadrature between the noise in data and simulation and is found to be 100~\MeV\ in terms
of transverse energy.
Finally, the energy resolution's constant term is derived from the data-to-simulation comparison
of the energy resolution for electrons from $Z\rightarrow ee$ decays, as described in Section~\ref{sec:alpha}.

A formalism similar to that for the energy scale uncertainty can be used to describe the resolution uncertainties.
If $\Delta\sigma_i^{e,\gamma}(\ET,\eta)$
is the uncertainty in the relative energy resolution for a given particle from a given uncertainty variation $i$, the residual
uncertainty after the adjustment of the resolution based on the $Z$ decays can be written as
\begin{equation*}
\delta\Sigma_i^{e,\gamma}(\ET,\eta) = \Delta\Sigma_i^{e,\gamma}(\ET,\eta) - \Delta\Sigma_i\left(\left<\ET^{e(Z\rightarrow ee)}\right>,\eta\right),
\label{equation:resolution_syst}
\end{equation*}
where $\Sigma$ denotes the square of the relative energy resolution $\sigma$.

The uncertainty in the energy resolution comparison between data and simulation for $Z\rightarrow ee$ decays
is described by an additional uncertainty in the constant term of the
energy resolution.

\begin{figure}[htb]
          \centering
          \mbox{
            \begin{tabular}{cc}
            \includegraphics[width=0.5\figwidth]{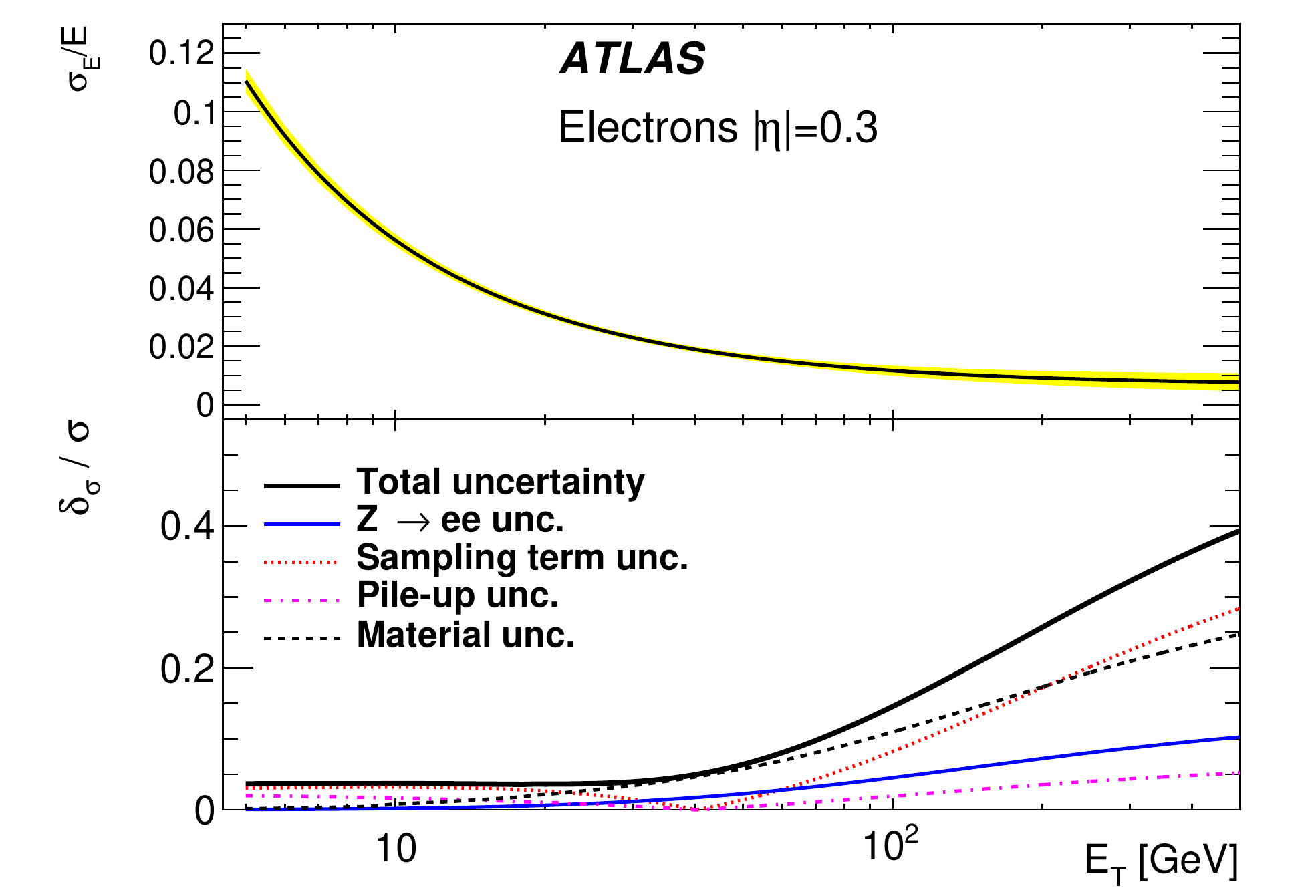}  &
            \includegraphics[width=0.5\figwidth]{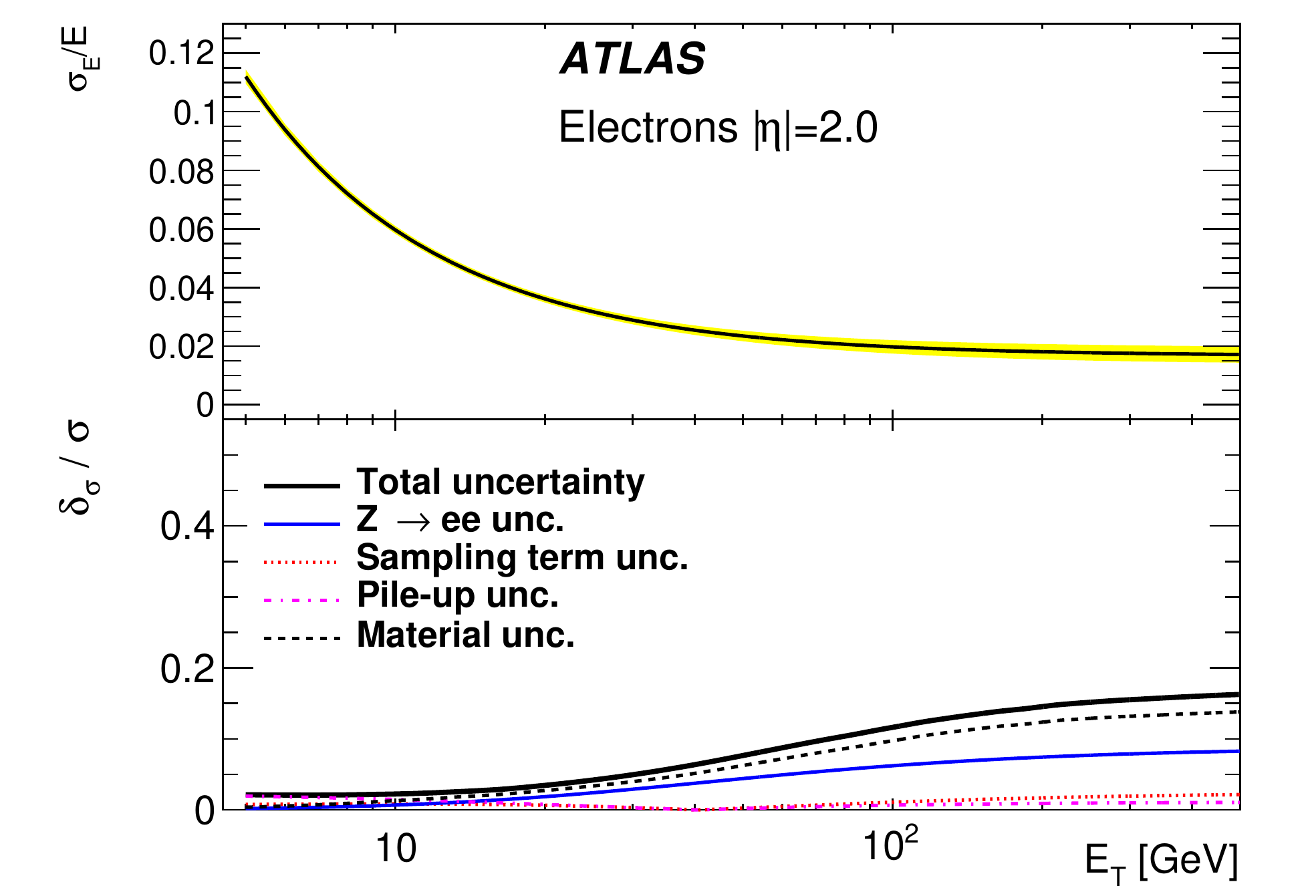} \\
            \includegraphics[width=0.5\figwidth]{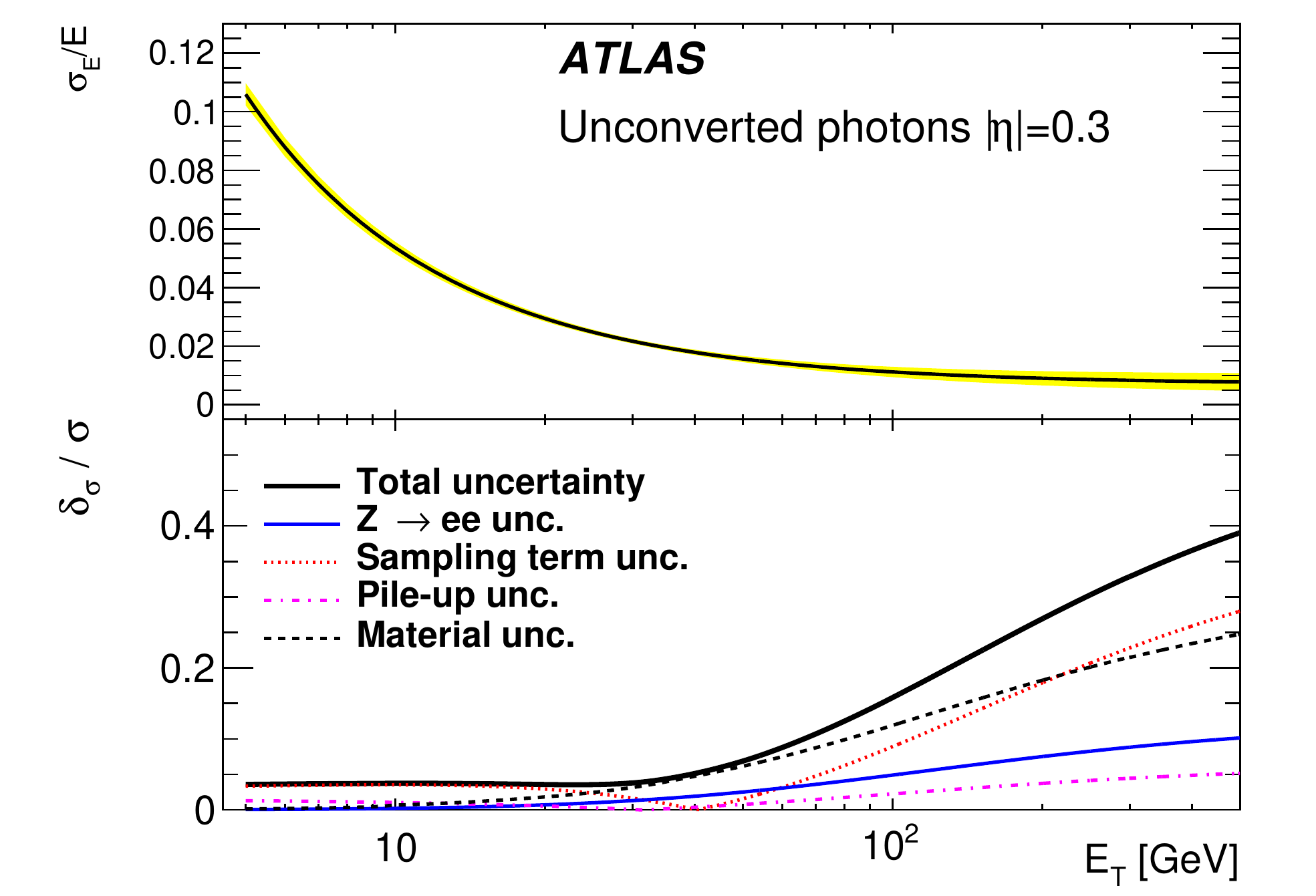} &
            \includegraphics[width=0.5\figwidth]{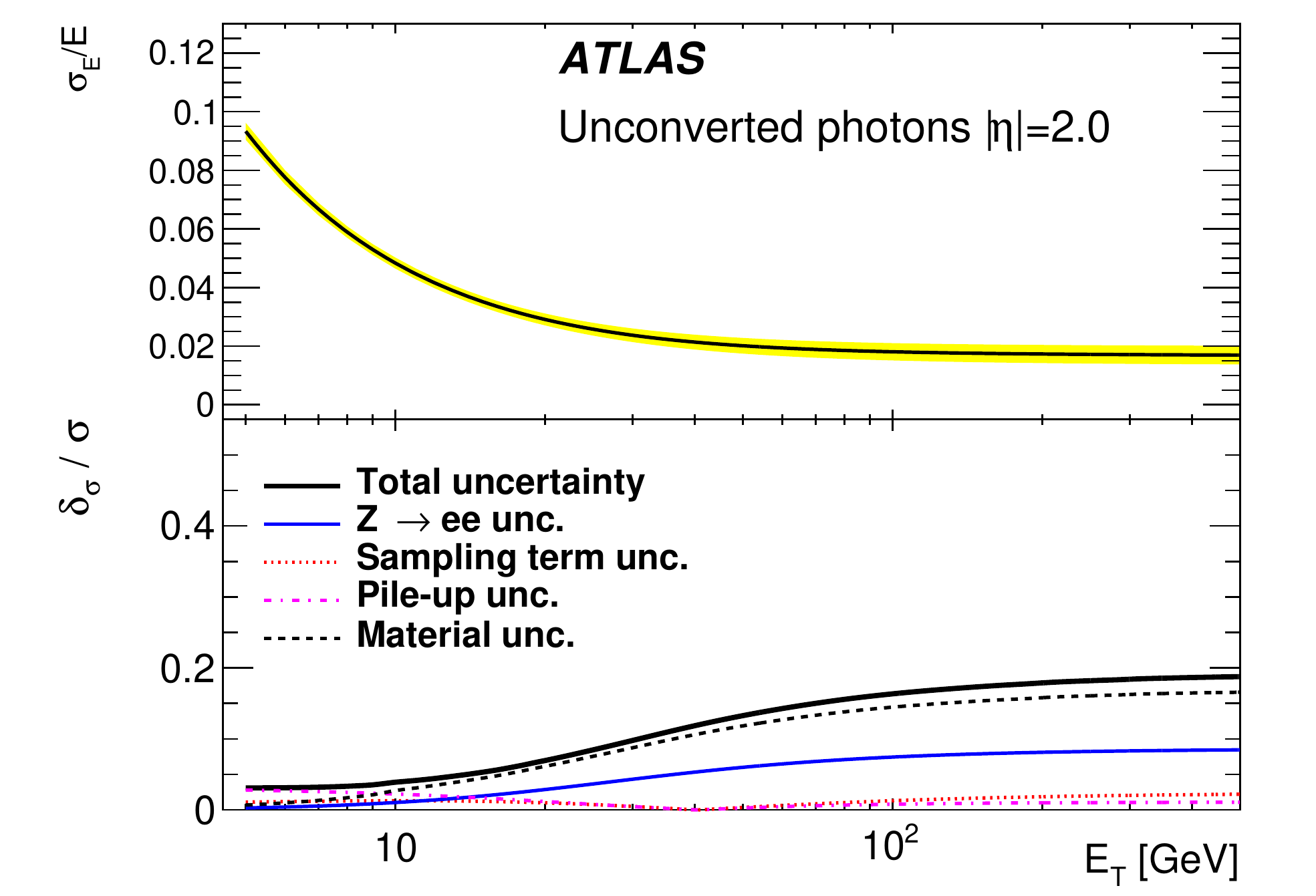} \\
            \end{tabular}
               }
        \caption{Relative energy resolution, $\sigma_E/E$, as a function of \ET for electrons and unconverted photons at $|\eta|=0.3$ and
 $|\eta|=2.0$. The yellow band in the top panels shows the total uncertainty in the resolution. The breakdown
of the relative uncertainty in the energy resolution, $\delta_{\sigma}/\sigma$ is shown in the bottom panels.}
        \label{fig:resolution_uncertainty}
\end{figure}

Figure~\ref{fig:resolution_uncertainty} shows the energy resolution, its total uncertainty and the different contributions to the
total relative uncertainty in the resolution as a function of transverse energy for electrons and unconverted photons at two
different $\eta$ values.
The uncertainty $\Delta\Sigma_i^{e,\gamma}(\ET,\eta)$ due to the material in front of the calorimeter
is estimated as the change of the core Gaussian
component of the energy resolution in simulated single-particle samples
with different amounts of material in front of the calorimeter.
The term $\Delta\Sigma_i\left(\left<\ET^{e(Z\rightarrow ee)}\right>,\eta\right)$ is computed from simulated $Z\rightarrow ee$ samples.
Energy resolution corrections are derived by comparing samples simulated with additional material with the nominal geometry simulation, following the same
procedure as used for the data and discussed in Section~\ref{sec:alpha}.

For electrons or photons in the transverse energy range 30--60~\GeV, the energy resolution is known to a precision of the
order of 5\% to 10\%. For high-energy electrons or photons, where the resolution is better, the relative uncertainty in the
energy resolution reaches 20\% to 50\%.  Compared with the results reported in Ref.~\cite{PERF-2013-05}, the main change is the smaller
uncertainty in the constant term of the energy resolution extracted from the $Z\rightarrow ee$ samples.
This uncertainty reduction is mainly due to an improvement of the validation step performed on pseudo-data as discussed in 
Section~\ref{sec:zee_method} and from better agreement between the two methods considered.

\section{Cross-checks with $J/\psi \rightarrow ee$ and $Z\rightarrow \ell\ell\gamma$ decays}
\label{sec:checks}
\subsection{$J/\psi \rightarrow ee$ decays}

The energy scale of low-energy electrons (average transverse energy around 10~\GeV) is probed using $J/\psi \rightarrow ee$ events.
The known mass of the $J/\psi$ resonance provides a completely independent check
of the energy calibration for low-energy electrons.
The full calibration procedure discussed in the previous sections,
including the energy scale derived from $Z\rightarrow ee$ events,
is applied.
The difference between data and simulation for  $J/\psi \rightarrow ee$ events is then quantified
using residual energy scale differences, $\Delta\alpha$, extracted from the peak positions of the reconstructed
invariant mass. The formalism is very similar to the one used for the $Z\rightarrow ee$ data-to-simulation
energy scale corrections, but fewer $|\eta|$ regions are defined, given the smaller size of the collected
 $J/\psi \rightarrow ee$ sample.
If the energy calibration is correct, $\Delta\alpha$ should be consistent with zero within the combined
uncertainties of the $J/\psi \rightarrow ee$ measurement and the systematic uncertainty of the energy
calibration.  The event selection and data and MC samples are introduced in Section~\ref{sec:samples}.

To compare data and simulation, the relative fraction of $J/\psi$ produced in $b$-hadron decays compared
to promptly produced $J/\psi$ is determined, since the electrons from $J/\psi$ produced in $b$-hadron decays are
less isolated. This fraction is extracted from a fit to the proper decay-time distribution, following
the procedure discussed in Ref.~\cite{BPHY-2010-01}. The fraction of prompt $J/\psi$
is found to be between 68\% and 83\% depending on the \ET requirement imposed on the electron
with highest \ET, with uncertainties between 3\% and 14\%.

To extract the energy scale differences between data and simulation from $J/\psi$ events, a procedure similar
to the simultaneous fit used for $Z\rightarrow ee$ events, described in Section~\ref{sec:zee_method}, is applied. The significant
contributions from the continuum background and the $\psi(2S)$ resonance have to be taken into account.
The typical signal-to-background ratio integrated over the 2.6--3.4~\GeV\ mass range, which contains
most of the signal, is around 10 to 1.

The dielectron invariant mass distribution in the range 2.1 to 4.1~\GeV\ is
described by the following function:
\begin{equation}
f(m_{ee}) = f^{\text{DSCB}}_{J/\psi}(m_{ee}) + f^{\text{DSCB}}_{\psi(2S)}(m_{ee}) + f^{\text{bkg}}(m_{ee})
\end{equation}
with the $J/\psi$ and $\psi(2S)$ mass distributions described by a double-sided Crystal Ball function ($f^{\text{DSCB}}$)
and the background mass distribution ($f^{\text{bkg}}$) described by a second-order Chebyshev  polynomial.
Since in the simulation the continuum background is not considered, the last term is used only when fitting data.

The parameters describing the $\psi(2S)$ mass distribution are related to the ones
describing the $J/\psi$ mass distribution by a scaling factor equal to the ratio of the masses of these
two resonances.
All the parameters but the  $m_{ee}$ peak position are fixed to the DSCB parameters extracted
from the MC samples. This free parameter is expressed as a function of $\Delta\alpha$.
The $\Delta\alpha$ factors are extracted
from a simultaneous fit of the different $i$-$j$ data regions in $\eta$. The normalizations of the $J/\psi$,
$\psi(2S)$ and background yields as well as the parameters describing the background shape are also free in the fit.

The systematic uncertainties affecting the extraction  of $\Delta\alpha$ include the uncertainties
in the shape of the signal mass distribution (choice of DSCB function and parameters
of the DSCB functions), in the modelling of the background mass distribution, in the
results of the proper-time fit and in the modelling of the $\eta$
distribution of the electrons in the simulation.
These systematic uncertainties are significantly smaller than the statistical uncertainties.

\begin{figure}[htb]
 \centering
 \includegraphics[width=0.50\textwidth]{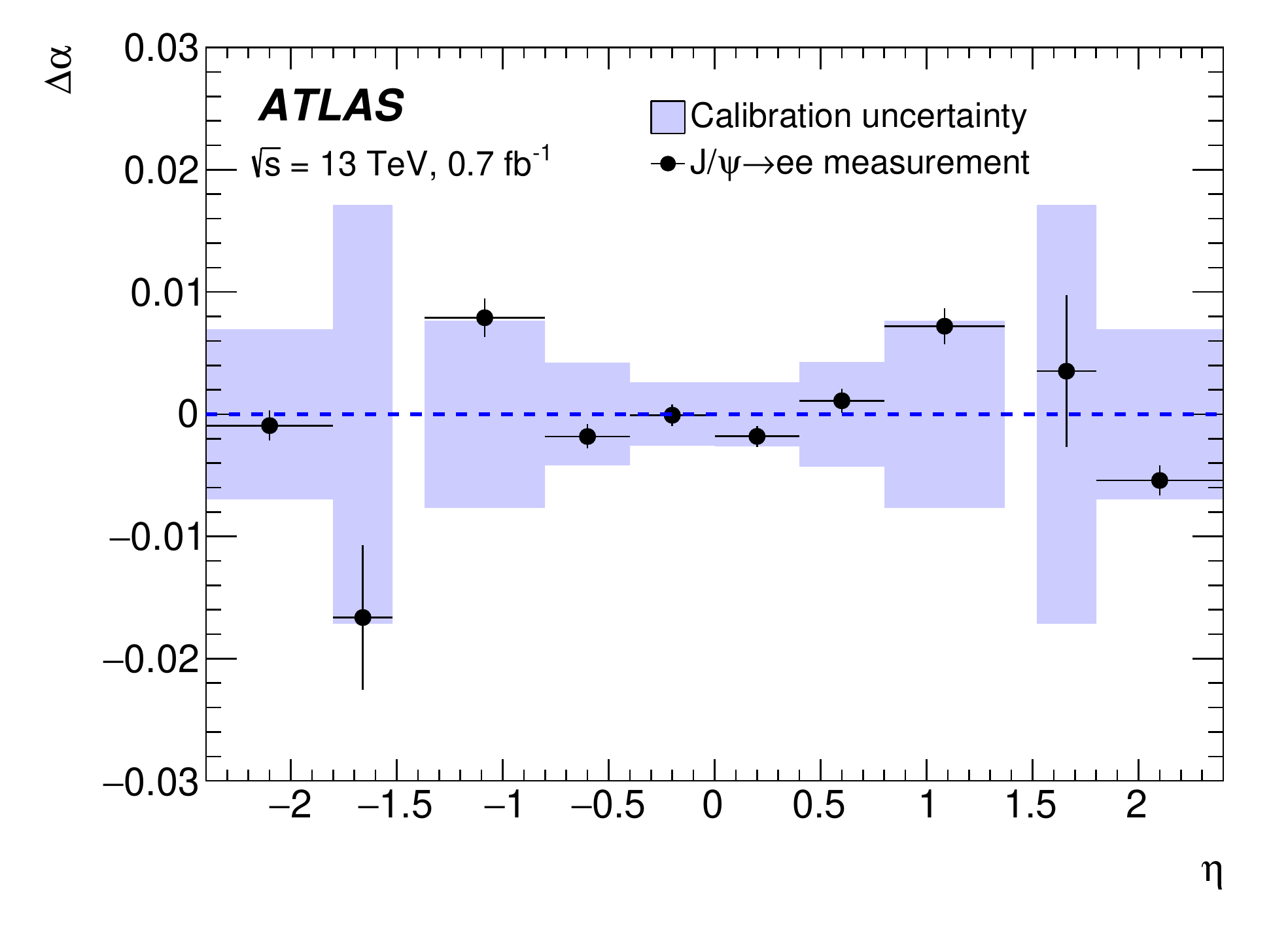}
  \caption{Residual energy scale differences, $\Delta\alpha$, between data and simulation extracted
from $J/\psi \rightarrow ee$ events as a function of $\eta$. The points show the measurement
with its total uncertainty. The band shows the uncertainty of the energy calibration for the energy range
of the $J/\psi \rightarrow ee$ decays.}
 \label{fig:jpsi-scales}
\end{figure}

Figure~\ref{fig:jpsi-scales} shows the extracted  $\Delta\alpha$ values with their uncertainties as
a function of $\eta$.  
They are compared with the systematic uncertainty of the calibration procedure
for electrons with the \ET  distribution of those observed in the $J/\psi$ sample.
No measurement is reported in the transition region between the barrel and
endcap calorimeters due to the limited measurement accuracy in this region.
The uncertainty in the calibration for low-\ET electrons arises mostly from uncertainties in the
amount of material in front of the calorimeter
and in the relative calibration of the different calorimeter layers.
Good agreement is observed between the residual energy scale differences
and the calibration described in this paper.
This agreement confirms that the method to extract the nominal scales and the estimate of the systematic uncertainties
are valid over a wide range of electron energies.

The width of the reconstructed  $J/\psi \rightarrow ee$ mass distribution can also be used to probe the energy resolution
for low-energy electrons. The observed width in the data is consistent with the predicted resolution within its uncertainties.

\subsection{$Z\to\ell\ell\gamma$ decays}
The energy scale correction of photons is assumed to be the same as the one extracted from $Z\to ee$ decays, as described in Section~\ref{sec:alpha},
within the uncertainties described in Section~\ref{sec:systematics}.
The radiative decays of the $Z$ boson can be used to check the energy scale of photons, in the low-energy region in particular.
Converted and unconverted photons are studied separately. The electron and muon channels are treated independently and then combined.
All the corrections previously described are applied to electrons and photons,
and residual energy scale factors for photons are derived by comparing the data with simulations.
The samples of simulated events and the selection are described in Section~\ref{sec:samples}.

The residual photon energy scale difference is parameterized as an additional correction to the photon energy 
$\Delta\alpha$, similarly to the $J/\psi$ study.
The mass distribution of the $\ell\ell\gamma$ system in the simulation is modified by applying $\Delta\alpha$ to the photon energy and
the value of  $\Delta\alpha$ that minimizes the $\chi^2$ comparison between the data and the simulation is computed.
A second method based on an analytic function adjusted to the simulation to describe the shape of the mass distribution is also
investigated. The two methods give consistent results.

The measurement is limited by the statistical accuracy.
The considered systematic uncertainties are from the lepton energy scale and the background contamination, and they 
are negligible compared with the statistical uncertainty.

\begin{figure}[htbp]
 \subfloat[\label{fig:alpha_photon_unconv}]{\includegraphics[width=0.5\textwidth]{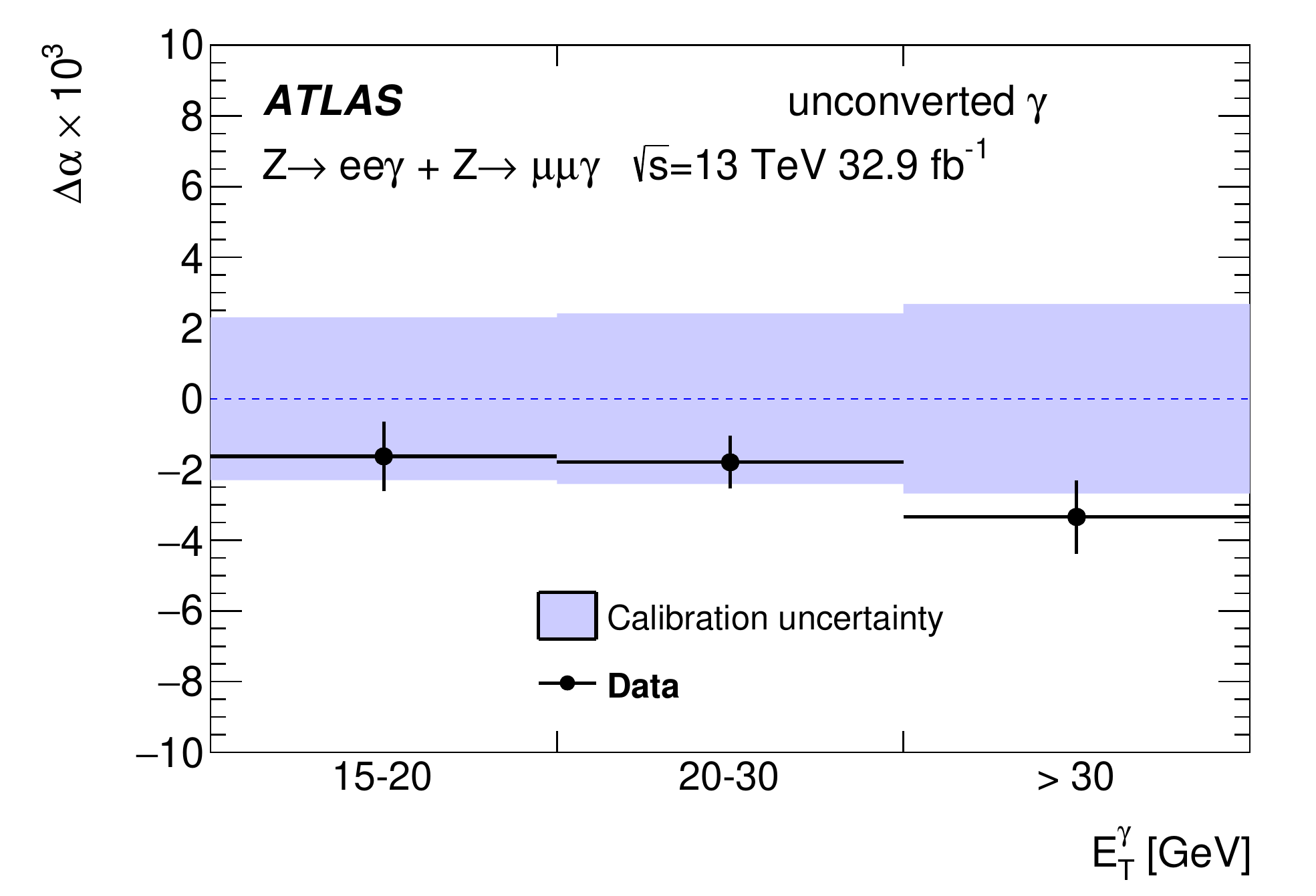}}
 \subfloat[\label{fig:alpha_photon_conv}]{\includegraphics[width=0.5\textwidth]{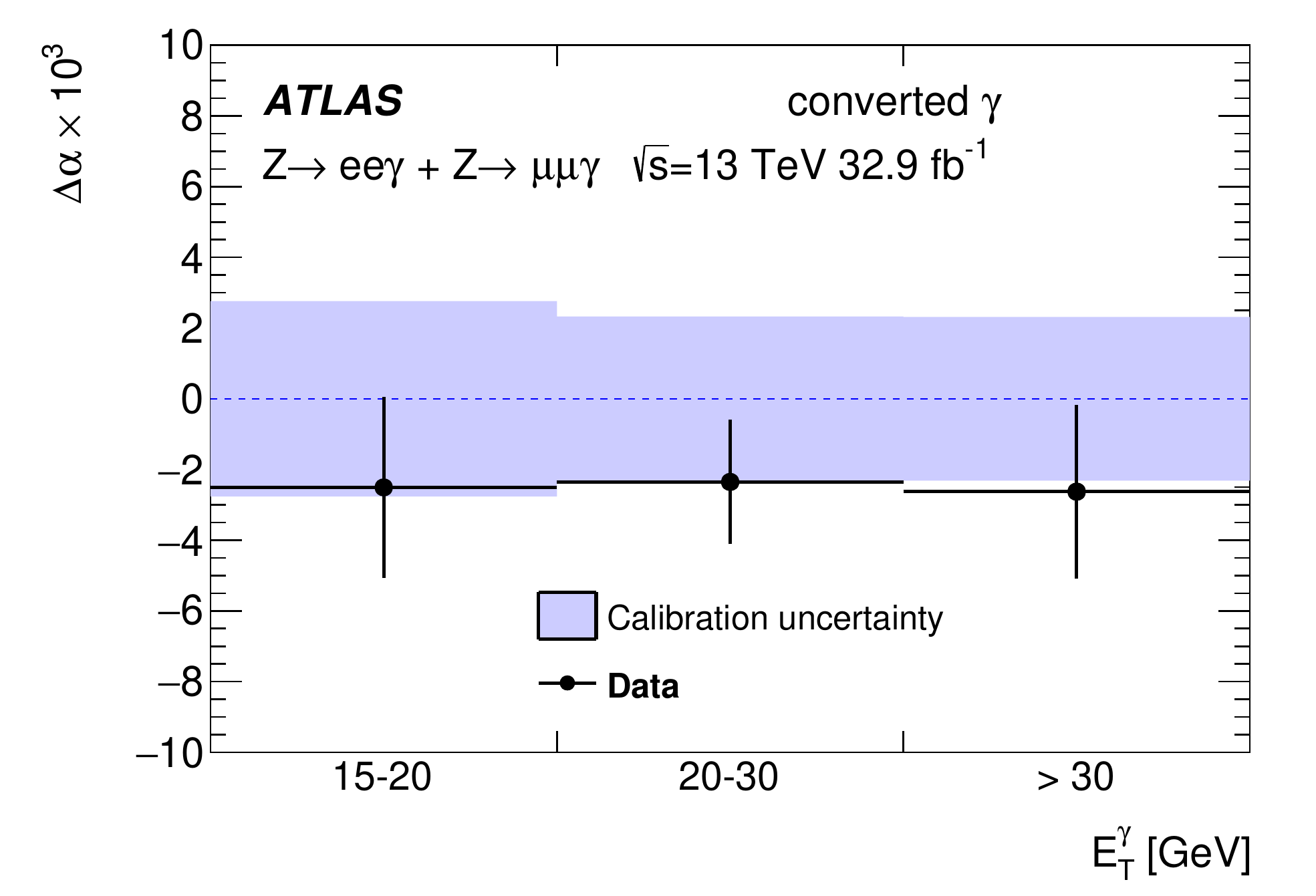}}
 \caption{Residual energy scale factor, $\Delta\alpha$, for (a) unconverted and (b) converted photons with their uncertainties. 
 The points show the measurement with its total uncertainty.
 The superimposed band represents the full energy calibration uncertainty for photons from $Z\to\ell\ell\gamma$ decays.}
 \label{fig:alpha_photon}

\end{figure}

Figure~\ref{fig:alpha_photon} shows the measured $\Delta\alpha$ as a function of \ET , separately for converted and unconverted photons.
The value of $\Delta\alpha$ is consistent with zero within the uncertainties in the measurement and in the photon energy scale.

\section{Summary}
\FloatBarrier 
\label{sec:conclusions}
The calibration of the energy measurement of electrons and photons
collected with the ATLAS detector during 2015 and 2016 using about 36~fb$^{-1}$ of LHC
proton$-$proton collisions at $\sqrt{s}=13$~\TeV\ is presented.
The estimate of the energy is optimized in simulation using variables
related to the shower development in the calorimeter and the properties
of the photon conversions. In the transition region between the
barrel and endcap calorimeters, the energy resolution is improved
using information provided by gap scintillators.

To achieve good linearity of the energy response, an accurate intercalibration
of the different longitudinal layers of the calorimeter is required.
The intercalibration of the first two layers of the calorimeter is derived
from a study of muon energy deposits. Despite the moderate signal-to-noise
ratio for muon energy deposits at the luminosity where the data are collected, an accuracy of 0.7\% to 2.5\%
is achieved for this measurement. The calibration of the presampler layer is
derived from a detailed study of electron and photon showers, with an accuracy
varying between 1.5\% and 3\%. The impact of pile-up on the energy measurement
is investigated and small effects are corrected.

The overall calorimeter energy scale is set from a large
sample of $Z\rightarrow ee$ events, comparing the invariant mass distribution
in data and simulation. 
Differences between data and simulation for the energy resolution are derived and energy scale corrections are extracted.
The accuracy of the energy scale measurement varies from
0.03\% to 0.2\% depending on  $|\eta|$. The constant term of the energy resolution in the
data is less than 1\% in the barrel calorimeter and typically 1--2\% in the endcap calorimeter.

The calorimeter energy scale is found to be stable with time and with changes in luminosity, with
effects of up to few per mille observed in the endcap calorimeter and less than one per mille
in the barrel calorimeter.

Uncertainties in the amount and location of material in front of the calorimeter are mostly the same as in the Run~1
studies. The impact of the detector components that were changed before data taking started in 2015 is
investigated and small additional uncertainties in the energy response are derived.

Uncertainties in the linearity of the measurement of the energy deposited in each
calorimeter readout cell are estimated using data collected with special settings in 2015
and 2017 and found to be at the few per mille level for most of the calorimeter acceptance.

From these measurements, the energy calibration and its total uncertainty are
derived for electrons and photons at all energies.
The systematic uncertainty in the energy scale
calibration is about 0.03\% to 0.2\% in most of the detector acceptance
($|\eta|<2.5$) for electrons with transverse momentum
close to 45~\GeV. For electrons with transverse momentum of 10~\GeV\ the typical uncertainty is
0.3\% to 0.8\% and it is about 0.25\% to 1\% for photons with transverse momentum around 60~\GeV.
This energy calibration was used for the Higgs boson mass measurement performed by the ATLAS Collaboration
using the two-photon and four-lepton decay channels with data collected in 2015 and 2016~\cite{HIGG-2016-33}.

The accuracy of this calibration is probed with low-energy electrons from $J/\psi \rightarrow ee$
events and with photons from radiative $Z$ boson decays and good agreement is found.

\section*{Acknowledgements}

We thank CERN for the very successful operation of the LHC, as well as the
support staff from our institutions without whom ATLAS could not be
operated efficiently.

We acknowledge the support of ANPCyT, Argentina; YerPhI, Armenia; ARC, Australia; BMWFW and FWF, Austria; ANAS, Azerbaijan; SSTC, Belarus; CNPq and FAPESP, Brazil; NSERC, NRC and CFI, Canada; CERN; CONICYT, Chile; CAS, MOST and NSFC, China; COLCIENCIAS, Colombia; MSMT CR, MPO CR and VSC CR, Czech Republic; DNRF and DNSRC, Denmark; IN2P3-CNRS, CEA-DRF/IRFU, France; SRNSFG, Georgia; BMBF, HGF, and MPG, Germany; GSRT, Greece; RGC, Hong Kong SAR, China; ISF and Benoziyo Center, Israel; INFN, Italy; MEXT and JSPS, Japan; CNRST, Morocco; NWO, Netherlands; RCN, Norway; MNiSW and NCN, Poland; FCT, Portugal; MNE/IFA, Romania; MES of Russia and NRC KI, Russian Federation; JINR; MESTD, Serbia; MSSR, Slovakia; ARRS and MIZ\v{S}, Slovenia; DST/NRF, South Africa; MINECO, Spain; SRC and Wallenberg Foundation, Sweden; SERI, SNSF and Cantons of Bern and Geneva, Switzerland; MOST, Taiwan; TAEK, Turkey; STFC, United Kingdom; DOE and NSF, United States of America. In addition, individual groups and members have received support from BCKDF, CANARIE, CRC and Compute Canada, Canada; COST, ERC, ERDF, Horizon 2020, and Marie Sk{\l}odowska-Curie Actions, European Union; Investissements d' Avenir Labex and Idex, ANR, France; DFG and AvH Foundation, Germany; Herakleitos, Thales and Aristeia programmes co-financed by EU-ESF and the Greek NSRF, Greece; BSF-NSF and GIF, Israel; CERCA Programme Generalitat de Catalunya, Spain; The Royal Society and Leverhulme Trust, United Kingdom. 

The crucial computing support from all WLCG partners is acknowledged gratefully, in particular from CERN, the ATLAS Tier-1 facilities at TRIUMF (Canada), NDGF (Denmark, Norway, Sweden), CC-IN2P3 (France), KIT/GridKA (Germany), INFN-CNAF (Italy), NL-T1 (Netherlands), PIC (Spain), ASGC (Taiwan), RAL (UK) and BNL (USA), the Tier-2 facilities worldwide and large non-WLCG resource providers. Major contributors of computing resources are listed in Ref.~\cite{ATL-GEN-PUB-2016-002}.



\printbibliography

 \newpage
 
\begin{flushleft}
{\Large The ATLAS Collaboration}

\bigskip

M.~Aaboud$^\textrm{\scriptsize 34d}$,    
G.~Aad$^\textrm{\scriptsize 99}$,    
B.~Abbott$^\textrm{\scriptsize 125}$,    
D.C.~Abbott$^\textrm{\scriptsize 100}$,    
O.~Abdinov$^\textrm{\scriptsize 13,*}$,    
B.~Abeloos$^\textrm{\scriptsize 129}$,    
D.K.~Abhayasinghe$^\textrm{\scriptsize 91}$,    
S.H.~Abidi$^\textrm{\scriptsize 164}$,    
O.S.~AbouZeid$^\textrm{\scriptsize 39}$,    
N.L.~Abraham$^\textrm{\scriptsize 153}$
% ...
\bigskip
\\

% ...
$^{13}$Institute of Physics, Azerbaijan Academy of Sciences, Baku; Azerbaijan.\\
% ...
$^{34}$$^{(a)}$Facult\'e des Sciences Ain Chock, R\'eseau Universitaire de Physique des Hautes Energies - Universit\'e Hassan II, Casablanca;$^{(b)}$Centre National de l'Energie des Sciences Techniques Nucleaires (CNESTEN), Rabat;$^{(c)}$Facult\'e des Sciences Semlalia, Universit\'e Cadi Ayyad, LPHEA-Marrakech;$^{(d)}$Facult\'e des Sciences, Universit\'e Mohamed Premier and LPTPM, Oujda;$^{(e)}$Facult\'e des sciences, Universit\'e Mohammed V, Rabat; Morocco.\\
% ...
$^{39}$Niels Bohr Institute, University of Copenhagen, Copenhagen; Denmark.\\
% ...
$^{91}$Department of Physics, Royal Holloway University of London, Egham; United Kingdom.\\
% ...
$^{99}$CPPM, Aix-Marseille Universit\'e, CNRS/IN2P3, Marseille; France.\\
$^{100}$Department of Physics, University of Massachusetts, Amherst MA; United States of America.\\
% ...
$^{125}$Homer L. Dodge Department of Physics and Astronomy, University of Oklahoma, Norman OK; United States of America.\\
% ...
$^{129}$LAL, Universit\'e Paris-Sud, CNRS/IN2P3, Universit\'e Paris-Saclay, Orsay; France.\\
% ...
$^{153}$Department of Physics and Astronomy, University of Sussex, Brighton; United Kingdom.\\
% ...
$^{164}$Department of Physics, University of Toronto, Toronto ON; Canada.\\
% ...
$^{a}$ Also at Borough of Manhattan Community College, City University of New York, NY; United States of America.\\
$^{b}$ Also at California State University, East Bay; United States of America.\\
$^{c}$ Also at Centre for High Performance Computing, CSIR Campus, Rosebank, Cape Town; South Africa.\\
$^{d}$ Also at CERN, Geneva; Switzerland.\\
$^{e}$ Also at CPPM, Aix-Marseille Universit\'e, CNRS/IN2P3, Marseille; France.\\
% ...
$^{*}$ Deceased

\end{flushleft}


\end{document}